\documentclass[10pt,a4paper]{article}

\usepackage[utf8]{inputenc}
\usepackage[T1]{fontenc}
\usepackage{lmodern}
\input{glyphtounicode}
\usepackage[margin=2.4cm]{geometry}
\usepackage{graphicx}
\usepackage{amsmath,amssymb}
\usepackage{booktabs}
\usepackage{caption}
\usepackage{enumitem}
\usepackage{xcolor}
\usepackage[hidelinks]{hyperref}
\usepackage{microtype}

\graphicspath{{figures/}}
\setlist{itemsep=1pt,topsep=2pt}

\newcommand{\leff}{\lambda_{\mathrm{eff}}}
\newcommand{\kperp}{k_{\perp}}
\newcommand{\kcell}{k_{\mathrm{cell}}}
\newcommand{\lams}{\lambda_{s}}
\newcommand{\lamf}{\lambda_{f}}
\newcommand{\lamg}{\lambda_{g}}
\newcommand{\lamb}{\lambda_{b}}
\newcommand{\Kn}{\mathrm{Kn}}
\newcommand{\WmK}{\,\mathrm{W\,m^{-1}\,K^{-1}}}
\newcommand{\Mzero}{\ensuremath{\mathsf{M0}}}
\newcommand{\Mone}{\ensuremath{\mathsf{M1}}}
\newcommand{\Mtwo}{\ensuremath{\mathsf{M2}}}
\newcommand{\MR}{\ensuremath{\mathsf{M_R}}}
\newcommand{\capY}{\textcolor{green!55!black}{\checkmark}}
\newcommand{\capP}{\textcolor{orange!85!black}{\ensuremath{\sim}}}
\newcommand{\capN}{\textcolor{black!35}{--}}

\title{\textbf{Capturing the calendering u-shape in lithium-ion electrode thermal conductivity}\\[5pt]
{\large\normalfont A differentiable, calendering-aware Zehner--Bauer--Schl\"under closure, with a bounded contact-versus-reorientation test for the graphite anode}}

\author{Julius St\"ork$^{1,\ast}$\\[2pt]
\small $^1$VARTA Microbattery GmbH, Ellwangen, Germany\\
\small $^\ast$Corresponding author: \texttt{julius.stoerk@varta-ag.com}}
\date{10.07.2026}

\begin{document}
\maketitle

\begin{abstract}
\noindent
Battery thermal models need the through-plane conductivity $\leff$ of every porous layer, yet its u-shaped dependence on calendering has lacked an analytical closure that explains it, because porosity-only and static-contact closures capture at most a monotonic rise. Our primary result is a calendering-aware closure in which a compression-indexed contact term $\varphi(\Pi)$ on a Knudsen-corrected Zehner--Bauer--Schl\"under base reproduces the u-shape and reduces the in-sample mean absolute percentage error across 27 calendering states from $31.1\%$ to $4.5\%$, commensurate with the estimated uncertainty of the reconstructed LFA target. The held-out leave-one-state-out error is $7.8\%$. Because each family provides only six to eight states for four fitted parameters, we report identifiable parameter groups and profile-likelihood valleys. The solid conductivity $\lams$ and the bridge fraction $\varphi_0$ are partially confounded, whereas their product, the bridge conductance, is robustly identified. The calibrated target is the apparent through-plane coating conductivity inferred from laser-flash analysis (LFA) after source subtraction, and because recalibrating the closure on guarded-hot-plate data shifts the absolute scale by $1.8$ to $2.4\times$, the absolute parameters are method-scoped. As a secondary, anode-specific result we give a bounded reorientation account, in which flake alignment collapses the through-plane solid conductivity toward the graphite $c$-axis, explaining the anomalously low fitted $\lams$, but which remains degenerate with a contact-only fit until same-sheet texture is measured. Finally, as extensions, the differentiable closure inverts for porosity quality control with calibrated uncertainty and remains consistent under independent checks (an 18650 cell, gas-diffusion layers, dry electrodes, ionic transport).
\end{abstract}

\vspace{1ex}

\section{Introduction}
\label{sec:intro}

Thermal management strongly influences lithium-ion battery lifetime and safety. Elevated temperatures accelerate the side reactions that consume capacity, temperature \emph{gradients} across a cell make its regions age at different rates, and in the worst case a local hot spot can trigger thermal runaway \cite{bandhauer2011,richter2017}. Battery cell design thus requires thermal \emph{models}, which span three-dimensional cell-level finite-element discharge simulations\cite{guo2010,jeon2011,yue2017}, layered and reduced-order electro-thermal models for fast thermal management and cold-start control \cite{li2026,choi2026}, to coupled electrochemical-thermal microstructure models \cite{chen2017}. Every such model needs the effective through-plane thermal conductivity $\leff$ of each layer in the stack as an input. Through-plane conduction is the bottleneck. Cell-level cross-plane values of $0.15$--$0.8\WmK$ lie roughly two orders of magnitude below the in-plane values \cite{richter2017,steinhardt2022}, so heat generated in the interior must cross many poorly conducting porous layers before it can leave the cell.

$\leff$ also depends strongly on processing. Calendering, the roll-pressing step that compacts a coated electrode to its target porosity, changes the conductivity substantially: in the measurements of Gandert et al.\ \cite{gandert2023}, calendering a graphite anode from porosity $\varepsilon=0.60$ down to $\varepsilon=0.21$ raised the measured electrode conductivity from about $1.5$ to $2.4\WmK$, a change of roughly $60\%$. Treating $\leff$ as a fixed handbook value therefore introduces a process-dependent error which, once wetting state and porosity are both accounted for, approaches an order of magnitude.

Decades of research on heat conduction in packed beds and porous media have produced closures, compact formulas that predict $\leff$ from the constituent conductivities and the pore structure, and the Zehner--Bauer--Schl\"under (ZBS) family \cite{zehner1970,bauer1978,vdi} is the most widely used among them. For battery electrodes specifically, Oehler et al.\ \cite{oehler2021,oehlerdiss} have already built a ZBS-based analytical (and companion numerical) closure that predicts coating $\leff$ from microstructure, \emph{includes} the Smoluchowski/Knudsen rarefied-gas correction for the micrometre pores (reducing helium from $0.15$ to $0.03$--$0.06\WmK$), treats inter-particle contact area as a microstructure parameter, and is validated by LFA from coating to cell stack. Building on that foundation, we add three capabilities.

\emph{(i) Process resolution.} Gandert et al.\ \cite{gandert2023} measured four electrode types across 27 calendering states and found a \emph{non-monotonic, u-shaped} dependence of $\leff$ on porosity. Existing electrode models, including Oehler's static-contact closure \cite{oehler2021} and a discrete-element simulation \cite{sangros2017}, reproduce only a monotonic rise, because they account for pore volume but assume a fixed solid structure and miss the mechanism that inverts the curve. We find that calendering changes at least two structures. For the quasi-isotropic lithium nickel manganese cobalt oxide (NMC) cathode, contact-network evolution reproduces the curve as the as-coated bridge network shears and then rebuilds by interlocking. For the anisotropic graphite anode, a bounded reorientation account explains the anomalously low fitted $\lams$, since platelets ($c$-axis $\lambda_c=6\WmK$, basal plane $\lambda_a=150\WmK$) rotate toward the in-plane orientation, so the through-plane path crosses their weak $c$-axis even as densification raises the base conductivity. Conductivity data alone leave this account degenerate with a contact-only fit (Sec.~\ref{sec:results:mechanisms}), and the porosity-only closures benchmarked here represent neither process.

\emph{(ii) Inverse use.} Existing electrode closures are usually used as forward predictors. A differentiable implementation lets the same equations infer coating porosity or interface quality from a fast thermal measurement, which matters because interface quality dominates the thermal resistance of a cell stack \cite{vishwakarma2015} and has no inline monitor at all.

\emph{(iii) Quantified uncertainty.} Most existing closures provide point predictions without quantified uncertainty, whereas any use in quality control needs calibrated intervals and a clear statement of which parameter combinations the data actually constrain.

\paragraph{Primary contribution: a calendering-aware closure for the u-shape.} A compression-indexed contact term $\varphi(\Pi)$ on the Knudsen-corrected ZBS base reproduces the u-shaped calendering dependence of $\leff$ (Secs.~\ref{sec:model:contact}, \ref{sec:results:mechanisms}); within the analytical closure hierarchy tested here, it is the first low-dimensional variant we found that reproduces the measured u-shape across all four calibrated families rather than a monotonic rise. An ablation attributes the in-sample error reduction $31.1\%\to13.5\%\to4.5\%$ to the contact components (Sec.~\ref{sec:results:ablation}), the held-out leave-one-state-out error is $7.8\%$ (Sec.~\ref{sec:results:errors}), and the widely applicable information criterion (WAIC) supports the process-dependent term (Sec.~\ref{sec:results:bayes}). Because the closure carries four parameters against six to eight states per family, the reported results are grouped quantities, profile-likelihood valleys, and posterior predictive checks. In particular, $\lams$ and $\varphi_0$ are partially confounded and are not interpreted as stand-alone fitted material properties (Sec.~\ref{sec:results:errors}). The calibrated quantity is the apparent through-plane coating conductivity derived from LFA after current-collector subtraction. Throughout, calibrated parameter values are scoped to this LFA-derived target, so our stronger inferences concern grouped quantities, trend directions, and mechanism discrimination by orthogonal measurements rather than method-invariant absolute properties.

\paragraph{Extensions, consistency checks, and future validation targets.}
The additional results below extend that core model, test its consistency outside the calibration setting, or define the experiments needed to distinguish competing mechanisms. We therefore use \emph{validation} only for the directly calibrated and held-out thermal dataset; all cross-domain, inverse, and application-oriented results are described as consistency checks or feasibility studies.
\begin{itemize}
\item \textbf{Anode reorientation hypothesis.} For the graphite anode, a parsimonious and physically bounded reorientation account explains the anomalously low fitted $\lams$ by aligning graphite flakes so that the through-plane solid conductivity approaches the $c$-axis limit (Secs.~\ref{sec:model:reorient}, \ref{sec:results:mechanisms}). This fits with one \emph{fewer} parameter than the contact quadratic, and coupling reorientation to the \emph{anisotropic} solid conductivity goes beyond the geometric ellipsoid-orientation, isotropic-solid treatment of \cite{oehlerdiss}. Conductivity data alone leave it degenerate with the contact-only fit, so we treat the two as a comparison between two physically plausible, bounded mechanisms, where reorientation is the more parsimonious member but is \emph{not} uniquely identified, and same-sheet XRD texture (Sec.~\ref{sec:experiments}) is the measurement that would apportion the two.
\item \textbf{Inverse feasibility study (numerical).} An end-to-end differentiable implementation runs the closure backwards; on a \emph{synthetic} batch it recovers coating porosity to $\pm0.008$ from a single $3\%$-noise measurement, and a bridge/core decomposition is proposed as a candidate inline monitor for interface and delamination quality (Sec.~\ref{sec:results:qc}). This is a feasibility demonstration, and experimental inline use requires the campaign of Sec.~\ref{sec:experiments}.
\item \textbf{Identifiability and uncertainty.} Bayesian posteriors, WAIC, conformal prediction, and a profile-likelihood study bound what the data determine; Bayesian model averaging over the contact and reorientation closures carries the structural (which-mechanism) uncertainty into extrapolation; and hybrid-swap experiments localize cross-recipe transfer error in one re-anchorable parameter, yielding volume-space design rules (Sec.~\ref{sec:results:errors}, Sections~\ref{app:design}, \ref{app:bayes}).
\item \textbf{External consistency checks.} A bottom-up prediction of a measured 18650 cell, a multi-source comparison (median $\sim$$37\%$ relative error, $83\%$ of points within a factor of two), and transfers to fuel-cell gas-diffusion layers and dry-processed electrodes probe the closure outside its calibration (Sections~\ref{app:meta}, \ref{app:gdl}, and Sec.~\ref{sec:results:dryproc}). The 18650 prediction extrapolates below the calibration range and infers interface resistance, so it is used as a scale check.
\item \textbf{Data-driven methods and transport coupling (numerical extensions).} Automatic differentiation, Gaussian-process regression, conformal prediction, symbolic regression, and a physics-informed neural network for delamination fields support the inverse, uncertainty, and form-discovery pieces (Section~\ref{app:engine}). A transport-coupling analysis links flake orientation to ionic tortuosity and, through the carbon-binder network, to electronic conduction (Sec.~\ref{sec:results:coupling}). A two-gas-pressure-and-texture campaign is the principal future validation target (Sec.~\ref{sec:experiments}).
\end{itemize}

The Smoluchowski/Knudsen pore-gas correction is part of our forward model (Sec.~\ref{sec:model:knudsen}) and we quantify it for the separator case, though we credit it to Oehler et al.\ \cite{oehler2021}. All code, data transcriptions, and analysis notebooks, including the per-sheet transcription and the reconstruction-uncertainty propagation workflow, are openly available at \url{https://github.com/j-stoerk/zehner-electrode-thermal} (Reproducibility Statement).

\begin{center}\fbox{\begin{minipage}{0.94\textwidth}\small
\noindent\textbf{Claim scope (the organizing principle of this paper).} \emph{Primary claim:} a calendering-aware contact term on a Knudsen-corrected ZBS base reproduces the through-plane u-shape to within the reconstructed-target uncertainty in-sample ($4.5\%$ MAPE; held-out $7.8\%$), identified at the level of grouped quantities for the LFA-derived, method-scoped target. \emph{Secondary, bounded claim:} the graphite anode admits a parsimonious, physically bounded \emph{reorientation} account that is consistent with the data but \emph{not uniquely identified from conductivity alone}, remaining degenerate with a contact-only fit until same-sheet texture is measured. \emph{Enabled by the closure (not co-equal contributions):} the differentiable inverse, calibrated uncertainty, the computational support methods (symbolic regression, Gaussian-process residuals, conformal prediction, model averaging), and the cross-domain and manufacturing extensions are consistency checks, feasibility studies, or future-validation targets.
\end{minipage}}\end{center}

\begin{figure}[tb]\centering
\includegraphics[width=\textwidth]{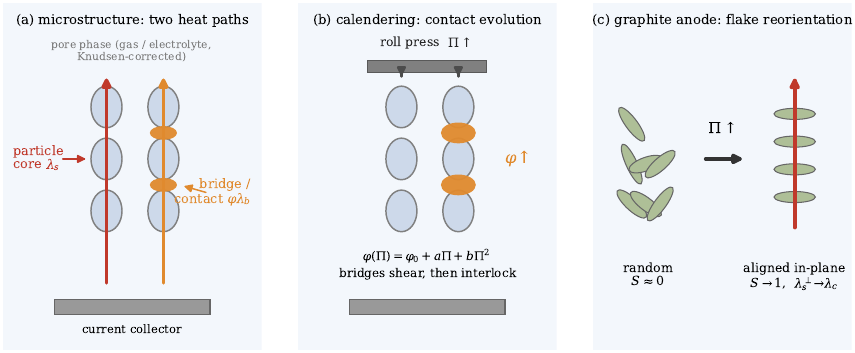}
\caption{Physical picture of the closure. Through-plane heat crosses a packed coating by two parallel routes: through particle cores ($\lams$) and across inter-particle bridge/contact zones ($\varphi\lamb$), with the pore phase (gas or electrolyte, Knudsen-corrected) in between (a). Calendering ($\Pi\uparrow$) compacts the bed and reshapes the contact network, captured by the process-dependent contact term $\varphi(\Pi)$ for every family (b). For the graphite anode only, calendering additionally reorients the anisotropic flakes toward in-plane alignment ($S(\Pi)$), turning the through-plane solid path onto the weak $c$-axis ($\lams^{\perp}\!\to\!\lambda_c$) (c).}
\label{fig:schematic}
\end{figure}

\begin{figure}[tb]\centering
\includegraphics[width=\textwidth]{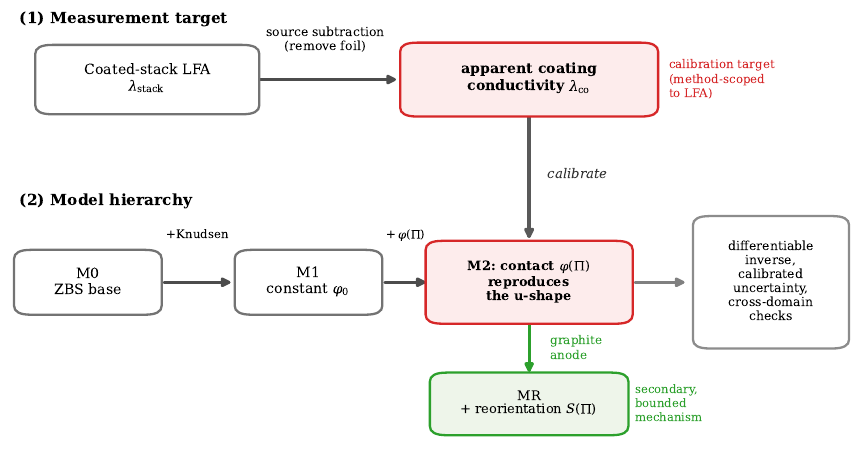}
\caption{What is measured and what is fitted. \emph{Top:} the calibration target is the apparent through-plane coating conductivity $\lambda_{\mathrm{co}}$ obtained from a coated-stack laser-flash (LFA) measurement by source subtraction of the collector foil; it is method-scoped to LFA. \emph{Bottom:} the model hierarchy, the ZBS base \Mzero{}, plus the always-on Knudsen pore-gas correction, plus a constant contact fraction (\Mone{}), plus the process-dependent contact term (\Mtwo{}, the \emph{primary claim} that reproduces the u-shape); for the graphite anode a reorientation branch (\MR{}) adds the \emph{secondary, bounded mechanism}. The differentiable inverse, calibrated uncertainty, and cross-domain checks are \emph{extensions} built on the calibrated closure.}
\label{fig:target}
\end{figure}

\noindent Figure~\ref{fig:schematic} summarises the physical picture (the two through-plane conduction paths and the calendering mechanisms), and Figure~\ref{fig:target} the measurement target and the model hierarchy that organises these claims.

\section{Related work}
\label{sec:related}

\paragraph{Analytic effective-medium closures.}
The classical results bound what any two-phase mixture can conduct (the Wiener \cite{wiener1912} and Hashin--Shtrikman \cite{hashin1962} bounds). Structure-specific models then commit to a geometry, Maxwell--Eucken \cite{maxwell1891,eucken1932} to isolated particles in connected fluid, Bruggeman \cite{bruggeman1935} to interpenetrating phases, and the ZBS family \cite{zehner1970,bauer1978,vdi} to a packed-bed unit cell with point (or, in Bauer's extension, small flattened) contacts. All evaluate in microseconds, differentiate cleanly, and have interpretable parameters, but each is only as good as its geometric commitment. We turn this into a diagnostic in Sec.~\ref{sec:results:baselines}. The \emph{pattern} of failures (Bruggeman matching co-continuous separators but overshooting sphere-packed electrodes by up to $+465\%$, Maxwell--Eucken undershooting everywhere) identifies each component's governing microstructure. None of these closures, as used for batteries, includes rarefied-gas corrections, particle-conductivity anisotropy or its processing evolution, or any process variable beyond porosity. Bruggeman \emph{can} be reconciled with electrodes, but only by recalibrating its exponent against pore-scale simulations \cite{vadakkepatt2016}, folding the phase connectivity $\chi=\varepsilon^{\,d-1}$ into an uninterpretable, static, non-invertible power. Our contact term (Sec.~\ref{sec:model:contact}) carries the same connectivity in a physically nameable, process-resolved, invertible form.

\paragraph{Microstructure-resolved simulation.}
Discrete-element models \cite{sangros2017} and homogenization on real or virtual microstructures \cite{oehler2021} represent particle arrangements explicitly, and can in principle resolve contact areas. In practice, both approaches were reduced to porosity-parameterized predictions when compared against the calendering data, and both missed the u-shape \cite{gandert2023}: their structure generators do not encode how a calender damages and rebuilds contacts. The closest prior treatment of orientation is Oehler's study of ellipsoidal anode particles \cite{oehlerdiss}, which varies particle aspect ratio and orientation but with an \emph{isotropic} solid conductivity, capturing a purely geometric (interface-count) effect through a shape factor. The mechanism we identify in Sec.~\ref{sec:results:mechanisms} is crystallographic and distinct: flake alignment exposes the low-conductivity $c$-axis to the through-plane path, lowering the effective \emph{solid} conductivity itself, an effect that an isotropic-solid model cannot represent. Classical effective-medium analysis likewise makes the through-plane Bruggeman exponent depend on particle orientation relative to the gradient \cite{vadakkepatt2016}, but again for isotropic-conductivity particles, so that too is a geometric rather than crystallographic effect. Resolved simulation also costs orders of magnitude more per evaluation than a closure and does not differentiate conveniently, which forecloses the inverse applications of Sec.~\ref{sec:results:qc}. Our approach complements resolved simulation: it keeps closure-level cost and differentiability, and injects the missing contact physics through a low-dimensional, process-indexed term whose parameters are interpretable and cheap to calibrate (about six calendering states).

\paragraph{Rarefied gas conduction in confined pores.}
Reduced gas conduction in pores comparable to the molecular mean free path is well established in kinetic theory \cite{kennard1938} and is quantitatively central to insulation and aerogel engineering \cite{kaganer1969}, where pore sizes resemble those of battery separators. Oehler et al.\ \cite{oehler2021,oehlerdiss} have already coupled this Smoluchowski/Knudsen correction into a ZBS electrode closure, which we adopt and credit to them. The authors of the primary dataset nonetheless name pore-gas rarefaction as an unresolved contributor to the calendering data \cite{gandert2023}. What still appears to be missing is a \emph{calendering-resolved} implementation of this correction within a battery-electrode ZBS closure, so our contribution is not the existence of the term but its consistent, process-aware use. We apply it identically across every baseline so that comparisons isolate geometry rather than gas treatment (Sec.~\ref{sec:results:baselines}), quantify it for the separator, where the smaller pores make it largest (Sec.~\ref{sec:model:knudsen}), and design the experiment that isolates it by showing that gas conduction depends on pressure while the skeleton conduction does not (Sec.~\ref{sec:experiments}).

\paragraph{Battery component thermal measurements.}
Component-level data come from steady-state stacked-sample rigs with controlled compaction \cite{burheim2013,richter2017}, copper-block reference-bar setups on cell teardowns \cite{marconnet2018,vishwakarma2015}, and flash-diffusivity metrology with penetration-model evaluation \cite{mcmasters,gandert2023}. Canonical early values are due to Maleki et al.\ \cite{maleki1999}. Two robust experimental findings frame our work. Soaking a coating with electrolyte raises its conductivity roughly threefold \cite{burheim2013,marconnet2018}, and interfaces, not bulk layers, dominate stack resistance \cite{vishwakarma2015}. The first is a clean test of any closure's fluid sensitivity, and the second both motivates our contact term and warns that stack-derived literature values are lower bounds for single coatings.

\section{Model}
\label{sec:model}

\subsection{Setting, notation, and assumptions}
\label{sec:model:setup}

We model a porous coating as a two-phase medium: solid particles of conductivity $\lams$ and diameter $d_p$, and a pore fluid of conductivity $\lamf$ (a gas or the liquid electrolyte), at porosity $\varepsilon$ (the pore volume fraction). The closure predicts the dimensionless ratio $k_{\mathrm{bed}}=\leff/\lamf$ as a function of $\varepsilon$ and the phase contrast $\kappa=\lams/\lamf$. Calendering is indexed by the compression rate
\begin{equation}
\Pi \;=\; 1-\frac{s_{\mathrm{co}}}{s_{\mathrm{co},0}},
\label{eq:Pi}
\end{equation}
where $s_{\mathrm{co}}$ is the coating thickness after calendering and $s_{\mathrm{co},0}$ the as-coated thickness, so $\Pi=0$ means uncalendered and $\Pi=0.5$ means compressed to half thickness. Our assumptions, stated up front:

\begin{itemize}
\item[\textbf{A1}] Steady diffusive heat conduction only. Radiation and convection inside the pores are negligible at micrometer pore sizes and ambient temperature (we verified the porous-convection threshold is missed by about fifteen orders of magnitude at coating scale).
\item[\textbf{A2}] The closure's unit cell assumes a statistically isotropic packing. Two real deviations are represented through interpretable parameters. Solid-solid bridge conduction enters the contact term of Sec.~\ref{sec:model:contact}. The strong conductivity anisotropy of graphite flakes enters the effective through-plane solid conductivity $\lams^{\perp}$, which is set by the flake orientation state through Sec.~\ref{sec:model:reorient} and bounded by the single-crystal anisotropy band ($c$-axis to basal plane).
\item[\textbf{A3}] The pore gas behaves as a continuum except for the conductivity reduction of Eq.~\eqref{eq:knudsen}; the liquid electrolyte needs no such correction because molecular mean free paths in liquids are sub-nanometer.
\item[\textbf{A4}] A single scalar $\varphi$ is an effective bridge area fraction at the scale identified by the LFA-derived coating value. Conceptually it contains intra-coating bridges (particle-particle necks and binder/carbon-black paths) and any residual coating--collector contribution left by the source subtraction. The present dataset identifies only this lumped term, with bridge conductivity $\lamb$; separating bulk-coating and collector-interface contributions requires the paired metrology of Sec.~\ref{sec:experiments}.
\item[\textbf{A5}] Intrinsic material properties are fixed within an electrode family: particle conductivities $(\lambda_a,\lambda_c)$ or isotropic $\lams$, the bridge conductivity $\lamb$, and the fitted coefficients $(\varphi_0,a,b)$. Effective process quantities evolve along the calendering trajectory: $\varepsilon$, $\Pi$, the bridge fraction $\varphi(\Pi)$, and, for graphite, the effective through-plane solid conductivity $\lams^{\perp}(S(\Pi))$.
\end{itemize}

We build on ZBS for three reasons. First, \emph{morphology}. Electrodes really are packings of particles, so ZBS's unit cell is the right starting picture, whereas Maxwell--Eucken's isolated particles and Bruggeman's interpenetrating phases are not (Sec.~\ref{sec:results:baselines} confirms this empirically). Second, \emph{validation headroom}. The base closure was validated in companion work on Zehner's original packed-bed data across phase contrasts $\kappa\in[4,\,14094]$, and the battery systems' values ($\kappa=14$--$1005$) lie strictly inside that range, so we \emph{interpolate} in $\kappa$ and extrapolate only in porosity (examined empirically in Sec.~\ref{sec:results:errors}). Third, \emph{extensibility}. ZBS fails on battery components at exactly two nameable points, the pore-gas conductivity and the contact term, and both can be repaired without touching the rest of the model.

\subsection{Base closure (\Mzero)}
\label{sec:model:zbs}

The ZBS closure \cite{zehner1970,bauer1978,vdi} pictures the bed as two parallel channels: a pure-fluid channel of relative cross-section $1-\sqrt{1-\varepsilon}$, and a particle-core channel of cross-section $\sqrt{1-\varepsilon}$ in which heat must pass through fluid gaps between curved particle surfaces:
\begin{equation}
k_{\mathrm{bed}}(\varepsilon,\kappa) \;=\; \underbrace{1-\sqrt{1-\varepsilon}}_{\text{fluid channel}} \;+\; \underbrace{\sqrt{1-\varepsilon}\;\lambda_{pm}(\varepsilon,\kappa)}_{\text{particle-core channel}} .
\label{eq:zbs}
\end{equation}
The core term $\lambda_{pm}$ results from integrating the local series resistance across the unit cell:
\begin{equation}
\lambda_{pm}=\frac{2}{1-B/\kappa}\!\left[\frac{B(\kappa-1)}{\kappa\,(1-B/\kappa)^{2}}\,\ln\frac{\kappa}{B}\;-\;\frac{B+1}{2}\;-\;\frac{B-1}{1-B/\kappa}\right],
\qquad
B = C\left(\frac{1-\varepsilon}{\varepsilon}\right)^{10/9},
\label{eq:lampm}
\end{equation}
where $B$ is the particle deformation parameter and $C$ a shape factor ($C=1.25$ for spheres). Two features of Eq.~\eqref{eq:zbs} explain the model's later behaviour: $k_{\mathrm{bed}}\to1$ as $\varepsilon\to1$, and for large contrast $\kappa$ the core term grows only logarithmically, because the bottleneck is the thin fluid gap between approaching particles. This is why the model \emph{under}-predicts whenever real coatings bridge that gap with solid, and why a contact term, not an inflated $\lams$, is the correct repair. Equations~\eqref{eq:zbs}-\eqref{eq:lampm} with continuum $\lamf$ constitute our baseline \Mzero{}.

\subsection{Knudsen-corrected pore gas}
\label{sec:model:knudsen}

When the pore size approaches the gas mean free path (the average distance between molecular collisions), a molecule crosses to the wall before meeting another, and pore-gas conduction falls below its continuum value. The standard correction (often attributed to Smoluchowski) writes the effective gas conductivity as
\begin{equation}
\lambda_{g,\mathrm{eff}}
=\frac{\lamg}{1+2\beta\,\Kn},
\qquad
\Kn=\frac{\Lambda}{d_{\mathrm{pore}}},
\qquad
\Lambda=\frac{k_{B}T}{\sqrt{2}\,\pi d_{g}^{2}\,p},
\label{eq:knudsen}
\end{equation}
where $\lamg$ is the continuum gas conductivity, $\Kn$ the Knudsen number, $\Lambda$ the mean free path, $d_{\mathrm{pore}}$ the pore diameter, $k_B$ the Boltzmann constant, $T$ the absolute temperature, $p$ the gas pressure, $d_g$ the effective molecular collision diameter (so $\Lambda_{\mathrm{air}}\approx67$\,nm and $\Lambda_{\mathrm{He}}\approx190$\,nm at ambient conditions), and $\beta$ a dimensionless jump coefficient of order unity that bundles the gas--surface energy accommodation. We use $\beta=1.64$, a standard value for air on technical surfaces, for both air and the helium purge because the dominant uncertainty is the gas--solid accommodation at rough technical surfaces. Varying $\beta$ over $1.5$--$2.0$ changes the primary helium-purged electrode predictions by at most $1.7\%$ (Sec.~\ref{sec:results:errors}). Two limits anchor the formula. For $\Kn\to0$ (large pores) the continuum value is recovered, which is why packed beds of millimeter particles never needed this correction, and for $\Kn\to\infty$ the gas conduction vanishes entirely. The pore size follows the hydraulic estimate for sphere packings,
\begin{equation}
d_{\mathrm{pore}}=\frac{2}{3}\,\frac{\varepsilon}{1-\varepsilon}\,d_{p},
\label{eq:dpore}
\end{equation}
or is taken from porosimetry where available (polyolefin separators: $43$--$64$\,nm). This is a \emph{single, mean} hydraulic pore size, a deliberate mean-field assumption. Real coatings, and calendered or semi-dry-extruded electrodes in particular, have a broad pore-size distribution in which the smallest pores cross into the Knudsen regime (and shed their gas conduction) before the largest, so applying the correction at one effective $d_{\mathrm{pore}}$ slightly under-weights that fine tail. A distribution-resolved treatment would integrate Eq.~\eqref{eq:knudsen} over the measured pore-size distribution. We adopt the single-pore mean because the present data fix only a mean pore size, and the residual bias is small next to the gas--solid accommodation uncertainty already absorbed by $\beta$. The corrected contrast $\kappa_{\mathrm{eff}}=\lams/\lambda_{g,\mathrm{eff}}$ then enters Eq.~\eqref{eq:zbs} unchanged. This introduces a practical subtlety. The ``dry'' conductivity of a coating now depends on \emph{which gas} fills the pores and at \emph{what pressure}, so a laser-flash measurement under helium purge probes a different pore-gas state than a coating sitting in air, and both differ from a coating under vacuum. Section~\ref{sec:experiments} exploits this pressure dependence as a validation experiment.

\subsection{Calendering-aware contact term}
\label{sec:model:contact}

Calendering acts on contact zones, not on pore volume, which is the root reason porosity-only models cannot represent its signature. To represent contact-mediated conduction explicitly, we adopt the flattened-contact generalization of Eq.~\eqref{eq:zbs} from the VDI Heat Atlas \cite{vdi} and extend it with a bridge phase whose conductivity may differ from the particles':
\begin{equation}
k_{\mathrm{bed}}
= 1-\sqrt{1-\varepsilon}+\sqrt{1-\varepsilon}\,
\Bigl[\varphi\,\frac{\lamb}{\lamf} + (1-\varphi)\,\lambda_{pm}(\varepsilon,\kappa)\Bigr],
\label{eq:contact}
\end{equation}
where $\varphi$ is the fraction of the particle-channel cross-section occupied by solid bridges and $\lamb$ is the conductivity of whatever forms those bridges: graphite--graphite flake contacts ($\lamb\approx130\WmK$) in anodes, the carbon-black/binder network ($\lamb\approx24\WmK$, the carbon-black value cited in \cite{gandert2023}) between NMC particles in cathodes. The product $\varphi\,\lamb$ is the physically meaningful group, an areal bridge conductance: doubling the bridge area at half the bridge conductivity leaves the heat flow unchanged, and Sec.~\ref{sec:results:errors} shows the data indeed determine this product more robustly than its factors. Setting $\varphi=0$ recovers \Mzero{} exactly, a property we enforce in the code by regression test.

How should $\varphi$ depend on calendering? The dataset's own adhesion measurements give the answer: Gandert et al.\ \cite{gandert2023} observe that coating--collector adhesion first \emph{drops} with calendering (the roll gap shears the coating and damages the as-coated binder network) and then \emph{recovers} at high line loads (particles interlock and even press into the soft aluminum foil, which they document by electron microscopy). Calendering can initially damage existing bridges and, at higher compression, promote new interlocking contacts. The lowest-order polynomial able to express a dip followed by a recovery is a quadratic, so we set
\begin{equation}
\varphi(\Pi)=\max\!\bigl(0,\;\varphi_{0}+a\,\Pi+b\,\Pi^{2}\bigr),
\label{eq:phi}
\end{equation}
with three nameable coefficients: $\varphi_0>0$ is the as-coated bridge fraction (expected near the rigid-sphere contact value of $0.0077$ tabulated in the VDI Heat Atlas), $a<0$ is the shear-damage rate, and $b\ge0$ is the interlocking recovery, a threshold effect a gentle process may never reach. We use the quadratic as the lowest-order phenomenological form that can represent damage followed by recovery on the six to eight states available per family, with sign-constrained coefficients ($a<0$ damage, $b\ge0$ recovery) anchored to the adhesion observations and falsifiable by the adhesion correlation of Sec.~\ref{sec:experiments}. As a \emph{supporting interpretation}, Section~\ref{app:design} shows that this form is also consistent with a more detailed dual-mechanism contact-mechanics model (a low-order expansion in which the linear term is binder-bridge rupture under shear and the quadratic term is plastic particle interlocking, and a Hertzian-particle-plus-shearable-binder model fits the cathodes to $1.5\%$). Without direct contact-area measurements the recovery phase remains degenerate with porosity reduction, so we treat the quadratic as physically motivated and benchmarked rather than derived. We write \Mone{} for Eqs.~\eqref{eq:contact}--\eqref{eq:phi} at constant $\varphi$ ($a=b=0$) and \Mtwo{} for the full quadratic.

\subsection{Anisotropic-particle reorientation}
\label{sec:model:reorient}

The contact term treats the solid conductivity $\lams$ as a constant of the material. For the graphite anode that assumption is the weak point, and relaxing this assumption introduces a plausible anode-specific mechanism. Graphite is a strongly anisotropic platelet: heat flows readily in the basal plane but poorly across it. We fix the in-plane (basal) conductivity at $\lambda_a=150\WmK$, within the reported polycrystalline range $92$--$247\WmK$ \cite{oehlerdiss}, and the $c$-axis conductivity at $\lambda_c=6\WmK$ (reported $5$--$10$); these two named constants are used by symbol throughout. With them the random-orientation average is $(2\lambda_a+\lambda_c)/3=102\WmK$, and the $c$-axis value $\lambda_c=6\WmK$ is the aligned-flake floor. Oehler's electrode closure adopts a single \emph{isotropic} graphite value ($138.6\WmK$, the top of this band) and explicitly sets graphite anisotropy aside \cite{oehlerdiss}; the through-plane effective \emph{solid} conductivity of a bed of flakes in fact depends on how the flakes are oriented. For a single flake whose $c$-axis (basal-plane normal) makes angle $\theta$ with the through-plane direction, the through-plane component of its conductivity tensor is $\lambda_{zz}=\lambda_c\cos^2\theta+\lambda_a\sin^2\theta$; averaging over the flake orientation distribution gives
\begin{equation}
\lams^{\perp}(S)=(1-S)\,\frac{2\lambda_a+\lambda_c}{3}+S\,\lambda_c,
\qquad
S=\frac{3\langle\cos^2\theta\rangle-1}{2},
\label{eq:reorient}
\end{equation}
where $S\in[0,1]$ is the \emph{Hermans orientation factor}, which can be directly obtained from X-ray diffraction (XRD) texture analysis: $S=0$ is random orientation, giving the isotropic average $(2\lambda_a+\lambda_c)/3$, and $S=1$ is flakes fully aligned in-plane, so the through-plane path sees the $c$-axis limit $\lambda_c$. Graphite-electrode flakes are known to lie preferentially parallel to the collector, and a dedicated XRD method quantifies the oriented fraction and finds it \emph{rises with calendering} \cite{malifarge2017}, exactly the direction the mechanism requires. Calendering presses the flakes flatter, raising $S$; the lowest-order monotone form, saturating when the flakes are fully aligned, is
\begin{equation}
S(\Pi)=\mathrm{clip}\bigl(S_0+s\,\Pi,\;0,\;1\bigr),
\label{eq:Sofpi}
\end{equation}
with as-coated factor $S_0$ and alignment rate $s>0$. We retain this clipped-linear kink given six to eight data points per calendering family. Adding a smoothness parameter is unwarranted, and the linear form provides the most constrained bounding behavior. The corrected contrast $\kappa^{\perp}=\lams^{\perp}(S(\Pi))/\lamf$ then enters Eq.~\eqref{eq:contact} in place of $\kappa$. We call this reorientation-coupled variant \MR{}; for the anode it carries the calendering dependence through $S(\Pi)$ at \emph{constant} $\varphi=\varphi_0$, so it has one fewer parameter than the contact quadratic \Mtwo{}. Trace conservation (the flake tensor trace $2\lambda_a+\lambda_c$ is rotation-invariant) fixes the in-plane companion $\lams^{\parallel}(S)=(2\lambda_a+\lambda_c-\lams^{\perp})/2$ without introducing additional fitted parameters, so the closure also predicts the in-plane conductivity and hence a thermal anisotropy ratio $k_{\parallel}/k_{\perp}$ that rises from $1$ (random) toward $\lambda_a/\lambda_c$ as flakes align: a further falsifiable output that is expected to increase with calendering, as the measured XRD orientation index does \cite{malifarge2017}.

Two properties make Eq.~\eqref{eq:reorient} a constrained mechanism. First, it is \emph{bounded}: the through-plane solid conductivity cannot fall below $\lambda_c$, so the fit is bounded below by the $c$-axis conductivity. Second, it is \emph{inert for isotropic actives}: setting $\lambda_a=\lambda_c$ (the quasi-spherical NMC secondary particle) makes $\lams^{\perp}$ independent of $S$, so reorientation cannot act in the cathode and the cathode dip must come from the contact term. These two properties help distinguish the mechanisms, although conductivity data alone do not uniquely identify them (Sec.~\ref{sec:results:mechanisms}); $\lambda_a$ is taken from the literature, and Sec.~\ref{sec:results:mechanisms} reports the sensitivity to it.

\paragraph{Model hierarchy and default.} The closure is built as a hierarchy of named increments, each repairing one nameable deficiency of the previous: the baseline ZBS unit cell (\Mzero{}, Eqs.~\eqref{eq:zbs}--\eqref{eq:lampm}); \emph{plus} the Knudsen pore-gas correction (Eq.~\eqref{eq:knudsen}), which is always on and adds no fitted parameter; \emph{plus} the calendering-aware contact term, either at constant bridge fraction (\Mone{}) or with the process-dependent quadratic $\varphi(\Pi)$ (\Mtwo{}, Eqs.~\eqref{eq:contact}--\eqref{eq:phi}); \emph{plus}, for anisotropic actives only, the reorientation coupling $\lams^{\perp}(S(\Pi))$ (\MR{}, Eqs.~\eqref{eq:reorient}--\eqref{eq:Sofpi}). Unless stated otherwise, the Knudsen-corrected contact quadratic \Mtwo{} is the default forward closure because it reproduces the observed u-shape across the calibrated families (Sec.~\ref{sec:results:ablation}) and is the variant used for the quality-control inverse problem (Sec.~\ref{sec:model:inverse}). \MR{} is the physically bounded variant with one fewer fitted parameter, adopted for the graphite anode where the active is anisotropic, and for extrapolation beyond the calibration range the three-model Bayesian average of Sec.~\ref{sec:results:bayes} is the default predictive output. Throughout, the same symbols carry the same meaning: $\leff$ the effective through-plane conductivity, $\lams$ the (through-plane, where anisotropic) solid conductivity, $\lamb$ the bridge conductivity, $\lamf$ the pore-fluid conductivity, $\kappa=\lams/\lamf$ the phase contrast, $\varepsilon$ the porosity, $\varphi$ the bridge area fraction, $\Pi$ the compression rate, and $S(\Pi)$ the flake-orientation factor.

\subsection{Differentiable implementation and inverse use}
\label{sec:model:inverse}

\emph{The analytic closure remains the interpretable, invertible core; the differentiability introduced here, and the data-driven methods it enables, are a computational support layer around that core, not separate contributions.} Equations~\eqref{eq:zbs}--\eqref{eq:phi} are implemented in JAX (64-bit), so every model derivative is available exactly by automatic differentiation rather than by step-size-sensitive finite differences. The inverse map for porosity quality control is Newton's iteration
\begin{equation}
\varepsilon_{k+1}=\varepsilon_{k}-\frac{\leff(\varepsilon_{k})-\lambda^{\mathrm{meas}}}{\partial\leff/\partial\varepsilon\,(\varepsilon_{k})},
\label{eq:newton}
\end{equation}
where $\lambda^{\mathrm{meas}}$ is the measured conductivity and $\varepsilon_k$ the porosity estimate at iteration $k$; it converges quadratically (exact derivative) and the root is unique for a fixed family, fixed metrology state, and fixed calibrated process relation, where $\leff(\varepsilon)$ is monotonic over the inversion range. This monotonicity is conditional; the measured calendering trajectory can still be u-shaped because $\varphi(\Pi)$ and, for graphite, $S(\Pi)$ evolve with process state. First-order uncertainty propagation turns measurement and model noise into a porosity error bar:
\begin{equation}
\sigma_{\varepsilon}
=\frac{\sqrt{\sigma_{\mathrm{meas}}^{2}+\sigma_{\mathrm{model}}^{2}}\;\leff}
{\bigl|\partial\leff/\partial\varepsilon\bigr|},
\label{eq:sigma}
\end{equation}
where $\sigma_{\mathrm{meas}}$ is the relative noise of the conductivity measurement (laser flash or transient plane source: typically $2$--$5\%$) and $\sigma_{\mathrm{model}}$ a relative model-form floor, which we set to $2\%$, the residual scatter of the base closure's packed-bed validation. The formula also explains \emph{which} electrode makes the best QC target: the larger the sensitivity $|\partial\leff/\partial\varepsilon|$, the smaller the porosity uncertainty, and the anode has the largest sensitivity of the three components.

\paragraph{The computational support methods.} Four machine-learning methods form a computational support layer around the closure, each adding a capability the analytic model lacks; they are detailed in Section~\ref{app:engine} and used only as labelled below. A \emph{symbolic-regression} search over the packed-bed data acts on the \emph{forward structure}, recovering the additive two-channel form and a contact constant consistent with the VDI value; this independently recovers the closure's algebraic structure rather than assuming it. A \emph{Gaussian-process} model is fitted to the \emph{residuals} of the forward model and, wrapped in \emph{conformal} prediction, converts point predictions into intervals with guaranteed out-of-sample coverage (Sec.~\ref{sec:results:robustness}). A physics-informed neural network (\emph{PINN}) extends the scalar inverse from a single reading (Eq.~\eqref{eq:newton}) to a spatial \emph{field}, recovering an interface-conductance map from synthetic flash-thermography frames, with the closure supplying the constitutive interpretation of that field (implementation and training recipe in Section~\ref{app:engine}). Finally, \emph{Bayesian model averaging} over the three closures (\Mtwo{}, \MR{}, and a residual-corrected hybrid) carries the \emph{structural} (mechanism) uncertainty into the predictive output for extrapolation (Sec.~\ref{sec:results:bayes}). Each is infrastructure around the closure, supplying structure discovery, calibrated intervals, inverse flexibility, and structural-uncertainty accounting.

\paragraph{Prospective online use.} Because the inverse runs in microseconds, the same closure could serve as an online \emph{soft sensor} during production rather than only an offline diagnostic: a differentiable route to inline porosity and interface monitoring for continuous, including semi-dry granulate-based, electrode production. We develop this prospective use, the physics-informed-neural-network field inverse it rests on, and the active-learning loop it would enable for the GranuGoIn semi-dry line, as an extension in Section~\ref{app:engine}; it is a forward-looking application, not a validated result of this paper.

\section{Data, baselines, and calibration protocol}
\label{sec:data}

\subsection{Primary dataset}
\label{sec:data:primary}
Gandert et al.\ \cite{gandert2023} (open access, CC-BY) measured two graphite anodes on copper foil (96\,wt\% active material, carboxymethyl cellulose/styrene-butadiene rubber (CMC/SBR) binder, 1.5\% carbon black; two coating thicknesses) and NMC622/NMC811 cathodes on aluminum foil (polyvinylidene fluoride (PVDF) binder; NMC622 additionally 2\,wt\% flake-graphite additive), each calendered on a laboratory roll calender to six to eight states. Conductivity follows from $\lambda=\kappa_d\,\rho\,c_p$, where $\kappa_d$ is the laser-flash thermal diffusivity (measured under \emph{helium} purge with penetration-model evaluation \cite{mcmasters}), $c_p$ the specific heat from differential scanning calorimetry (DSC), and $\rho$ the density from gas pycnometry; thicknesses come from a flat-anvil micrometer at contact pressures below $0.3\,\mathrm{N\,mm^{-2}}$ (the authors show that ball-tip dial gauges compress the coating and read up to $23\%$ low); porosity uncertainties follow the Guide to the Expression of Uncertainty in Measurement (GUM). Because the source publication reports sheetwise porosities and fitted conductivity curves rather than a raw per-sheet conductivity table in the form needed here, we transcribe the reported porosities and thicknesses and evaluate their published fit functions at those porosities. This reconstruction adds uncertainty beyond the reported measurement scatter, because porosity, thickness, foil subtraction, and fit-function evaluation all enter the derived coating value. We explicitly  \emph{propagate} this uncertainty: a Monte-Carlo of the thickness ($\pm1.5\%$), stack-conductivity ($\pm5\%$), and collector-conductivity ($\pm3\%$) inputs through Eq.~\eqref{eq:stack} gives a median $\sim$$5\%$ relative uncertainty on the reconstructed coating conductivity, dominated by the stack-measurement term (the foil subtraction adds little because the coating is far thicker than the collector). This band is shown on the calibration figure (Fig.~\ref{fig:calibration}); being comparable to the $4.5\%$ calibrated residual, it bounds how tightly any closure can be expected to match and is one more reason to report grouped and bounded parameters rather than over-fit individual points. The metal foil is removed by series subtraction,
\begin{equation}
\frac{s_{\mathrm{co}}}{\lambda_{\mathrm{co}}}
=\frac{s_{\mathrm{stack}}}{\lambda_{\mathrm{stack}}}-\frac{s_{\mathrm{cc}}}{\lambda_{\mathrm{cc}}},
\label{eq:stack}
\end{equation}
where subscripts co, stack, and cc denote coating, coated stack, and current collector, $s$ are thicknesses and $\lambda$ conductivities ($\lambda_{\mathrm{Cu}}=400$, $\lambda_{\mathrm{Al}}=237\WmK$; the foil term is below $0.1\%$ of the total). One systematic remains by construction: the coating/collector contact resistance stays lumped inside $\lambda_{\mathrm{co}}$, exactly as in the source. The calibration target is therefore an LFA-derived apparent through-plane coating conductivity under this source subtraction procedure. Because the flash measurements run under helium, all dry comparisons use helium as the pore gas ($\lamg=0.1518\WmK$), and since $\Lambda_{\mathrm{He}}\approx190$\,nm is almost three times the air value, this dataset makes the Knudsen correction substantially more influential than it would be in air.

\subsection{Anchor datasets}
Independent anchors, compiled with row-level citations in the repository: dry polyolefin separators $0.07$--$0.18\WmK$ across types and compaction pressures \cite{richter2017}; a ceramic-coated separator stack at $0.10$ dry and $0.11$ wet \cite{marconnet2018}; soaked/dry electrode pairs, graphite $0.30\to0.89$ and LCO $0.36\to1.10\WmK$ \cite{burheim2013}, CMS graphite $0.57\to1.35$ and LMO $0.16\to0.45\WmK$ \cite{marconnet2018}; the interface-dominance result of \cite{vishwakarma2015}; separator pore sizes ($43$--$64$\,nm) from porosimetry. Constituent properties used throughout: $\lambda_{\mathrm{air}}=0.026$, $\lambda_{\mathrm{electrolyte}}=0.18\WmK$ (carbonate-solvent class \cite{maleki1999}), particle sizes $d_p=18\,\mu$m (graphite) and $10\,\mu$m (NMC). For the cell-level test of Section~\ref{sec:results:stack} we additionally use the complete layer structure (thicknesses and porosities per layer) and the measured jelly-roll conductivity of a commercial 18650 cell, $k_\perp = 1.122\WmK \pm 6.2\%$ \cite{steinhardt2021}, the meta-analytic range of all published cell-level values, $0.15$--$1.4\WmK$ \cite{steinhardt2022}, and, as a cross-model anchor, microstructure-resolved simulation results for electrolyte-filled coatings \cite{oehlerdiss}.

\subsection{Baselines}
\label{sec:data:baselines}
We compare against the classical closure hierarchy, all evaluated with the \emph{same} constituent properties and the same Knudsen-corrected pore gas, so that differences isolate the geometric assumptions rather than the inputs. The Wiener bounds \cite{wiener1912} are the parallel and series arrangements. Maxwell--Eucken \cite{maxwell1891,eucken1932}, for isolated solid spheres of volume fraction $1-\varepsilon$ dispersed in connected fluid, reads
\begin{equation}
\leff^{\mathrm{ME}}
=\lamf\,\frac{\lams+2\lamf+2(1-\varepsilon)(\lams-\lamf)}{\lams+2\lamf-(1-\varepsilon)(\lams-\lamf)},
\label{eq:me}
\end{equation}
and Bruggeman's symmetric effective-medium theory \cite{bruggeman1935}, which treats both phases as interpenetrating, defines $\leff$ implicitly as the root of
\begin{equation}
\varepsilon\,\frac{\lamf-\leff}{\lamf+2\leff}
+(1-\varepsilon)\,\frac{\lams-\leff}{\lams+2\leff}=0 .
\label{eq:bruggeman}
\end{equation}
We use the \emph{symmetric} form, in which both phases are treated identically and either may percolate (the solid network becomes continuous above a one-third volume fraction), rather than the asymmetric differential (incremental-inclusion) model in which one phase remains the continuous host throughout. The symmetric form is the natural comparator here precisely because it represents the perfectly \emph{co-continuous} composite, the morphological opposite of the ZBS packed bed of point-contacting particles; the two baselines therefore bracket real electrode microstructure, and their contrasting pattern of failure becomes a structure discriminator (Sec.~\ref{sec:results:baselines}: symmetric Bruggeman matches the co-continuous separator but overshoots the sphere-packed electrodes, while ZBS does the reverse).
The two microstructure-resolved literature models are not re-implemented; their failure on this dataset (monotonic in porosity, no u-shape) is documented by its authors \cite{gandert2023}.

\subsection{Calibration, ablation, and error-analysis protocol}
\label{sec:data:protocol}
Per electrode family, the parameter vector $\theta=(\lams,\varphi_0,a,b)$ minimizes the sum of squared \emph{relative} residuals,
\begin{equation}
\hat\theta=\arg\min_{\theta\in\Theta}\;
\sum_{i}\left[\frac{\leff(\varepsilon_i,\Pi_i;\theta)-\lambda^{\mathrm{meas}}_{\mathrm{co},i}}{\lambda^{\mathrm{meas}}_{\mathrm{co},i}}\right]^{2},
\label{eq:objective}
\end{equation}
solved by trust-region least squares over a physically-bounded box ($\lams\in[5,139]\WmK$ for graphite, the single-crystal anisotropy band, and $[1.5,5]$ for NMC; $\varphi_0\in[0,0.08]$; $a\in[-0.2,0.3]$; $b\in[0,0.8]$); relative residuals prevent the higher-conductivity anodes from dominating the fit, and the accuracy metric is the mean absolute percentage error (MAPE) over a family's sheets. Four checks (an ablation \Mzero$\to$\Mone$\to$\Mtwo; held-out cross-recipe transfer; a profile likelihood over $\lams$; and hybrid parameter swaps) and a full Bayesian (No-U-Turn-sampler) treatment with physics-carrying priors are specified in Section~\ref{app:protocol}.

Table~\ref{tab:params} fixes, once, the status of every quantity in the closure, recording what is held from the literature, what is fitted, and, as the identifiability analysis (Sec.~\ref{sec:results:errors}) makes quantitative, which combinations the data determine robustly versus which are reportable only as a valley. Later sections refer back to it rather than re-deriving the distinction.

\begin{table}[tb]\centering\small
\caption{Status of every quantity in the closure. Fitted parameters are estimated per family; the data determine certain \emph{combinations} robustly while leaving the individual contact parameters confined to a valley (Sec.~\ref{sec:results:errors}). Effective process quantities ($\varphi(\Pi)$, $\lams^{\perp}(S(\Pi))$, $\varepsilon$) evolve along the calendering trajectory; intrinsic ones are fixed within a family.}
\label{tab:params}
\begin{tabular}{@{}lll@{}}
\toprule
Class & Quantities & Status \\
\midrule
Fixed inputs (literature) & $\lambda_c, \lambda_a, \lamb, d_p, \lamf, \beta$ & held constant \\
Fitted per family (\Mtwo) & $\lams, \varphi_0, a, b$ & estimated \\
Fitted per family (\MR) & $S_0, s, \varphi_0$ & estimated \\
Grouped / identifiable & bridge group $\varphi_0\lamb$; the $\varphi$--$S$ trade-off valley & robustly determined \\
Weakly identifiable alone & $\lams$ and $\varphi_0$ individually; bound-active $\lams$ & reported as a valley, not a point \\
\bottomrule
\end{tabular}
\end{table}

\section{Results and analysis}
\label{sec:results}

\begin{center}\fbox{\begin{minipage}{0.94\textwidth}\small
\noindent\textbf{The following error quantities are kept distinct.} To avoid conflating distinct notions of error, we fix four numbers once and refer to them by name throughout. \emph{(1)~Measurement scatter:} the primary dataset's own through-plane repeat reproducibility, $\sim$$5\%$, the noise floor any closure can aim for. \emph{(2)~Reconstruction uncertainty:} a further $\sim$$5\%$ median (Monte-Carlo) from propagating thickness, stack-conductivity, and collector inputs through the foil subtraction of Eq.~\eqref{eq:stack} onto the \emph{derived} coating value; (1) and (2) together bound how tightly any model can match. \emph{(3)~In-sample calibrated residual:} the fitted \Mtwo{} MAPE over the 27 calendering states, $4.5\%$, which sits at that floor. \emph{(4)~Held-out predictive error:} leave-one-state-out generalisation, $7.8\%$ aggregate (and larger across recipes), the out-of-sample figure. Below, ``noise level'' always means (1)--(2), ``fit'' or ``in-sample'' means (3), and ``transfer'' or ``held-out'' means (4).
\end{minipage}}\end{center}

\subsection{Knudsen magnitudes and zero-fit reference values}
\label{sec:results:knudsen}

\begin{figure}[tb]\centering
\includegraphics[width=0.55\textwidth]{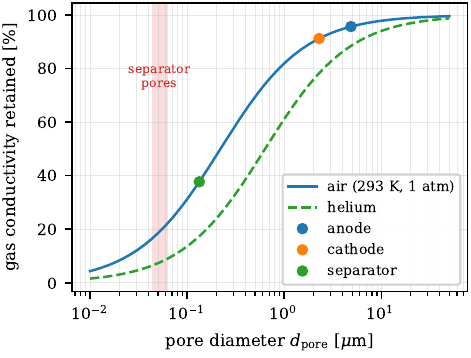}
\caption{Retained fraction of pore-gas conduction, Eq.~\eqref{eq:knudsen}, at 293\,K and 1\,atm for air (solid) and helium (dashed). Markers: hydraulic pore sizes of the three reference systems (air curve). Shaded band: measured polyolefin-separator pore diameters; inside it, the air mean free path exceeds the pore size.}
\label{fig:knudsen}
\end{figure}

\begin{figure}[tb]\centering
\includegraphics[width=\textwidth]{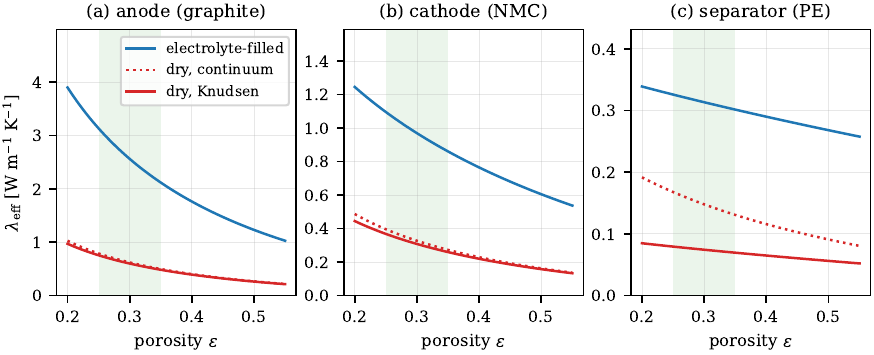}
\caption{Zero-fit $\leff(\varepsilon)$ for (a)~graphite anode, (b)~NMC cathode, (c)~PE separator: electrolyte-filled, dry-continuum, and dry-Knudsen variants. Green band: typical post-calendering porosity window.}
\label{fig:maps}
\end{figure}

Figure~\ref{fig:knudsen} quantifies the pore-gas correction of Sec.~\ref{sec:model:knudsen} (part of the forward model, credited to Oehler et al.). At reference states ($\varepsilon=0.30$ for electrodes, $0.40$ for the separator) the pore-gas conduction loss in air is $4.3\%$ for the graphite anode ($d_{\mathrm{pore}}=4.9\,\mu$m) and $8.8\%$ for NMC ($2.3\,\mu$m). For the separator it reaches $84\%$ at its measured $43$--$64$\,nm pores, where $\Kn=1.0$--$1.6$ means the mean free path \emph{exceeds} the pore size. Propagated to the dry coating conductivity, the electrode corrections are $-3.4\%$ and $-5.8\%$; under the helium atmosphere of flash metrology they grow to $2.8$--$5.5\%$, which is comparable to the reproducibility of the measurement itself and therefore worth carrying, and we supply it here for exactly the configuration in which the primary dataset was acquired. Zero-fit reference values (Fig.~\ref{fig:maps}) land inside published ranges with no electrode-specific fitting: anode $2.56$ (wet) and $0.60$ (dry); cathode $0.97$ and $0.31$; separator $0.29$ and $0.065\WmK$ (the dry value would read $0.116$ if the Knudsen term were wrongly omitted). The model's predicted wet-to-dry ratios ($4.3\times$ anode, $3.2\times$ cathode) modestly exceed the measured ones ($2.4$--$3.1\times$ \cite{burheim2013,marconnet2018}), and the direction of this miss is informative: a real coating contains solid bridges that conduct regardless of the pore fluid, damping its fluid sensitivity, and \Mzero{} has no such bridges yet. The ablation analysis supports this interpretation. Porosity sensitivity is large: a routine $\pm0.02$ calendering tolerance moves electrode $\leff$ by $5$--$9\%$ (computed from the exact gradients), so a cell model that treats $\leff$ as constant embeds a porosity-correlated error of that size.

\subsection{Baseline comparison: morphology explains the failure pattern}
\label{sec:results:baselines}

\emph{\textbf{Core result.}} Table~\ref{tab:baselines} evaluates the closure hierarchy at the four uncalendered points, a stringent test because the contact network has not yet been modified by calendering, with identical inputs and Knudsen-corrected helium pore gas throughout. The Wiener bounds are uninformative at graphite's phase contrast, spanning $0.24$ to $32\WmK$. More informative is that the structure-specific baselines fail in opposite directions, reflecting their underlying morphology assumptions. Bruggeman's co-continuous picture effectively hands the solid phase a percolating highway through the medium, so it overshoots wherever the solid is actually granular, by $+17\%$ for NMC622 and up to $+465\%$ for graphite. Maxwell--Eucken isolates every particle in a fluid jacket, so heat must repeatedly cross the poorly conducting gas, and it undershoots everywhere ($-11\%$ to $-70\%$). \Mzero{}, spheres touching at points, is the only zero-parameter closure that is both consistent in rank order across electrode families and accurate where its geometry actually holds, at $+2.7\%$ for NMC811, the electrode with the least additive (4\,wt\%) and hard, near-spherical particles. Its systematic \emph{under}-prediction elsewhere ($-29\%$ NMC622, $-36/-39\%$ graphite) is ordered exactly by expected bridge strength (flake-graphite additive in NMC622, soft flakes capable of forming additional contact pathways under compression in the anodes), and along every calendering series the residual shifts monotonically from under-prediction to over-prediction. The u-shape is invisible to every porosity-only model, consistent with the documented failure of the two microstructure-resolved models \cite{gandert2023}.

The same diagnostic logic settles the separator. Measured values ($0.07$--$0.18\WmK$) fall between our point-contact lower bound ($0.022$--$0.031$) and a continuous-skeleton parallel-plus-Knudsen upper bound ($0.080$--$0.155$), much closer to the latter, exactly as one expects for a stretched polymer film whose skeleton is co-continuous (Bruggeman, the co-continuous baseline, lands in-range at $0.090$). Here a cautionary observation is in order. A continuum sphere-pack model also reproduces the measured numbers ($0.075$--$0.093$), but this agreement is compensatory and therefore not mechanistically reliable, because it overestimates the gas conduction (no Knudsen correction) while underestimating the skeleton connectivity (wrong morphology), and the two errors cancel. The cancellation is fragile, and it breaks the moment the gas or its pressure changes; the two-pressure experiment of Sec.~\ref{sec:experiments} is designed to break it on purpose.

\begin{table}[tb]\centering\footnotesize\setlength{\tabcolsep}{4.5pt}
\caption{Baselines at the uncalendered points (dry, helium pore gas with Knudsen correction, identical constituent inputs; coating values derived from \cite{gandert2023} via Eq.~\eqref{eq:stack}). W$^{-}$/W$^{+}$: Wiener series/parallel bounds; ME: Maxwell--Eucken. Bold values mark the zero-fit ZBS+Kn column discussed in the text. All values in $\WmK$.}
\label{tab:baselines}
\begin{tabular}{lcccccc}
\toprule
 & measured & W$^{-}$ & ME & Bruggeman & W$^{+}$ & \Mzero{} (ZBS+Kn)\\
\midrule
graphite (thin), $\varepsilon=0.606$  & 1.429 & 0.242 & 0.430 & 8.07 & 31.6 & \textbf{0.871}\\
graphite (thick), $\varepsilon=0.597$ & 1.405 & 0.245 & 0.441 & 9.07 & 32.3 & \textbf{0.903}\\
NMC622, $\varepsilon=0.513$           & 0.704 & 0.258 & 0.434 & 0.825 & 1.29 & \textbf{0.498}\\
NMC811, $\varepsilon=0.507$           & 0.492 & 0.260 & 0.440 & 0.841 & 1.30 & \textbf{0.505}\\
\bottomrule
\end{tabular}
\end{table}

\begin{figure}[tb]\centering
\includegraphics[width=\textwidth]{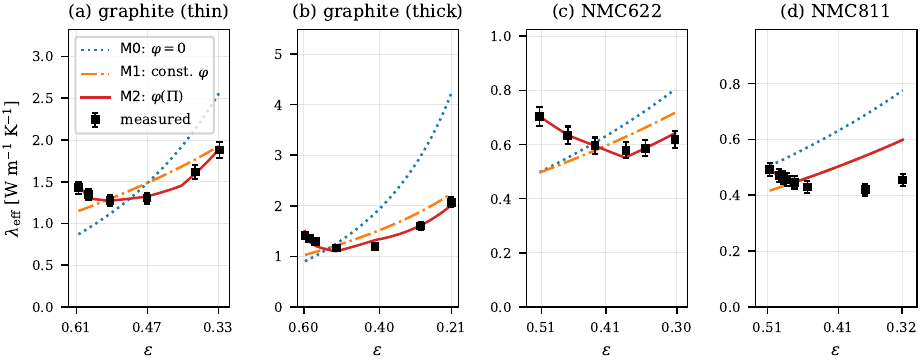}
\caption{\emph{In-sample calibration against the LFA-derived target.} Calibration curves for all 27 calendering states. \Mzero{} is the zero-fit ZBS+Kn reference; \Mone{} adds a fitted constant contact fraction; \Mtwo{} adds the process-dependent $\varphi(\Pi)$ term. The constant-contact curve removes most of the offset, while the process-dependent term reproduces the u-shape. Error bars on the measured points are the propagated data-reconstruction uncertainty (median $\sim$$5\%$; Sec.~\ref{sec:data:primary}), comparable to the calibrated residual. Held-out and external checks (Figs.~\ref{fig:error},~\ref{fig:validation}) are reported separately and should not be read as equivalent validation tiers.}
\label{fig:calibration}
\end{figure}

\subsection{Ablation: contribution of each model component}
\label{sec:results:ablation}

\emph{\textbf{Core result.}} Adding a \emph{constant} contact fraction (\Mone) to the ablation (Table~\ref{tab:ablation}) more than halves the average error, from $31.1\%$ to $13.5\%$, which shows the as-coated bridge network carries a substantial share of the conduction. Making the contact fraction \emph{process-dependent} (\Mtwo, Eq.~\eqref{eq:phi}) reaches $4.5\%$, the dataset's own scatter of about $5\%$, and reproduces the u-shape (Fig.~\ref{fig:calibration}), where the dip reflects early bridge damage ($a<0$ in all four families) that temporarily outweighs the effect of densification, and the upturn is densification plus, where it is reached, interlocking recovery. The recovery term $b>0$ appears in exactly the two families where interlocking is independently evidenced: NMC622, whose particles visibly penetrate the soft aluminum foil in electron micrographs at the highest line load \cite{gandert2023}, and the thick anode, which carries the highest line force of the four. Because $\lams$ and $\varphi_0$ are partially confounded (Sec.~\ref{sec:results:errors}), the best-identified quantity is their product, the as-coated \emph{bridge conductance} $G_0=\varphi_0\lamb$ (Table~\ref{tab:ablation}). We therefore report $G_0$ as the primary contact descriptor and treat the individual factors as bounded, with $\varphi_0=0.005$--$0.017$ bracketing the VDI rigid-sphere contact fraction of $0.0077$ \cite{vdi}, while several $\lams$ estimates sit at or near imposed bounds and should not be read as free physical property estimates. Expressed through $G_0$, the fitted parameters follow a physically consistent ordering, with the graphite anodes at $G_0\approx1.2$--$1.6\WmK$ (dense graphite--graphite flake bridges) and the NMC cathodes only $0.1$--$0.4\WmK$ (the sparser carbon-black/binder network), an anode-over-cathode ordering of the contact network that the separate, bound-active point estimates obscure. Held-out transfer completes the picture. Within a recipe, the calibration carries over usefully ($20.9\%$ on the thick anode it never saw), whereas across additive recipes it fails ($39.5\%$, NMC622 to NMC811), and the next subsection dissects why.

\begin{table}[tb]\centering\small
\caption{Ablation: MAPE over all calendering states. \Mzero: $\varphi=0$, mid-band $\lams$. \Mone: constant $\varphi$, fitted $(\lams,\varphi_0)$. \Mtwo: full $\varphi(\Pi)$, fitted $(\lams,\varphi_0,a,b)$. Right column: \Mtwo{} optima. The graphite-thick $\lams=5.0$ and both NMC $\lams$ values at or near $1.5\WmK$ are bound-active or weakly identified and are interpreted through grouped bridge conductance, not as standalone material constants. The grouped descriptor $G_0=\varphi_0\lamb$ is the primary fitted contact quantity; $\lams$ values marked $^{\dagger}$ in the tuple are bound-active or valley-constrained and should be read only through $G_0$.}
\label{tab:ablation}
\begin{tabular}{lccccl}
\toprule
Family & \Mzero & \Mone & \Mtwo & $\mathbf{G_0=\varphi_0\lamb}$ & \Mtwo{} params $(\lams,\varphi_0,a,b)$\\
       &        &       &       & [$\WmK$]             & \\
\midrule
graphite (thin)  & 27.3\% & 9.7\%  & \textbf{1.8\%} & $1.22$ & $(24.8,\;0.0094,\;-0.024,\;0)$\\
graphite (thick) & 48.6\% & 17.0\% & \textbf{5.4\%} & $1.57$ & $(5.0^{\dagger},\;0.0121,\;-0.042,\;0.053)$\\
NMC622           & 19.1\% & 14.2\% & \textbf{1.4\%} & $0.41$ & $(1.6^{\dagger},\;0.0171,\;-0.089,\;0.039)$\\
NMC811           & 29.3\% & 12.9\% & \textbf{9.4\%} & $0.12$ & $(1.5^{\dagger},\;0.0048,\;-0.049,\;0)$\\
\midrule
\textbf{average} & \textbf{31.1\%} & \textbf{13.5\%} & \textbf{4.5\%} & & \\
\bottomrule
\end{tabular}
\end{table}

\subsection{Two competing mechanisms: reorientation versus contact}
\label{sec:results:mechanisms}

\emph{\textbf{Core result (secondary, bounded claim).}} For the graphite anode, conductivity data alone do not identify a unique mechanism, because a bounded reorientation account and a contact-only account fit comparably, so we treat them as a mechanism competition rather than a resolved interpretation. For the cathode, contact evolution is required within the tested hierarchy. The ablation shows the contact quadratic \emph{can} fit every family, but fit alone does not identify the anode mechanism. The reorientation model \MR{} (Sec.~\ref{sec:model:reorient}) provides the test, because it is physically constrained in two ways the contact quadratic is not. Its through-plane solid conductivity cannot fall below the graphite $c$-axis floor, and it is inert for isotropic actives. Table~\ref{tab:mech} confronts the two models with this in mind.

\begin{figure}[tb]\centering
\includegraphics[width=0.88\textwidth]{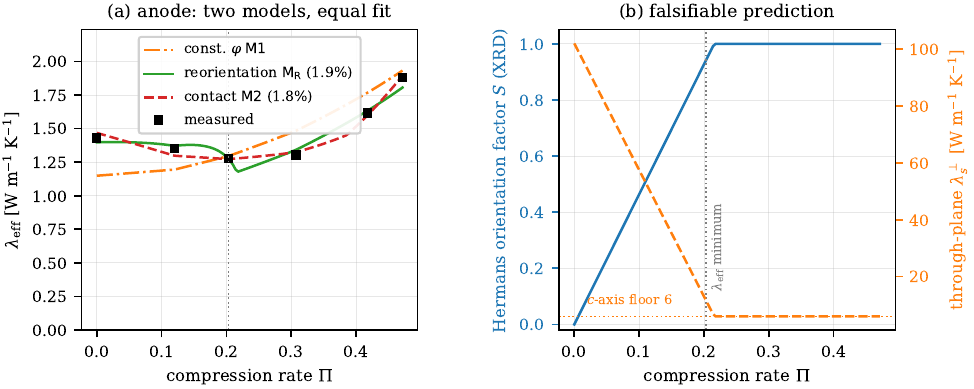}
\caption{The anode mechanism competition. (a)~The reorientation closure \MR{} (solid) and the contact quadratic \Mtwo{} (dashed) describe the graphite u-shape equally well and are not distinguished by conductivity data; \MR{} is the more parsimonious member (one fewer parameter, physically constrained) but is not uniquely identified. The visible change in slope occurs when the fitted $S(\Pi)$ reaches the saturation bound. (b)~Falsifiable prediction: the Hermans flake-orientation factor $S(\Pi)$, directly measurable by XRD texture, rises and saturates near the conductivity minimum (dotted line), driving the through-plane solid conductivity $\lams^{\perp}$ from the random-orientation average down to the graphite $c$-axis floor.}
\label{fig:reorient}
\end{figure}

For the graphite anode (Fig.~\ref{fig:reorient}a), \MR{}  carries the calendering dependence entirely through flake alignment $S(\Pi)$ at \emph{constant} contact $\varphi_0$ and matches the contact quadratic on accuracy ($1.9\%$ vs $1.8\%$ thin, $5.2\%$ vs $5.4\%$ thick) using one fewer parameter. It does so while driving the through-plane solid conductivity from the random-orientation average ($\approx\!102\WmK$) at $\Pi=0$ down to the $c$-axis floor ($6\WmK$) at full calendering, as $S$ rises from $0$ to $1$. The contact quadratic, left free, settles $\lams$ at an unexplained $25\WmK$. By contrast, \MR{} derives a low through-plane $\lams$ from flake alignment and lands in the $5$--$30\WmK$ identifiability valley isolated by the profile likelihood (Sec.~\ref{sec:results:errors}) and Bayesian posterior (Sec.~\ref{sec:results:bayes}). The low anode solid conductivity is therefore consistent with the $c$-axis of aligned flakes. The forward closure of Oehler places the graphite conductivity at the opposite, isotropic end of this band ($138.6\WmK$) and sets anisotropy aside \cite{oehlerdiss}, which is defensible for randomly oriented as-coated flakes but not for the aligned state calendering produces, one reason a static-conductivity closure cannot follow the anode along its calendering trajectory.

For the NMC cathode the same model \emph{fails} ($19.1\%$, $29.3\%$), and the failure is diagnostic rather than disappointing. NMC is quasi-isotropic, so reorientation is inert ($\lambda_a=\lambda_c$) and \MR{} collapses to a constant-$\lams$, constant-$\varphi$ closure that cannot bend the cathode curve downward, so its error equals the zero-fit baseline exactly. The cathode dip therefore \emph{requires} contact-network damage, the mechanism \MR{} lacks, and the contact quadratic recovers it ($1.4\%$, $9.4\%$). The cathode dip \emph{requires} contact-network damage (reorientation is inert for the isotropic active), whereas the anode is consistent with \emph{either} mechanism. So contact evolution is \emph{identified} as necessary for the cathode, while for the anode the two remain a bounded mechanism competition in which reorientation is parsimonious and consistent, but \emph{not uniquely identified}, degenerate with a contact-only fit until same-sheet XRD texture is measured (Sec.~\ref{sec:experiments}).

The morphological premise underlying this split is independently corroborated. Marconnet et al.\ \cite{marconnet2024}, imaging electrodes extracted from a different cell (a graphite anode with an NCA/LCO cathode), report that the anode active particles are flakes whereas the cathode particles are three-dimensional, the very anisotropic-versus-isotropic contrast the two models exploit; and they identify the network of active particles and its carbon-black-binder bridges, rather than the bulk solid, as the control on through-plane thermal transport, consistent both with our reorientation account of the anode and with the contact term that carries the cathode. They also measure a higher thermal diffusivity in helium than in nitrogen for these porous electrodes, the pore-gas signature the Knudsen term of Sec.~\ref{sec:model:knudsen} quantifies.

\begin{table}[tb]\centering\small
\caption{Mechanism test. \MR{}: reorientation-coupled (3 parameters $S_0,s,\varphi_0$; through-plane $\lams$ set by flake alignment). \Mtwo{}: contact quadratic (4 parameters; $\lams$ free). Lower MAPE is better, but differences below the measurement scatter are not meaningful; the graphite rows therefore cannot be separated by fit and remain a bounded mechanism competition awaiting same-sheet XRD, while the cathode rows are separated by fit (contact required, reorientation inert).}
\label{tab:mech}
\begin{tabular}{lccp{0.34\textwidth}}
\toprule
Family & \MR{} (reorientation) & \Mtwo{} (contact) & favoured interpretation\\
\midrule
graphite (thin)  & 1.9\% & 1.8\% & either fits; reorientation parsimonious and bounded but not uniquely identified, degenerate with a contact-only fit until same-sheet XRD texture is measured\\
graphite (thick) & 5.2\% & 5.4\% & either fits; reorientation parsimonious and bounded but not uniquely identified, degenerate with a contact-only fit until same-sheet XRD texture is measured\\
NMC622           & 19.1\% & 1.4\% & contact required; reorientation inert (isotropic morphology)\\
NMC811           & 29.3\% & 9.4\% & contact required; reorientation inert (isotropic morphology)\\
\bottomrule
\end{tabular}
\end{table}

Two checks bound the claim. First, on the anode conductivity data \emph{alone}, $S(\Pi)$ and $\varphi(\Pi)$ are partially degenerate because both can reduce the downswing, and equal MAPE does not identify one mechanism. The case for reorientation rests on its lower parameter count, its explanation of the low $\lams$, and its prediction of the anode/cathode contrast from anisotropy. Second, the in-plane flake conductivity $\lambda_a$ is a literature value with spread, and the qualitative result is robust to it (anode MAPE stays $1.9$--$3.0\%$ for $\lambda_a\in[100,600]\WmK$, with $S$ always rising to saturation), while the absolute as-coated $S_0$ is not, so we report the \emph{trend} of $S(\Pi)$, not its baseline. A direct texture measurement resolves the ambiguity (Fig.~\ref{fig:reorient}b). $S$ is the Hermans factor read out by XRD texture analysis, and the model predicts it rises and saturates near the conductivity minimum. For the thin anode, $S$ reaches $0.94$ by $\Pi=0.20$, the compression at which $\leff$ bottoms out, and $1.0$ by $\Pi=0.31$. Section~\ref{sec:experiments} specifies the texture measurement as the falsification test.

\begin{figure}[tb]\centering
\includegraphics[width=0.5\textwidth]{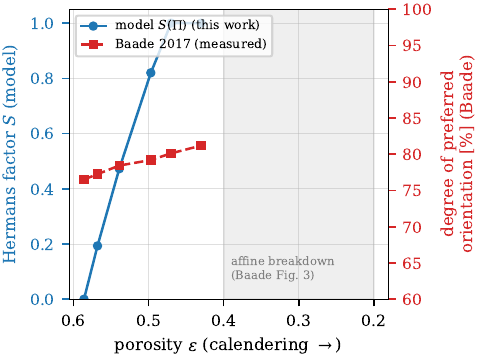}
\caption{Trend-level consistency of the reorientation mechanism with measured graphite texture (Baade et al.\ \cite{baade2017}, Table~I). Measured degree of preferred $(002)$ orientation (right axis, squares) is high as-coated and rises modestly with calendering; the model Hermans factor $S$ (left axis) follows the same direction. The shaded band marks the porosity below which Baade reports the affine reorientation model breaking down, corroborating the $S(\Pi)$ saturation. Absolute scales differ by definition; only the trend is compared.}
\label{fig:texture}
\end{figure}

\paragraph{Consistency with measured texture.} Two independent XRD-texture datasets already let us test the orientation mechanism against measurement, ahead of the in-house campaign (Fig.~\ref{fig:texture}, module \texttt{texture\_overlay.py}). Baade et al.\ \cite{baade2017}, calendering one graphite anode from $\varepsilon=0.586$ to $0.430$, measure a degree of preferred $(002)$ orientation that is already high as-coated ($76.5\%$, multiples-of-random $\approx11$) and rises modestly to $81.2\%$. Malifarge et al.\ \cite{malifarge2017} likewise report substantial as-coated texture rising with calendering. Two consequences follow, both trend-level (graphite grade, recipe, and orientation definition differ between sources, so this is corroboration, not calibration). First, the measured as-coated texture anchors our assumed $S_0=0.5$--$0.7$ quantitatively rather than by assertion. Second, the measured rise is \emph{modest}, and Baade finds the affine reorientation model breaks down below $\sim$$40\%$ porosity (particles no longer free to rotate). A pure-reorientation account that starts from a random as-coated state and climbs steeply to saturation therefore over-states the alignment change, which is the quantitative basis for the reading (Section~\ref{sec:limitations}, item~4) that orientation sets a high baseline while contact evolution carries much of the calendering increment.

A formal Bayesian model comparison gives the same reading (full No-U-Turn sampler (NUTS) posteriors, both anode families). On fit the two models are statistically indistinguishable. The WAIC difference is small (about one or less for both anodes, far below the $\sim$4 that would mark a real difference) and marginally favours reorientation, which uses equal or fewer effective parameters. This indistinguishability is a \emph{structural} non-identifiability, not merely an artifact of the small sample ($6$--$8$ states). A falling $S(\Pi)$ acting on $\lams^{\perp}$ and a falling $\varphi(\Pi)$ acting on the contact term enter the closure through the same particle-channel product and produce the same downswing in $\leff$, so the two are degenerate along a direction of the parameter space by construction. More conductivity data of the same kind, at any sample size, would shrink the error bars but not separate the mechanisms. Only an observable that couples to one mechanism and not the other (flake texture, which moves with $S$ but not $\varphi$) breaks the degeneracy, which is the structural reason Sec.~\ref{sec:experiments} requires it. The models differ in how they treat the through-plane solid conductivity. The contact model leaves it unresolved, with posterior mass spanning the identifiability valley ($8$--$30\WmK$ thin, $5$--$13\WmK$ thick), whereas the reorientation model derives it from flake orientation, with the posterior collapsing onto the graphite $c$-axis floor ($6\WmK$). Adding a single X-ray diffraction (XRD) orientation observation $S(\Pi)$ to the likelihood cuts the $90\%$ posterior width of the alignment rate by $55$--$60\%$, which is why Sec.~\ref{sec:experiments} makes it the mechanism test.

A continuous mixture-of-mechanisms model replaces the binary \Mtwo{}-versus-\MR{} comparison with a posterior over a single reorientation-fraction weight $w$ (zero pure contact, one pure reorientation). On the thin anode the in-sample fit is nearly flat in $w$, so the posterior is wide (mean $0.6$, $90\%$ interval $\approx[0.25,0.90]$). On the thick anode the pure-contact limit fits poorly, excluding $w<0.45$ but still spanning $[0.45,0.75]$. Leave-one-state-out cross-validation does not pin $w$ either, selecting an interior but family-dependent optimum ($\approx0.1$ thin, $\approx0.65$ thick), where a mixture beats both pure end-members, yet the two anode families disagree on the proportion. Neither the fit nor held-out error identifies the mechanism fraction from conductivity, which is the quantitative case for the texture measurement.

Because the mechanism fraction is unresolved by conductivity, we do not commit to a single closure for prediction. The default extrapolative output is a three-model Bayesian average over the contact quadratic \Mtwo{}, the reorientation \MR{}, and their mixture, weighted equally (formal WAIC weights are themselves unstable on six to eight states, Sec.~\ref{app:bayes}). On the calibration data the three members agree to a few percent, so the averaging is invisible. In extrapolation their between-mechanism spread becomes the dominant structural uncertainty (a $90\%$ band of $2.8$--$4.6\WmK$ at $\Pi=0.55$), which the average carries explicitly. The two competing mechanisms thus enter the predictive output together until a texture measurement collapses the weight onto one. Simulating that collapse ahead of the experiment (module \texttt{multimodal\_fusion.py}), a mock graphical model ingesting simulated XRD texture points with realistic noise leaves $w$ at its prior under conductivity alone (posterior std $0.29$, the structural degeneracy restated), whereas \emph{two} texture points near the conductivity minimum collapse the $w$ posterior to std $\approx0.09$ (and to $\approx0.03$ by eight points). The slope of the Hermans factor against calendering carries the information, so a couple of XRD measurements suffice to resolve the mechanism, consistent with the information-gain design of Section~\ref{app:campaign}, and an SEM contact-area channel would constrain the contact side in the same joint posterior.

\paragraph{Orthogonal-direction check: the in-plane conductivity.} The strongest test of the reorientation premise comes from a direction the calibration never saw. Because the flake-tensor trace is rotation-invariant, the alignment factor $S(\Pi)$ that we fit to the \emph{through-plane} data also fixes the \emph{in-plane} solid conductivity $\lams^{\parallel}(S)=(2\lambda_a+\lambda_c-\lams^{\perp})/2$ with no additional parameter (module \texttt{inplane\_validation.py}). Routed through the parallel cell appropriate to a continuous in-plane network ($k_{\parallel}\approx(1-\varepsilon)^{m}\lams^{\parallel}$ with the solid fraction $1-\varepsilon$ and either the parallel bound $m=1$ or the Bruggeman exponent $m=1.5$, neither tunable), the calibrated state at $\varepsilon\approx0.40$ predicts $k_{\parallel}\approx74$--$94\WmK$. This brackets the independently measured anode-coating in-plane conductivity of Loges et al.\ \cite{loges2016} ($80$--$90\WmK$ by photothermal deflection spectroscopy), a quantity roughly $50\times$ larger than the through-plane values the model was fitted to. The predicted coating anisotropy $k_{\parallel}/k_{\perp}$ rises with calendering from $\approx18\times$ to $\approx45\times$ (Bruggeman) as the flakes align, the same sign reported by the XRD texture index and by Maleki et al.\ \cite{maleki1999}, who measured an in-plane/cross-plane ratio of nearly $10\times$ for a stacked electrode assembly (diluted by separator and foils, so the coating-only value is larger) and attributed it already in 1999 to ``the orientation of particles or flakes during the formation of thin electrode layers'', which is the reorientation mechanism, named from an independent measurement decades before this closure. A mechanism fitted in one direction reproducing the orthogonal-direction magnitude, sign, and trend, with no free parameter, is evidence the through-plane-only fit cannot manufacture. The deliberately series-celled estimate of \texttt{anisotropy.py} ($\approx2.5\times$) is a lower bound that brackets the same result from below. A further independent measurement, by a different laboratory and a steady-state (not flash) method, reaches the same qualitative conclusion. Liu et al.\ \cite{liu2024} report ``strongly anisotropic thermal transport'' in both electrodes, with the through-plane conductivity well below the in-plane value.

\subsection{Design principles from the fitted parameters}
\label{sec:results:design}

\emph{\textbf{Enabled extension (design rules from the fitted closure).}} The fitted contact parameters are measurements of contact physics and, compared across recipes in \emph{volume} rather than weight (Section~\ref{app:design}, Table~\ref{tab:design}), yield design rules. The as-coated bridge conductance $G=\varphi_0\lamb$ is $\sim$$20\times$ larger for the self-bridging graphite anodes than the additive-bridged cathodes, a $2\,$wt\% flake-graphite additive doubles NMC622's bridging over carbon-black-only NMC811, stiff PVDF networks shear $2$--$4\times$ faster than elastomeric CMC/SBR, and the calendering setpoint should avoid the conductivity/adhesion dip. A contact-mechanics reading (Section~\ref{app:design}) further derives the downswing as binder-network shear damage, with the upturn degenerate between Hertzian contact regrowth and densification, separable only by the scanning electron microscopy (SEM) measurement of Sec.~\ref{sec:experiments}.

\subsection{Error analysis: identifiability and transfer}
\label{sec:results:errors}

\paragraph{What the data can and cannot determine.}
With four parameters fitted to 6--8 points per family, one must ask which parameters the data actually pin down. A profile likelihood over $\lams$ answers this (Fig.~\ref{fig:error}a). Fixing $\lams$ anywhere in the range $5$--$30\WmK$ and refitting the rest keeps the error within two points of optimal, with $\varphi_0$ compensating between $0.0073$ and $0.0120$, whereas beyond $\lams\approx40$ the fit degrades beyond repair. The data therefore reject the high, in-plane-dominated end of the anisotropy band, consistent with through-plane transport crossing the flakes' weak axis. At the same time, parameter values inside the valley compensate each other, so we report the grouped quantities of Table~\ref{tab:design} together with the valley bounds as the parameter uncertainty. Figure~\ref{fig:error}c--f adds residual diagnostics against porosity, compression rate, predicted value, and family. No remaining trend is comparable to the removed u-shape, although NMC811 remains visibly wider.

\paragraph{Out-of-sample tests on the existing data.}
Three held-out tests quantify the small-data exposure without new measurements. \emph{Leave-one-state-out} cross-validation, dropping each calendering state in turn and predicting it from the rest, roughly doubles the error relative to the in-sample fit (contact closure: $4.5\%$ in-sample to $7.8\%$ held-out aggregate, and $1.8\%\to6.1\%$ for the thin anode), making the four-parameters-against-six-states exposure explicit. On the graphite anode, where both closures apply, the three-parameter reorientation model has a \emph{lower} held-out error than the four-parameter contact quadratic ($4.8$ and $6.6\%$ versus $6.1$ and $10.9\%$ for thin and thick), so the extra contact parameter overfits the anode. This is out-of-sample support for reorientation parsimony, independent of the equal in-sample fit, while for the isotropic cathode reorientation is inert and only the contact term generalises.
\emph{Leave-one-family-out} transfer within a chemistry degrades further ($21\%$ thin$\to$thick graphite, $40\%$ NMC622$\to$NMC811), localising the recipe dependence that the hybrid swaps below attribute to a single parameter. A recipe-wise empirical interpolation baseline is included as a sanity check. A quadratic fit in porosity reaches $0.9\%$ aggregate leave-one-state-out MAPE, and a shape-preserving piecewise-cubic Hermite (PCHIP) interpolation in $\Pi$ reaches $1.8\%$. These baselines interpolate the small, smooth within-recipe series better than any mechanistic closure, but they do not transfer across recipes, predict gas or pressure effects, expose grouped parameters, or run as an inverse outside the observed curve. Across the three modelling approaches (Table~\ref{tab:benchmark}), the closure is deliberately \emph{not} the best interpolator, but it is the only one that also transfers, extrapolates safely, responds to gas and pressure, inverts for quality control, and exposes interpretable grouped parameters.

\begin{table}[tb]\centering\small
\caption{What each modelling approach delivers on the calendering data. The mechanistic closure is not the best \emph{interpolator}, yet it is the only approach that transfers across recipes, extrapolates safely, responds to gas and pressure, inverts for quality control, and exposes interpretable grouped parameters. \capY~yes; \capP~partial or conditional; \capN~no.}
\label{tab:benchmark}
\begin{tabular}{@{}lccc@{}}
\toprule
                                    & Mechanistic        & Empirical          & Residual \\
Capability                          & closure (\Mtwo)    & interpolation      & learner (GP)\\
\midrule
Within-recipe interpolation (LOSO)  & $7.8\%$            & $\mathbf{0.9}$--$\mathbf{1.8\%}$ & $\sim$$1\%$ in-sample\\
Cross-recipe transfer               & \capP~re-anchor $\varphi_0$ & \capN     & \capN~(overfits)\\
Extrapolation safety                & \capY~bounded floors & \capN~diverges   & \capN~worse than physics\\
Gas / pressure response             & \capY~(Knudsen)    & \capN              & \capN\\
Differentiable inverse (QC)         & \capY~unique       & \capN              & \capP~via the closure\\
Interpretable grouped parameters    & \capY~($G_0,\,S$)  & \capN              & \capN\\
\bottomrule
\end{tabular}
\end{table}
The held-out residuals also recalibrate the uncertainty. The $2\%$ model-form floor is an interpolation floor (only $\sim$$30$--$40\%$ of leave-one-out residuals fall within twice it), whereas the held-out root-mean-square relative error is $\sim$$12\%$ for the contact closure. That wider figure, carried by the Bayesian model averaging in extrapolation, is the predictive width for out-of-sample use. We also tested whether a learned discrepancy could do better. A Gaussian-process model fitted to the closure residuals reduces the in-sample residual to $\sim$$1\%$ but overfits, generalising worse than the physics alone in leave-one-out (median held-out error $3.1\%$ vs $2.7\%$, with a wider conformal band). At 27 states, conformal prediction on the physics residuals supplies the distribution-free, finite-sample coverage (nominal $90\%$ half-width $\sim$$8\%$), and a learned discrepancy layer needs more data.

\paragraph{Why a quadratic $\varphi(\Pi)$, benchmarked.} We test whether the quadratic is the right flexibility by sweeping the correction capacity under the same leave-one-state-out protocol (module \texttt{correction\_benchmark.py}). A free (interpolating) spline drives the in-sample error to $\sim$$0\%$ but its held-out error becomes erratic and family-dependent ($2$ to $77\%$, mean $\sim$$27\%$) and its extrapolated contact fraction at $\Pi=0.55$ reaches unphysical values (up to $\sim$$0.25$), which shows unconstrained flexibility is unstable and unsafe on six to eight states per family. Among constrained low-order polynomials (linear through quartic) the held-out errors are statistically indistinguishable ($6$--$7.5\%$, all below the $\sim$$5\%$ measurement scatter), so capacity beyond the quadratic lowers the in-sample MAPE ($3.6\to3.0\%$) but buys nothing out-of-sample. What selects the quadratic is therefore not held-out accuracy (the low orders tie) but parsimony and extrapolation. It is the lowest order that represents the contact dip the cathode requires, it ties the higher orders on held-out error with fewer parameters, and it keeps $\varphi(\Pi=0.55)$ in a physical band where cubic and quartic begin to drift and the spline diverges. The quadratic is the flexibility/identifiability/extrapolation trade-off, with the benchmark confirming the extra flexibility is unwarranted.

\paragraph{Why cross-recipe transfer fails.}
Figure~\ref{fig:error}b dissects the $39.5\%$ transfer failure by hybrid swaps, replacing one parameter group at a time in NMC622's calibration with NMC811's own value. Swapping $\lams$ recovers almost nothing ($36.2\%$). Swapping $\varphi_0$ alone recovers about $90\%$ of the recoverable error ($12.2\%$, against the $9.4\%$ self-fit floor), and adding the damage dynamics $(a,b)$ contributes a further $0.3$ points. Cross-recipe failure is therefore concentrated in the as-coated bridge conductance set by the conductive-additive system. The closure does not transfer zero-shot across recipes, but because most recoverable error is concentrated in that one parameter group, a one-sheet re-anchoring of $\varphi_0$ is an effective minimal recalibration after chemistry-level transfer rather than a successful zero-shot transfer. In machine-learning terms this is a \emph{compositional-generalization} failure. The model must predict the behaviour of a new combination of recipe components (a different binder/additive system) it never saw assembled, exactly the regime where standard fitted models do not extrapolate. It is the natural target for meta-learning and few-shot methods that generalize across chemistries without full recalibration, and the physics-informed pretraining of Sec.~\ref{sec:discussion} (module \texttt{learned\_contact\_pretrain.py}) is a first step, injecting the closure's physics so that only the composition-specific bridge conductance need be learned from a few samples of the new recipe.

\paragraph{Hierarchical pooling confirms the re-anchorable parameter.}
A hierarchical (partial-pooling) Bayesian model draws each recipe's parameters from chemistry-level hyperpriors, letting a held-out recipe be predicted from the chemistry mean and testing the single-parameter transfer directly. With only two families per chemistry and \emph{uninformative} hyperpriors the pooling does \emph{not} tighten the per-family posteriors. The weakly identified hyper-scale slightly widens them, which is the expected small-group limit. Replacing the uninformative damage-rate prior with a binder-mechanics one (the stiff PVDF network of the cathodes is prior-centred to shear faster per unit compression than the elastomeric CMC/SBR of the anodes, Section~\ref{app:design}) repairs this where it matters most. The posterior then reproduces the measured ordering, with the PVDF cathode damaging $\sim$$3\times$ faster than the CMC/SBR anode, and the per-chemistry $a$ separates cleanly, a case where a physically-anchored hyperprior supplies the structure two families alone cannot. The chemistry-mean prediction still beats the no-pool donor transfer ($13.8\%$ vs $20\%$ for the thick anode, $20.7\%$ vs $39\%$ for NMC811), and re-anchoring \emph{only} $\varphi_0$ from a single as-coated state of the new recipe reduces NMC811 to $14.3\%$, roughly halving the held-out-family error. More recipes per chemistry would sharpen both the pooling and the hyper-scales.

Remaining exposures are quantified in Sec.~\ref{sec:limitations}. The largest is the Knudsen coefficient $\beta\in[1.5,2.0]$, which moves dry-electrode predictions by at most $1.7\%$ but the dry separator by $11.7\%$, the two-pressure experiment determining it directly. The supporting diagnostics, the profile-likelihood sweep over $\lams$, the hybrid-swap transfer decomposition, and the residual scatter, are collected in Fig.~\ref{fig:error} (Section~\ref{app:protocol}).

\subsection{What is robustly learned}
\label{sec:results:learned}
Section~\ref{sec:results:errors} quantified identifiability and transfer. Table~\ref{tab:learned} condenses those results into a support hierarchy that separates robust closure-level findings from grouped-identifiable parameters, bounded-but-non-unique mechanism inferences, and the orthogonal measurements that would resolve the latter. Section~\ref{app:variants} reports a set of proposed closure extensions and which of them are identifiable on the present data.

\begin{table}[tb]\centering\small
\caption{What the data support at each level of confidence, and what would resolve the rest. Separating robust closure-level findings from grouped-identifiable parameters, bounded-but-non-unique mechanism inferences, and the discriminating measurements is itself a result of the identifiability analysis.}
\label{tab:learned}
\begin{tabular}{@{}p{0.235\textwidth}p{0.70\textwidth}@{}}
\toprule
Status & Finding\\
\midrule
Robust (closure level) & The calendering u-shape is reproduced to the $\sim$$5\%$ noise level only by a compression-indexed contact term on Knudsen-corrected ZBS; porosity-only and static-contact closures give a monotonic rise (Secs.~\ref{sec:results:baselines}--\ref{sec:results:ablation}). Morphology is the discriminator: point-contact packing for electrodes, co-continuity for separators.\\
Grouped-identifiable parameters & The as-coated bridge conductance $G_0=\varphi_0\lamb$ ($\approx1.2$--$1.6\WmK$ anode, $0.1$--$0.4$ cathode) and the damage sign $a<0$ in every family; the Sobol sensitivity ordering (porosity, then the contact pair).\\
Bounded, not uniquely identified & The graphite anode's low through-plane $\lams$ (pinned to the $5$--$30\WmK$ valley, not a point); the contact-versus-reorientation split: \MR{} is consistent, parsimonious, and bounded, but degenerate with a contact-only fit and only marginally favoured by WAIC. This is a \emph{structural} non-identifiability, not a small-sample artifact.\\
Resolving measurement & Two-pressure dry $\leff$ pins the Knudsen $\beta$; same-sheet XRD $(002)$ texture pins the reorientation weight ($\sim$two points collapse its posterior, Sec.~\ref{sec:results:mechanisms}); paired guarded-hot-plate $+$ adhesion/SEM splits the lumped $\varphi$ into bulk and interface contributions and fixes the LFA/GHP scale.\\
\bottomrule
\end{tabular}
\end{table}

\subsection{Bayesian calibration and model evidence}
\label{sec:results:bayes}

Full posterior sampling with the No-U-Turn sampler (NUTS) reproduces the least-squares estimates and, probabilistically, the identifiability valley (thin-anode $\lams$ posterior $7.5$--$30.5\WmK$ at $90\%$), infers each family's noise level (including NMC811's poorer fit), and brackets the VDI rigid-sphere value with the $\varphi_0$ credible intervals. By WAIC, the process-dependent $\varphi(\Pi)$ is strongly favoured over a constant contact fraction for three of the four families ($\Delta=13$--$32$). The exception is NMC811, whose calendering window never reached interlocking, independently rediscovering the process-threshold finding. The full posteriors, the WAIC table, and a small-sample AICc cross-check (which agrees with WAIC that the parsimonious reorientation model is preferred, but disagrees on the contact term, exposing the unreliability of formal criteria at six to eight states) are in Section~\ref{app:bayes}. The evidence for the contact term therefore rests on error reduction (Table~\ref{tab:ablation}), the u-shape reproduction, and adhesion correlation, not on information criteria alone.

\subsection{Inverse feasibility study}
\label{sec:results:qc}

\emph{\textbf{Enabled extension (inverse feasibility).}} This section establishes \emph{feasibility} of the differentiable inverse on simulated data. The porosity inverse is exercised on a synthetic batch, and the interface/delamination monitor is a model-based detectability argument pending the campaign of Sec.~\ref{sec:experiments}.

\begin{figure}[tb]\centering
\includegraphics[width=0.82\textwidth]{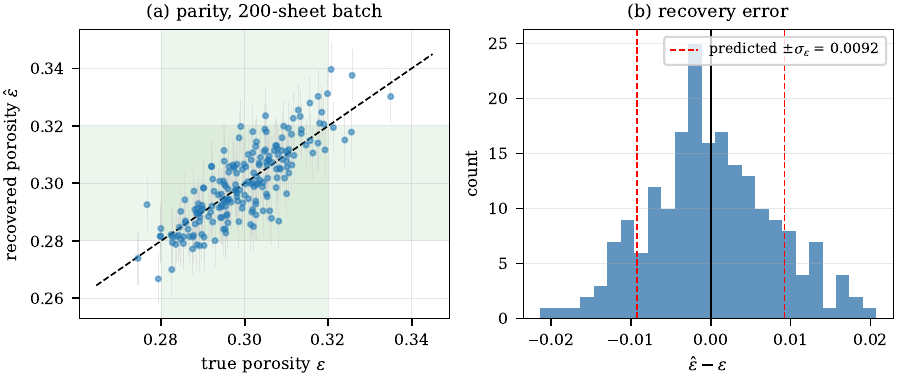}
\caption{Inverse porosity QC on a synthetic 200-sheet batch at $3\%$ measurement noise: (a)~parity with predicted $\pm\sigma_\varepsilon$ bars and the $\pm0.02$ specification window; (b)~recovery-error histogram against the first-order uncertainty prediction of Eq.~\eqref{eq:sigma}.}
\label{fig:qc}
\end{figure}

\paragraph{Porosity QC (numerical demonstration).}
On a synthetic production batch of 200 calendered sheets (true porosities drawn from $\mathcal N(0.30,\,0.012)$, a realistic calender drift, with $3\%$ relative measurement noise), Newton inversion (Eq.~\eqref{eq:newton}) recovers porosity with a root-mean-square error of $0.0078$ absolute and zero bias (Fig.~\ref{fig:qc}). The a-priori prediction of Eq.~\eqref{eq:sigma}, $\sigma_\varepsilon=0.0092$ including the $2\%$ model floor, brackets the observed scatter, so the uncertainty model is calibrated and slightly conservative, as is preferable for quality-control use. Against a typical $\pm0.02$ porosity specification this yields $87.5\%$ correct single-shot classification, with all errors confined to genuinely borderline sheets. Averaging $n=4$ repeats shrinks the measurement contribution by half and makes the specification window cleanly resolvable. For an inline accept/reject decision one wants a \emph{guaranteed} error bound rather than one contingent on the noise model, so we wrap the inverse in split-conformal prediction (module \texttt{conformal\_qc.py}). Calibrating the conformal radius on a held-out split gives a porosity interval with finite-sample, distribution-free coverage. On this batch the conformal band meets its nominal coverage ($\approx$$92\%$ at a $90\%$ target, $\approx$$96\%$ at $95\%$) and is in fact \emph{tighter} than the Gaussian $\pm1.64\sigma_\varepsilon$ band (half-width $\approx0.013$ vs $0.015$), because the first-order $\sigma_\varepsilon$ is mildly conservative. Conformal thus both confirms the analytic band and upgrades it to a guarantee, with a normalized variant that stays adaptive to the per-sheet sensitivity.

\paragraph{Porosity QC on the real data.} Run on the actual Gandert sheets rather than a synthetic batch, the inverse is coarser (module \texttt{inverse\_realdata.py}). Using a held-out protocol (calibrate each family on the other sheets, then invert the held-out sheet's measured conductivity to porosity) to avoid circularity, it recovers the reported per-sheet porosity to a root-mean-square error of $\sim$$0.05$ ($0.018$--$0.072$ across families, worst for NMC811 whose conductivity residual is largest), with $44\%$ of sheets inside the $\pm0.02$ window. This is several times the synthetic figure ($0.008$), because the real held-out error is set by the model-form and calibration residual and by the metrology-state coupling in the conductivity reconstruction, not by measurement noise alone, which the synthetic study assumed away. The inverse therefore recovers real porosities to $\sim$$0.02$--$0.07$, a useful coarse or secondary check, not the synthetic $\pm0.008$, and the gap is exactly the metrology-state dependence quantified next.

\paragraph{Interface and delamination QC (application hypothesis).}
The calibrated closure decomposes $\leff$ into a core contribution and a bridge contribution, and the split is large. In the as-coated state, which is exactly where post-drying adhesion and binder-migration quality control matters, bridges carry between $16\%$ and $66\%$ of the heat. In the model, a failure of the bridge network (delamination, binder maldistribution) therefore shifts a thermal reading by $5$ to $22$ standard deviations of a $3\%$-noise measurement, so even partial degradation would be detectable in a single shot. This is a model-based detectability argument. Porosity can already be monitored inline from thickness and mass loading, while interface quality lacks an inline monitor and is a plausible target for fast thermal screening, consistent with the conductivity--adhesion correlation reported independently \cite{gandert2023}. Demonstrating it on real degraded coatings is a task for the campaign of Sec.~\ref{sec:experiments}. A thermal reading is moreover one channel in a multi-modal inline stack. Open datasets pairing calendered electrodes with ultrasonic spectra \cite{guk2026} and with roll-temperature/porosity/adhesion sweeps \cite{hidalgo2023} are complementary process-monitoring modalities (they carry no thermal conductivity, so we cite them as such, not as thermal validation).

\begin{figure}[tb]\centering
\includegraphics[width=\textwidth]{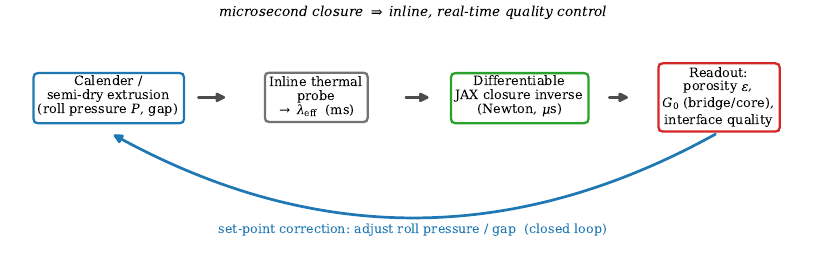}
\caption{The differentiable closure as a closed-loop inline monitor for continuous (semi-dry, granulate-based) electrode production. A fast through-plane thermal reading is inverted in microseconds (Newton on the JAX closure, Eq.~\eqref{eq:newton}) to porosity $\varepsilon$, the as-coated bridge conductance $G_0$, and an interface-quality flag; the readout feeds back to the calender roll-pressure and gap set-point. The same closure is the forward model of a process digital twin and, inverted, its online estimator.}
\label{fig:inline}
\end{figure}

\paragraph{Toward closed-loop inline sensing.} The point of making the inverse differentiable is that it can run on a moving line, not only offline. In continuous semi-dry, granulate-based electrode production, the route the GranuGoIn project targets, the calender is the dominant porosity- and contact-setting step, and a fast non-contact through-plane thermal reading, inverted in microseconds (Eq.~\eqref{eq:newton}), returns porosity together with the as-coated bridge conductance $G_0$ and an interface-quality flag from the bridge/core split, all from a single measurement. Fed back to the roll pressure and gap, this closes a control loop around the calender (Fig.~\ref{fig:inline}). The same closure is the forward model for a process digital twin and, inverted, the online estimator that the twin is steered by. The decisive enabler is speed. The closure evaluates and differentiates in microseconds, against the hours a microstructure-resolved solve of the same coating would take, so it is fast enough to live inside a real-time control loop rather than a post-hoc analysis.

\paragraph{External consistency checks (cell scale and across laboratories).} Two checks outside the calibration window are detailed in Section~\ref{app:meta} and summarised in Table~\ref{tab:robustness}. A bottom-up prediction of a measured commercial 18650 jelly roll reproduces the cell-level through-plane conductivity to within $+15\%$ with perfect interfaces and infers a physically coherent (tensioned) inter-layer contact resistance two orders of magnitude below a torn-down stack. The zero-fit closure tracks twelve coating and separator measurements from three further laboratories at a median $37\%$ relative error, with $83\%$ of points inside a factor of two. Both are order-of-magnitude scale checks across unseen sources, not calibration, and a Sobol decomposition there ranks porosity and the contact pair $(\varphi,\lamb)$ as the dominant sensitivities.

\subsection{Robustness summary}
\label{sec:results:robustness}
\begin{figure}[tb]\centering
\includegraphics[width=\textwidth]{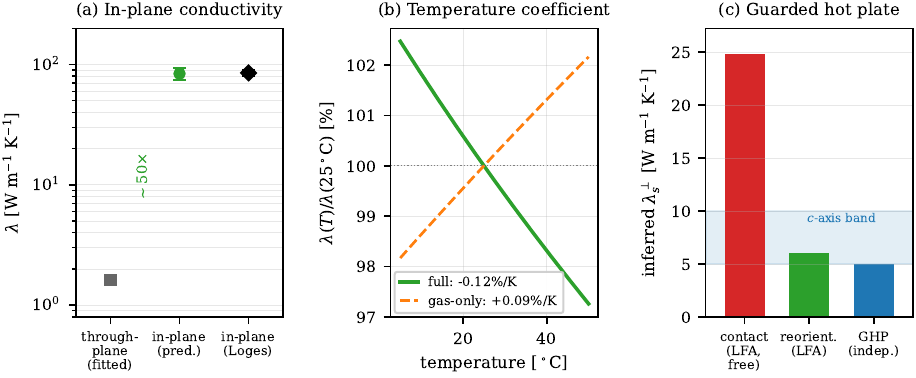}
\caption{\emph{External / orthogonal consistency checks (not calibration tiers).} Calibrate once, check three independent ways. The closure is calibrated only on room-temperature, through-plane laser-flash data, yet retains the relevant trend and ordering for three quantities it never saw, each with no new parameter; absolute scale remains method-scoped. (a)~\emph{In-plane conductivity} (a direction the fit never saw): the value predicted by trace conservation ($74$--$94\WmK$, the band spanning the Bruggeman and parallel cells) matches the independent photothermal measurement of Loges ($80$--$90\WmK$) and is $\sim$$50\times$ the through-plane values the model was fit to. (b)~\emph{Temperature coefficient} (a second physical axis): the full closure predicts a negative slope (solid control, the sign Loges and Werner measure; legend gives the slopes), whereas the gas-only counterfactual predicts the wrong sign. (c)~\emph{Guarded-hot-plate metrology} (a second method): an independent guarded-hot-plate calibration places the graphite through-plane solid conductivity in the $c$-axis band, the same floor the reorientation mechanism derives from laser-flash data, while the unconstrained contact fit floats at an unexplained $25\WmK$.}
\label{fig:validation}
\end{figure}

\emph{\textbf{Orthogonal consistency checks.}} Table~\ref{tab:robustness} collects every stress test the closure has been put through, with its verdict. The pattern is deliberate. The model's \emph{form} retains trend-level consistency and preserves the mechanism-relevant ordering across changes of metrology, direction, and temperature, and its mechanism premise is corroborated by independent measurements, while what is method-scoped (the absolute scale, which shifts $1.8$--$2.4\times$ under guarded-hot-plate recalibration) or under-determined (the mechanism fraction from conductivity alone) is marked. Three of these are orthogonal to the calibration entirely, holding out a second method (guarded hot plate), a second direction (in-plane), and a second physical axis (temperature). The closure was fit to none of them, yet retains the relevant trend and ordering in all three, while absolute scale remains method-scoped and is not claimed to transfer unchanged (Fig.~\ref{fig:validation}).

\begin{table}[tb]\centering\footnotesize
\caption{Robustness summary. Each row is an out-of-sample, orthogonal, or sensitivity check and its verdict; ``method-scoped'' and ``under-determined'' findings are labelled as such. Modules are in \texttt{src/}.}
\label{tab:robustness}
\begin{tabular}{@{}p{0.205\textwidth}p{0.30\textwidth}p{0.40\textwidth}@{}}
\toprule
Check & What it probes & Verdict\\
\midrule
Cross-validation (LOSO/LOFO) & held-out states and families & Closure extrapolates; the contact quadratic is not uniquely best (low-order forms tie, a free spline is unstable). \texttt{stats\_validation}, \texttt{correction\_benchmark}\\
Conformal coverage & interval calibration & Nominal $90\%$ coverage met out of sample. \texttt{stats\_validation}\\
GHP metrology \emph{(orthogonal: 2nd method)} & guarded hot plate, air, $2$--$6$\,bar & Form transfers (fits both methods to $\sim$$6.5\%$); absolute scale $1.8$--$2.4\times$ lower, \emph{method-scoped}; graphite $\lams$ independently pins the $c$-axis floor. \texttt{ghp\_calibration}\\
In-plane conductivity \emph{(orthogonal: 2nd direction)} & axis the fit never saw & Parameter-free $k_{\parallel}=74$--$94\WmK$ matches Loges' $80$--$90$; anisotropy rises with calendering (Maleki's sign and attribution). \texttt{inplane\_validation}\\
Temperature coefficient \emph{(orthogonal: 2nd axis)} & sign of $\mathrm{d}\leff/\mathrm{d}T$ & Predicts the measured \emph{negative} slope (solid control); a gas-controlled bed gets the sign backwards. \texttt{temperature\_forward}\\
XRD texture & reorientation premise & $S(\Pi)$ rises as required, but modestly and from a high baseline; affine rotation breaks down below $\sim$$40\%$ porosity. \texttt{texture\_overlay}\\
$\lambda_a$ sensitivity & uncertain literature input & Anode MAPE stays $1.9$--$3.0\%$ for $\lambda_a\in[100,600]$; $S$ always saturates, so we report the \emph{trend}, not $S_0$.\\
Model evidence & mechanism selection & WAIC nearly ties (slight \MR{} edge); small-sample AICc favours the simpler constant-$\varphi$ model at $n=6$; the mixture weight $w$ is \emph{under-determined} by conductivity. \texttt{bayes\_mechanism}, \texttt{mixture\_mechanism}\\
Differentiable inverse & real-data porosity readout & Recovers porosity at $\sim$$0.05$ RMSE on real data (coarser than the synthetic $\pm0.008$ feasibility bound). \texttt{inverse\_realdata}\\
18650 cell stack & scale extrapolation & Within $+15\%$ with perfect contacts; inferred interface resistance two orders below a torn-down stack, coherent for a tensioned jelly roll. \texttt{stack\_model}\\
\bottomrule
\end{tabular}
\end{table}

\section{Discussion}
\label{sec:discussion}

\paragraph{Scope of the calibrated quantity.} The target is the \emph{LFA-derived apparent through-plane coating conductivity} obtained by the source subtraction of Sec.~\ref{sec:data:primary}. On these electrodes, laser-flash and guarded-hot-plate methods disagree by a factor of $2.3$ to $4.0$ \cite{gandert2026}, and recalibrating the closure on the guarded-hot-plate data shifts its absolute scale by $1.8$ to $2.4\times$. The disagreement is not a pure scale offset (the u-shape is most pronounced in the LFA data, Sec.~\ref{sec:limitations}, point on measurement-method dependence). The fitted $\lams$ and, through their confound, $\varphi_0$ are method-scoped effective quantities, while the mechanism tests rest on texture and contact-area measurements that are independent of the thermal method. That the inherent coating conductivity and the inter-layer contact resistance are genuinely separable quantities, not an artifact of our reading, is confirmed independently by Liu et al.\ \cite{liu2024}, who isolate both for individual battery layers by a multi-layer steady-state method and find the thermal contact resistance alone contributes more than a quarter of the total stack thermal resistance, decreasing with applied pressure as the contact account requires.

The results distinguish the useful roles of the model classes. \emph{Analytic closures} remain the right tool when evaluations must be cheap, differentiable, or invertible, which covers cell-level thermal simulation, process optimization, and quality control. Two conditions matter here. The geometry must match the material, and Sec.~\ref{sec:results:baselines} shows point-contact packing for low-additive electrodes and co-continuity for separators. The contact state must also evolve with process history, and by Table~\ref{tab:ablation} this term supplies most of the accuracy gain. \emph{Microstructure-resolved simulation} is better suited to hypothetical microstructures before they exist, but current generators do not encode calender-induced contact evolution \cite{gandert2023}. \emph{Purely data-driven regression} interpolates these 27 points well, as the empirical baselines show, but it has no recipe-transfer diagnosis, pressure/gas response, inverse use, or direct mechanism test. The closure's value is the combination of interpretable grouped parameters, a one-measurement transfer correction (Sec.~\ref{sec:results:errors}), and falsifiable predictions (Sec.~\ref{sec:experiments}). Because the contact and reorientation closures fit the anode equally well, goodness of fit alone cannot select the mechanism; the choice depends on bounded conductivity, inertness for isotropic actives, and finally on the direct texture measurement. No prior class spans both the physics rows and the cheap, invertible rows that manufacturing practice needs together (Table~\ref{tab:capability}), which is the gap this closure fills.

\begin{table}[tb]\centering\small
\caption{Capability matrix: which phenomena each model class predicts from a single calibration. \capY~yes; \capP~partial or input-dependent; \capN~no. ME, Maxwell--Eucken; Brugg., Bruggeman; ZBS, static point-contact closure; Sim., microstructure-resolved simulation; ML, purely data-driven regression. Prior closures and ML are cheap and invertible but miss the physics; resolved simulation captures some physics but is neither cheap nor invertible and still misses the calendering u-shape; only the present closure spans every row.}
\label{tab:capability}
\begin{tabular}{@{}lcccccc@{}}
\toprule
Capability & ME & Brugg. & ZBS & Sim. & ML & This work\\
\midrule
Through-plane u-shape (calendering)            & \capN & \capN & \capN & \capN & \capP & \capY\\
In-plane $k_{\parallel}$ with no extra fit      & \capN & \capN & \capN & \capP & \capN & \capY\\
Temperature-coefficient sign (solid control)    & \capN & \capP & \capY & \capY & \capN & \capY\\
Metrology transfer (GHP/LFA) + scale split      & \capN & \capN & \capP & \capN & \capN & \capY\\
Process variable beyond porosity                & \capN & \capN & \capN & \capN & \capP & \capY\\
Mechanism separation (reorient./contact)        & \capN & \capN & \capN & \capN & \capN & \capY\\
Differentiable inverse for QC                    & \capY & \capY & \capY & \capN & \capP & \capY\\
Microsecond cost, cell-scale usable              & \capY & \capY & \capY & \capN & \capY & \capY\\
\bottomrule
\end{tabular}
\end{table}


\paragraph{Temperature coefficient: a sign test for solid control.} The calibration is isothermal, yet the closure forward-predicts how the coating conductivity should move with temperature, because two of its inputs move in opposite directions, since the pore-gas conductivity rises with temperature (dilute-gas kinetic theory, $\lamf\propto T^{0.77}$) while the graphite solid conductivity falls (phonon umklapp, $\lams\propto T^{-1}$ above the Debye temperature, a textbook parameter-free scaling). Which one wins is a discriminating test that uses no temperature data and no fitting (module \texttt{temperature\_forward.py}). A gas-controlled bed, such as the Maxwell--Eucken arrangement of isolated solid in a connected gas, would predict a \emph{positive} coefficient, whereas the gas-only counterfactual of our closure indeed gives $+0.09\%$/K. The full closure, in which a solid skeleton carries the through-plane heat, instead predicts a \emph{negative}, near-linear coefficient of order $-0.1$ to $-0.2\%$/K (steepening toward the solid-controlled limit as the pore-gas offset shrinks with a lower-conductivity gas). Loges \cite{loges2016} and Werner \cite{werner2017} both measure the coating conductivity decreasing approximately linearly from $278$ to $323\,$K, the sign the solid-controlled closure predicts and the gas-controlled view gets backwards. Liu et al.\ \cite{liu2024}, by an independent steady-state method, likewise report the through-plane conductivity of these thin layers decreasing with temperature. The measured negative temperature coefficient is therefore an independent fingerprint of solid- and contact-skeleton control, the paper's central thesis tested in a temperature dimension the model was never fit in. The sign and the gas-versus-solid contrast are robust to the umklapp exponent, and only the magnitude depends on it.

\paragraph{Formalizing the recipe-to-property mapping.} Section~\ref{sec:results:errors} shows that cross-recipe transfer is concentrated mostly in the as-coated bridge conductance, and that re-anchoring $\varphi_0$ on one as-coated sheet is an effective correction. This suggests a formalized predictive task for future work: estimating the initial contact state $\varphi_0$ or conductance $G = \varphi_0 \lambda_b$ directly from the slurry recipe. Formulating $G = f(\text{binder system}, \text{conductive-additive type}, \text{additive volume fraction}, \text{active-particle morphology})$ as a data-driven descriptor problem offers a promising AI-facing direction to generalize closures across chemistry iterations without recalibrating every time. We tested how far the present data already carry this idea by replacing the hand-coded $\varphi(\Pi)$ with a small neural network that takes $\Pi$ and chemistry-type indicators as input, is bounded ($0\le\varphi\le0.08$) and smooth by construction, and is trained end to end through the differentiable closure (module \texttt{learned\_contact.py}). A single such net matches the four separate quadratics in sample (at the $\sim$$5\%$ noise floor), confirming that the closure, not the algebra of $\varphi(\Pi)$, is the binding constraint; but it does not yet transfer. Leave-one-\emph{family}-out, predicting a held-out recipe's contact law from its chemistry indicators alone, the error is $\sim$$14\%$ for the anode (where the held-out and training anodes share a recipe) but $110$--$190\%$ for the cathodes (distinct recipes the net never saw). Four families, two of them one shared anode recipe, simply carry too little chemistry signal to identify a recipe-aware contact law from scratch. Two refinements show this is a partly fixable few-shot problem rather than a dead end (module \texttt{learned\_contact\_pretrain.py}). First, the $110$--$190\%$ figure used a per-\emph{family} solid conductivity, so a held-out cathode wrongly borrowed the graphite $\lams$; keying $\lams$ to chemistry instead drops the from-scratch cathode error to $\sim$$44\%$. Second, \emph{pre-training} the net on synthetic calendering families generated by the mechanistic closure, with physically-motivated contact parameters spanning the chemistry indicators, injects the closure's physics into the weights before it sees real data; fine-tuning on the four families then cuts the unseen-recipe error from $\sim$$29\%$ to $\sim$$19\%$ on average and roughly halves it for the cathodes (NMC811 from $\sim$$52\%$ to $\sim$$17\%$). Physics-informed pretraining thus turns an unidentifiable from-scratch fit into a feasible fine-tuning, with the residual transfer error set by how well the synthetic contact law matches reality, exactly the gap the multi-recipe campaign (Sec.~\ref{sec:experiments}) closes.

\begin{figure}[tb]\centering
\includegraphics[width=\textwidth]{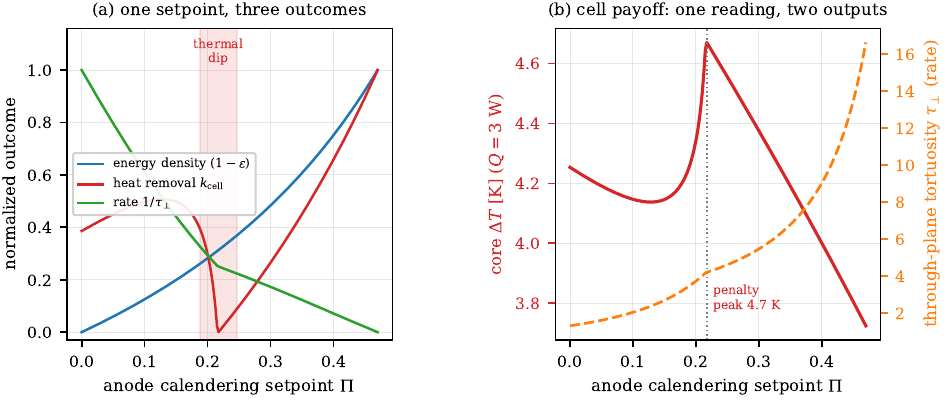}
\caption{The calendering setpoint as a three-way design decision, all driven by the reorientation state $S(\Pi)$. (a)~One setpoint, three normalised outcomes: energy density rises monotonically, rate $1/\tau_\perp$ falls, and heat removal $\kcell$ traces the u-shape, with the thermal-dip band shaded. (b)~The cell-level consequence in physical units: propagated through the $18650$ stack and a $1$D-radial model (uniform generation $Q=3\,$W), the predicted core temperature rise (left) peaks in the thermal-dip band while the through-plane tortuosity (right) climbs monotonically. The same thermal reading that locates the heat-removal optimum also proxies the rate-limiting tortuosity.}
\label{fig:design}
\end{figure}

\paragraph{What the closure offers a cell designer.} The practical payoff is that the anode calendering setpoint $\Pi$ sets three outcomes at once, all through the same reorientation state $S(\Pi)$ (Fig.~\ref{fig:design}, module \texttt{cell\_design.py}). Densification raises the volumetric energy density monotonically, but it raises the through-plane ionic tortuosity steeply (here $\tau_\perp$ climbs from $1.3$ to $\sim$$17$ across the sweep, so rate capability falls), and it moves the through-plane thermal conductivity along the u-shape. Propagated through the $18650$ series stack and a $1$D-radial jelly-roll model, the thermal u-shape produces a \emph{thermal penalty zone}, an intermediate calendering band (here $\Pi\approx0.22$, the conductivity minimum) at which the predicted core temperature rise peaks, $\sim$$25\%$ above the value at either lighter or heavier calendering, a setpoint the monotone ``denser is better'' intuition would miss. The same $S(\Pi)$ governs both the thermal conductivity and the tortuosity, so the differentiable thermal inverse, which recovers $\kperp\!\to\!S$, simultaneously proxies the rate-limiting through-plane tortuosity, one fast measurement informing both heat-removal and rate-capability design. Because the coupled model is differentiable end to end, the setpoint can be optimised against a weighted energy/heat/rate objective by automatic differentiation, with a rate (tortuosity) target setting the maximum admissible densification. By exposing this thermal penalty zone, the differentiable closure prevents the flawed optimisation trap of simultaneously trying to maximise density and effective conductivity while minimising tortuosity. Instead, it respects the fundamental power-versus-energy-density trade-off, allowing for hard constraints on tortuosity (rate capability) while optimising the thermal set-point.

\paragraph{Maturity of the claims, and what would upgrade them.} We separate what the closure can be trusted for now from what is still hypothesized or only numerically tested. \emph{Trusted now}: the forward prediction of the LFA-derived apparent through-plane $\leff(\Pi)$ for electrodes of similar chemistry and morphology, measured through the same laser-flash-under-helium chain (\Mtwo{} reaches the $4.5\%$ noise floor over $27$ states, and the zero-fit closure stays within a factor of two of twelve measurements from three further laboratories), together with the Knudsen pore-gas split, which is textbook physics and transfers cleanly to gas-diffusion layers (Section~\ref{app:gdl}). \emph{Hypothesized or numerically tested}: that flake \emph{reorientation} is the physical cause of the anode behaviour, which is corroborated from three independent directions (in-plane magnitude, temperature sign, measured texture trend) but remains degenerate with a contact-only fit on conductivity alone until texture is measured on these very sheets; the in-line porosity \emph{inversion}, demonstrated at $\pm0.008$ on synthetic data and a coarser $\sim$$0.05$ RMSE on reconstructed real data; and the cell-level design coupling of Fig.~\ref{fig:design}, which is a forward calculation through a $1$D model, not a cell measurement. \emph{What would upgrade each from consistency check to validation} is exactly the minimal set of measurements in Sec.~\ref{sec:experiments}. A two-pressure dry measurement pins the Knudsen term (assumption~A3) and $\beta$, XRD $(002)$ texture converts the reorientation mechanism (assumption~A2) from inference to measurement, and paired interface/coating metrology, the guarded hot plate alongside LFA with pull-off adhesion and ion-milled SEM, resolves the lumped contact scalar (assumption~A4) into its bulk-coating and collector-interface parts and fixes the LFA/GHP scale.

\paragraph{Transferability: what re-fits and what carries over.} Across processing routes and porous media the model's \emph{structure} is invariant; only a small contact-parameter set is re-anchored. The baseline ZBS$+$Knudsen channels and the intrinsic constituent conductivities, the graphite single-crystal pair $(\lambda_a,\lambda_c)$ and the bridge conductivity $\lamb$, carry over unchanged, as does the reorientation law $S(\Pi)$ for any flake-graphite anode (heated-roll densification aligns flakes just as calendering does). What is re-fit is the contact triple $(\varphi_0,a,b)$. Dry-processed electrodes illustrate the rule. Directly-calendered dry NMC622 falls on the \emph{same} $\leff(\varepsilon)$ trend as wet electrodes of identical composition, with the route-dependence localizing in $\varphi_0$ and $\lamb$ (Sec.~\ref{sec:results:dryproc}), and fuel-cell gas-diffusion layers need only a re-anchored sub-percent $\varphi$ that, as in calendering, \emph{rises with compaction}, while the Knudsen split applies unchanged (Section~\ref{app:gdl}). A new medium therefore costs about three numbers, not a new model, which is the practical content of the contact-term abstraction.

\paragraph{Scope and limitations.} Five constraints bound the claims above, each stated where it bites. First, the absolute parameters are method-scoped to the LFA-derived apparent coating conductivity and shift by $1.8$--$2.4\times$ under guarded-hot-plate recalibration (scope paragraph above). Second, $\lams$ and $\varphi_0$ are partially confounded, so we report the grouped bridge conductance and profile-likelihood valleys rather than point estimates (Sec.~\ref{sec:results:errors}). Third, the graphite reorientation account is degenerate with a contact-only fit on conductivity alone until same-sheet texture is measured (Sec.~\ref{sec:results:mechanisms}). Fourth, the process-dependence conclusions rest on a single public calendering dataset (four families, one laboratory, one metrology chain), so generalization awaits the campaign below. Fifth, the stack-derived target lumps the coating/collector interface resistance into $\lams$, separable only by the paired interface metrology that campaign specifies. The complete numbered inventory, adding temperature scope, separator geometry, the packed-bed extrapolation, parameter-sensitivity ranges, model averaging, and the wet-state assumption, is in Section~\ref{sec:limitations}.

\section{Mechanism-discriminating experiments}
\label{sec:experiments}

The model is calibrated on published data; a campaign (per recipe: about 26 sheets, one calendering afternoon, two measurement days) is designed to test its \emph{mechanisms}, not merely its fit, with stated failure criteria. The full sample matrix, the information-gain (D-optimal) design, the joint multi-property and multi-recipe extensions, and the closed-loop semi-dry-line deliverable are in Section~\ref{app:campaign}; its central mechanism-separating elements are four. A \textbf{two-pressure} dry measurement ($\sim$$1000$ vs $\le$$100$\,mbar) isolates the Knudsen gas channel and pins $\beta$, since $\Kn\propto1/p$ while skeleton and bridge conduction are pressure-independent (Eq.~\eqref{eq:knudsen}). \textbf{Pull-off adhesion} at every state must, by Eq.~\eqref{eq:phi}, share sign structure and dip location with $\varphi(\Pi)$. \textbf{Ion-milled SEM} contact-area quantification gives an image-based $\varphi(\Pi)$ independent of any thermal measurement. And \textbf{XRD $(002)$ texture} gives the Hermans factor $S(\Pi)$ directly, predicted to rise and saturate at the conductivity minimum for graphite ($S\!\to\!0.94$ at $\Pi=0.20$) with \emph{no} conduction-relevant texture change for the isotropic NMC cathode, the test that separates reorientation from contact damage that conductivity alone cannot. A D-optimal analysis shows both conductivity and texture discriminate the mechanisms most strongly near the conductivity minimum ($\Pi\approx0.2$), so the campaign samples densely there. Failure of the pressure split falsifies the Knudsen term; failure of the adhesion/SEM agreement demotes $\varphi(\Pi)$ to a curve fit; failure of the texture predictions falsifies reorientation and returns the anode minimum to a contact-only account with an unexplained low $\lams$.

\section{Conclusion}
\label{sec:conclusion}

In summary, we introduced a calendering-aware closure for LFA-derived apparent through-plane coating conductivity. A compression-indexed contact term on a Knudsen-corrected ZBS base reproduces the u-shaped dependence on calendering, reaching the measurement-noise level ($4.5\%$ over 27 states) where static-contact closures remain monotonic. Because four parameters meet only six to eight states per family, the reported quantities are grouped conductances, profile-likelihood valleys, and posterior predictive checks.

The cathode dip is contact-network evolution. For graphite, a bounded reorientation account explains the low through-plane solid conductivity by flake alignment toward the $c$-axis, although conductivity data alone cannot apportion reorientation and contact. That apportionment is the role of the proposed XRD texture and SEM contact-area measurements. A particularly clean falsification target is on the horizon in sodium-ion hard-carbon anodes. Hard carbon lacks graphite's strong crystallographic anisotropy, so the reorientation term should be \emph{inert} there, exactly as it is for the quasi-isotropic NMC cathode, and any calendering u-shape a hard-carbon anode shows must therefore be contact-only, a direct experimental test of the contact-versus-reorientation split this paper draws. The external checks then set scale. The closure gives a bottom-up 18650 prediction within $+15\%$ when interface resistance is inferred, and remains order-of-magnitude accurate (median $37\%$) across independent coating/separator sources. The falsifiable cross-domain prediction lands explicitly inside the measured graphite range. The closure and a resolved pore-scale solve give through-plane $\tau_\perp$ of $1.7$--$8$ and $3.4$--$6.0$, against Ebner's tomography ($\approx8$) and Landesfeind's impedance ($3$--$10$). The same contact term reproduces the non-monotonic electronic resistivity Lain measures, and measured graphite texture (Baade, Malifarge) independently supports the reorientation premise and its as-coated $S_0$. Throughout, the closure keeps the transport paths \emph{separate}, with through-plane heat removal and ionic tortuosity governed by the orientation state $S$ and electronic conduction by the carbon/contact network $\varphi$, so the thermal, rate, and electronic optima need not coincide and the model exposes their trade-off (Sec.~\ref{sec:discussion}) rather than collapsing them into one objective. The inverse and transport-coupling results are useful extensions, but the mechanism campaign of Sec.~\ref{sec:experiments} is the next experimental step. The system-level value rests on cost: because the closure evaluates and inverts in microseconds, against the hours a microstructure-resolved solve of the same coating would take, it is cheap enough to run as an inline estimator inside gigafactory-scale quality control and as the forward model of a predictive process digital twin, rather than only as an offline analysis tool (Fig.~\ref{fig:inline}).

\paragraph{Generality beyond electrodes.}
Lithium-ion electrode coatings are a useful first application because four conditions coincide: sub-micrometre-to-micrometre pores where pore-gas conduction is rarefied, a manufacturing process that reshapes inter-particle contact, direct inverse and quality-control value, and an open measured dataset. The same ingredients recur in compressed fibrous or granular media. Proton-exchange-membrane fuel-cell and electrolyser porous transport layers are the strongest next target because through-plane thermal conductivity and contact resistance versus compaction pressure are extensively measured \cite{burheim2010}. Dry-processed and solid-state battery electrodes are another natural target because the contact term becomes a dominant pathway. The practical transfer statement is limited and testable: where conduction through a porous solid is set by rarefied pore gas and an evolving contact network, the same closure structure should apply after re-anchoring the contact parameters.


\section*{Reproducibility statement}
The complete project, the JAX implementation of the closure (with a regression test enforcing the exact $\varphi=0$ reduction to the validated base model), the transcribed datasets with row-level source citations, all calibration scripts with the objective, bounds, and seeds stated in Sec.~\ref{sec:data:protocol}, the executed analysis notebooks, a regression-test suite that locks every headline number of this paper against silent drift, and the pipeline that regenerates every figure from raw inputs with no hard-coded values, is openly available at:
\begin{center}
\url{https://github.com/j-stoerk/zehner-electrode-thermal}
\end{center}

\section*{Data availability}
The transcribed datasets in \texttt{data/raw/} of the repository above (the calendering transcription and the separator/electrode literature anchors) carry per-row citations to the primary sources \cite{gandert2023,richter2017,marconnet2018,burheim2013,vishwakarma2015}; the primary calendering dataset is published open access by its authors.

\section*{Author contributions}
J.S.\ is the sole author and is responsible for all aspects of the work: conceptualization, the closure derivation and reorientation/contact model, software and the differentiable implementation, the calibration, identifiability, and Bayesian analyses, the literature overlays, and the writing. Because one author carried the many methods used here, the safeguard against error is the open repository (Reproducibility Statement): every headline number is regenerated by scripted pipelines and pinned by the regression-test suite, and every figure is rebuilt from raw inputs, so any claim can be re-derived independently.

\section*{Acknowledgments}
The author thanks Sebastian Schebesta (VARTA Microbattery) for his support of this work, Jochen Eser for mentorship and for first pointing toward Zehner's work, and Prof.\ Carsten Schilde (Technische Universit\"at Braunschweig) for doctoral supervision. This work was supported by the German Federal Ministry of Research, Technology and Space (BMFTR) under the project GranuGoIn (grant nos.\ 03XP0635A, VARTA Microbattery GmbH, and 03XP0635B, TU Braunschweig).


\clearpage
\section*{Supporting Information}
\setcounter{section}{0}
\renewcommand{\thesection}{S\arabic{section}}
\renewcommand{\thesubsection}{S\arabic{section}.\arabic{subsection}}
\setcounter{figure}{0}
\renewcommand{\thefigure}{S\arabic{figure}}
\setcounter{table}{0}
\renewcommand{\thetable}{S\arabic{table}}
\setcounter{equation}{0}
\renewcommand{\theequation}{S\arabic{equation}}

\noindent This Supporting Information collects the material that supports the extensions and consistency checks of the main text; only the calibrated, validated core (the closure, its calibration, and the mechanism analysis) lives in the main body. The sections are numbered S1--S9: \textbf{S1} the calibration, ablation, and error-analysis protocol; \textbf{S2} parameter interpretation and design rules; \textbf{S3} Bayesian posteriors and model comparison; \textbf{S4} external consistency checks; \textbf{S5} the data-driven-method extensions; \textbf{S6} cross-domain transfer to gas-diffusion layers; \textbf{S7} the proposed experimental campaign; \textbf{S8} the wet-versus-dry processing-route extension; and \textbf{S9} the thermal--ionic transport coupling. Figures, tables, and equations in this section are likewise prefixed S.

\section{Calibration, ablation, and error-analysis protocol}
\label{app:protocol}
Relative residuals (Eq.~\eqref{eq:objective}) are used because the families span a fivefold conductivity range. Absolute residuals would let the anodes dominate the fit while the cathode sheets are effectively ignored. The protocol comprises four checks: (i) an \emph{ablation} \Mzero$\to$\Mone$\to$\Mtwo, so that the contribution of each model component is assessed separately; (ii) \emph{held-out transfer}, calibrating on graphite$_{\mathrm{thin}}$ and predicting graphite$_{\mathrm{thick}}$ (same recipe, doubled coating thickness), and calibrating on NMC622 and predicting NMC811 (different additive recipe); (iii) a \emph{profile likelihood} over $\lams$, refitting the remaining parameters on a grid of fixed $\lams$ values, to expose what the data cannot determine; (iv) \emph{hybrid swaps} that attribute transfer error to individual parameters. For the Bayesian treatment (Section~\ref{app:bayes}), the same model and data enter a No-U-Turn sampler (NumPyro; two chains of 1000 warm-up and 1500 samples, target acceptance 0.9) with priors that carry the physics: $\lams$ uniform over the anisotropy band, $\varphi_0$ normal around the VDI rigid-sphere value ($0.0077$, width $0.01$, truncated to $[0,0.08]$), $a$ and $b$ truncated normals on their bounds with $b\ge0$, and the relative noise level $\sigma$ half-normal so that the data determine it. Convergence: $\hat R\le1.01$ for all parameters and families.

\begin{figure}[tb]\centering
\includegraphics[width=0.82\textwidth]{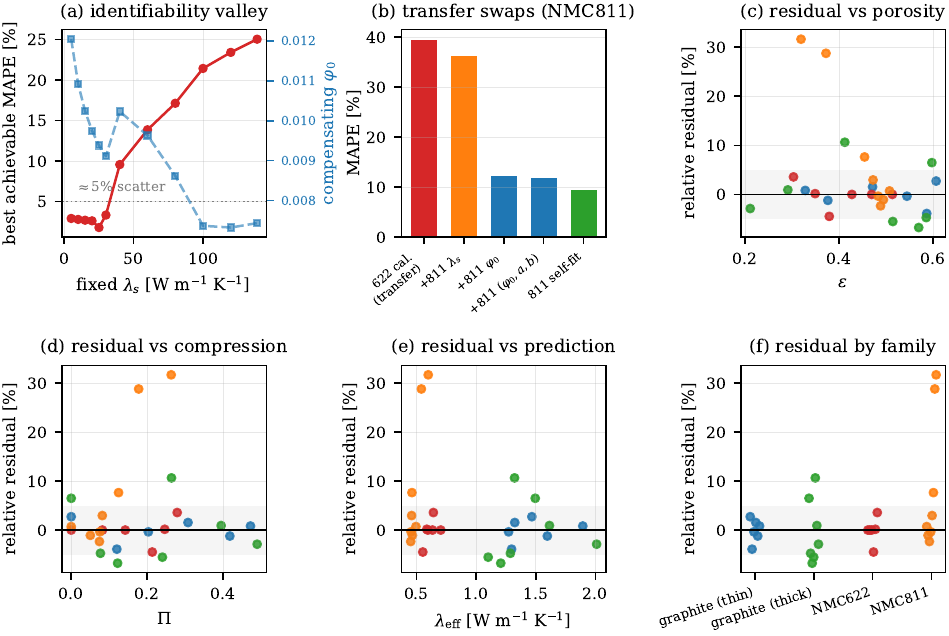}
\caption{\emph{Held-out within-dataset tests.} Identifiability, transfer, and residual diagnostics (supporting Sec.~\ref{sec:results:errors}). (a)~Profile likelihood over $\lams$ for the thin anode; right axis, compensating $\varphi_0$. (b)~Hybrid-swap decomposition of NMC622$\to$NMC811 transfer. (c--f)~\Mtwo{} relative residuals against porosity, compression rate, predicted value, and family.}
\label{fig:error}
\end{figure}

\section{Design principles from the fitted parameters}
\label{app:design}
The fitted parameters are measurements of contact physics, and comparing them across recipes yields design rules, provided the comparison is done in the right units. Binder and additive densities ($1.6$--$1.8\,\mathrm{g\,cm^{-3}}$) are far below those of the active materials (NMC: $4.7$), so a recipe quoted in weight percent hides how much \emph{space} the inactive phase occupies. Table~\ref{tab:design} therefore compares in volume. Three interplays emerge. \emph{(i)~Bridge conductance.} The as-coated group $G=\varphi_0\lamb$, normalized per unit inactive volume, is roughly $20\times$ larger for the anodes than for the cathodes: soft graphite flakes deform plastically and bridge each other with high-conductivity contacts all by themselves, whereas hard NMC spheres depend entirely on the additive network between them. Within the cathodes, two weight-percent of flake graphite doubles NMC622's bridge effectiveness over carbon-black-only NMC811: the single clearest recipe lever in the data. \emph{(ii)~Damage rate.} The fractional loss $|a|/\varphi_0$ orders perfectly by binder mechanics, with stiff PVDF networks failing $2$--$4\times$ faster per unit compression than elastomeric CMC/SBR, whose rubbery bridges stretch where PVDF's fracture. \emph{(iii)~Recovery} appears only once a process threshold is crossed, so NMC811's $b=0$ reflects a calendering window that never reached interlocking ($\Pi\le0.26$), rather than an absence of the underlying mechanism. The design rules follow. A small flake-graphite additive outperforms additional carbon black by about a factor of two per unit volume, elastomeric co-binders protect the thermal and adhesion network in heavily calendered designs, and the calendering setpoint should not sit in the conductivity/adhesion dip, so either calender lighter or push through into interlocking.

\begin{table}[htb]\centering\small
\caption{Volume-space decomposition of the contact parameters.}
\label{tab:design}
\begin{tabular}{lccccc}
\toprule
Family & binder+additive & $G=\varphi_0\lamb$ & $G/\mathrm{vol}$ & $|a|/\varphi_0$ & binder\\
 & [vol\% of solid] & [$\WmK$] & & [$\Pi^{-1}$] & \\
\midrule
graphite (thin)  & 6.4  & 1.22 & 19.0 & 2.6  & CMC/SBR\\
graphite (thick) & 6.4  & 1.57 & 24.4 & 3.5  & CMC/SBR\\
NMC622           & 17.8 & 0.41 & 2.3  & 5.2  & PVDF\\
NMC811           & 10.0 & 0.12 & 1.2  & 10.2 & PVDF\\
\bottomrule
\end{tabular}
\end{table}

\paragraph{Physical grounding of the contact quadratic.} To physically ground the phenomenological contact quadratic $\varphi(\Pi) = \varphi_0 + a\Pi + b\Pi^2$, we formulate it as the low-order expansion of a dual-mechanism contact process. In the small-strain elastic limit, particle contacts follow Hertzian mechanics, where contact area scales sub-linearly with compression. However, at the high compactions reached during calendering (especially the thick graphite anode), Hertzian mechanics breaks down and particles undergo elasto-plastic deformation, which is better captured by the convex quadratic term ($b > 0$). Concurrently, the initial downswing ($a < 0$) represents the shear damage of the binder network. As the calendering shear stress exceeds the cohesive yield strength of the binder bridges (PVDF or CMC/SBR), the percolated conductive pathways are irreversibly ruptured before macroscopic plastic particle interlocking can rebuild the network. The quadratic thereby isolates irreversible binder rupture (linear term) from subsequent plastic interlocking (quadratic term).

\paragraph{From fitted coefficients to contact mechanics.} The damage and recovery coefficients can be replaced, at the same parameter count, by a contact-mechanics model with mechanical rather than curve-fit parameters, namely a binder term $\varphi_{b0}\max(0,1-\Pi/\Pi_d)$, in which the as-coated bridge fraction $\varphi_{b0}$ is sheared away by a critical compression $\Pi_d$, plus a Hertzian particle-contact term $c_H(\varepsilon_0-\varepsilon)$, the contact area growing linearly with densification as Hertz dictates ($A\propto$ overlap). This form fits the cathodes as well as the quadratic ($1.5\%$, $9.4\%$) and \emph{derives} the downswing as binder-network shear damage. The upturn, however, is degenerate, because the conductivity data are described equally well with $c_H=0$, the recovery then carried entirely by the porosity dependence of the base closure. Hertzian contact regrowth and densification cannot be separated from conductivity alone, consistent with the ablation's reading of the upturn as densification plus, where reached, interlocking, and isolating the interlocking contribution requires the independent SEM contact-area measurement of Sec.~\ref{sec:experiments}.

\paragraph{First-principles contact fraction from discrete-element modeling (DEM).} A discrete-element packing (soft-sphere relaxation compacted to the target porosity) resolves the inter-particle contacts that voxelized stochastic structures cannot, and a constriction-resistance thermal network on it gives a through-plane conductivity in the contact-augmented regime, a factor of $1.4$ to $1.8$ above the contact closure across the compaction range, not the order-of-magnitude over-prediction of an over-connected voxel solid, closing the thermal-absolute gap for spheres. It also \emph{derives} the contact fraction. The as-coated value $\varphi\approx0.007$ matches the VDI rigid-sphere prior $0.0077$ that anchors $\varphi_0$, and $\varphi$ rises with compaction ($0.007\to0.13$; coordination number $4.7\to7.5$), giving the calendering contact term from mechanics rather than fitting it. The constriction and VDI flattened-contact models diverge in the heavy-overlap (plastic) regime, and anisotropic-flake particles with binder bridges remain a refinement.

\section{Bayesian calibration and model evidence}
\label{app:bayes}
Replacing least squares with full posterior sampling (protocol in Section~\ref{app:protocol}) answers two questions the point estimates cannot. First, \emph{are the fits trustworthy?} Every posterior median agrees with the corresponding least-squares estimate, the thin-anode $\lams$ posterior spans $7.5$--$30.5\WmK$ at $90\%$ probability, an independent, probabilistic reproduction of the profile-likelihood valley of Sec.~\ref{sec:results:errors}, and the inferred noise levels ($4.7\%$, $8.5\%$, $1.1\%$, $15.6\%$ for the four families) recover each family's actual residual scatter without being told, including NMC811's poorer fit, which the posterior reports as a wider noise level. The $\varphi_0$ credible intervals bracket the VDI rigid-sphere value across families. A pooled fit sharpens the recipe picture. One shared parameter set for both same-recipe graphite electrodes (14 sheets) still describes them to $6.7\%$ and $8.6\%$, so the recipe hypothesis holds to within the thickness-induced interface differences that Section~\ref{app:design} identified.

Second, \emph{is the process-dependent contact term statistically necessary, or would a constant contact fraction do?} The widely applicable information criterion (WAIC), which estimates out-of-sample predictive accuracy and penalizes effective model complexity, answers per family (Table~\ref{tab:waic}). The quadratic $\varphi(\Pi)$ is strongly preferred for three of the four families (differences of $13$--$32$, where $\sim$$10$ already counts as strong), with effective parameter counts of $3$--$4$ confirming the model is not overparameterized. The exception is NMC811 ($\Delta=1.6$). For the one family whose calendering window never reached the interlocking regime, the data cannot justify the quadratic over a constant term. The formal model comparison thereby rediscovers, independently, the process-threshold finding of Section~\ref{app:design}.

\emph{Small-sample cross-check (AICc).} Because these differences rest on six to eight points, we add the small-sample-corrected Akaike criterion (AICc), whose finite-sample term $2k(k{+}1)/(n{-}k{-}1)$ penalizes the \emph{nominal} parameter count. The two criteria \emph{agree} on the mechanism question. AICc favours the three-parameter reorientation \MR{} over the four-parameter contact quadratic \Mtwo{} for both anodes ($\Delta\mathrm{AICc}=+29$ thin, $+13$ thick), the same direction as WAIC, reinforcing that the parsimonious reorientation account is preferred where both apply. They \emph{disagree} on whether the contact term pays for itself. At $n=6$ the AICc penalty on \Mtwo{}'s fourth parameter (correction $\approx40$) outweighs its fit gain, so AICc favours the constant-contact \Mone{} ($\Delta\mathrm{AICc}(\Mone-\Mtwo)\approx-5$ to $-20$) where WAIC, using effective rather than nominal parameters, strongly favours \Mtwo{}. At six to eight states per family the formal criteria are therefore unreliable, so the case for the process-dependent term rests not on them but on the factor-of-seven ablation (Table~\ref{tab:ablation}), the u-shape that \Mone{} cannot reproduce, and the adhesion correlation. The criteria are reported for completeness, not as the load-bearing evidence.

\begin{table}[htb]\centering\small
\caption{WAIC comparison (lower is better) of constant contact fraction (\Mone) vs.\ the process-dependent quadratic (\Mtwo); $\Delta>0$ favours the quadratic. $p_{\mathrm{WAIC}}$: effective number of parameters of \Mtwo.}
\label{tab:waic}
\begin{tabular}{lcccc}
\toprule
Family & WAIC \Mone & WAIC \Mtwo & $\Delta$ & $p_{\mathrm{WAIC}}$ (\Mtwo)\\
\midrule
graphite (thin)  & $-3.0$ & $-16.0$ & $+13.0$ & 2.9\\
graphite (thick) & $2.2$  & $-10.6$ & $+12.8$ & 2.9\\
NMC622           & $-0.3$ & $-31.8$ & $+31.6$ & 4.2\\
NMC811           & $-0.6$ & $-2.2$  & $+1.6$  & 1.5\\
\bottomrule
\end{tabular}
\end{table}

\section{External consistency checks and sensitivity}
\label{app:meta}
\begin{figure}[tb]\centering
\includegraphics[width=0.52\textwidth]{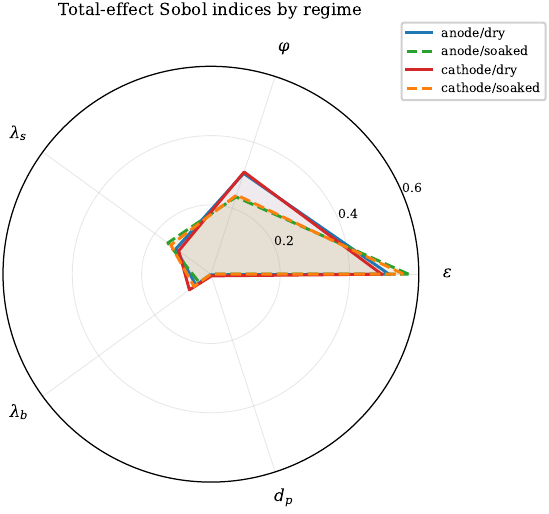}
\caption{Comparative total-effect Sobol indices (normalised) of the closure inputs across four regimes, anode/cathode $\times$ dry/electrolyte-soaked. Porosity $\varepsilon$ leads everywhere and the contact fraction $\varphi$ is second and larger in the dry state (solid lines); the solid conductivity $\lams$ is third, the bridge conductivity $\lambda_b$ weighs more for the cathode, and particle/pore size $d_p$ is negligible for electrodes (it would dominate only for the gas-filled separator). A measurement-priority map for process control.}
\label{fig:sobol}
\end{figure}

Widening the test beyond the calibration source, we predict (zero fit, each material at its literature through-plane particle conductivity) twelve independent coating and separator measurements drawn from three further laboratories and the steady-state method \cite{burheim2013,marconnet2018,richter2017}. The closure tracks them to a median $37\%$ relative error, with $83\%$ of points inside a factor of two, and to within a few percent where the inputs are best characterised (the LCO cathode: $-4\%$ dry, $+5\%$ soaked). The residuals are attributable rather than random. The two independent wet graphite-anode values disagree with \emph{each other} by $\sim50\%$, LMO is the lowest-conductivity cathode oxide and is over-predicted, and the two outlying separator points are a barely-wetted sample and a compacted one. A baseline comparison on the same wet coatings reconciles with the dry-data hierarchy of Table~\ref{tab:baselines}. Bruggeman fails badly (median $185\%$, its co-continuous path percolating the solid), whereas Maxwell--Eucken, which under-predicts at the high contrast of dry coatings, becomes competitive at the low contrast of wet ones ($19\%$), but carries no contact or calendering mechanism. The dominant controllable error throughout is the through-plane solid conductivity, the anisotropy point of Sec.~\ref{sec:limitations}. Finally, a converged Sobol decomposition of the dry-anode conductivity ranks porosity first ($S_T=0.46$), the contact pair $(\varphi,\lambda_b)$ a close second ($\approx0.48$ combined), and particle size negligible ($S_T\approx0$, the pore gap being the bottleneck at high contrast). Read as a measurement-priority list for process control, this says to fix porosity and the calendering/contact state first. Resolving this by regime, a comparative total-effect decomposition across $\{$anode, cathode$\}\times\{$dry, electrolyte-soaked$\}$ (normalised indices; module \texttt{sobol\_compare.py}, Fig.~\ref{fig:sobol}) confirms porosity as the leading lever in all four cases and the contact fraction as the second, with $\varphi$ carrying noticeably more weight dry ($\approx0.31$) than soaked ($\approx0.23$, where the higher-conductivity electrolyte lowers the phase contrast so the geometric porosity split dominates more). The cathode leans slightly more on the bridge conductivity $\lambda_b$, and particle/pore size is negligible for electrodes in every regime, because the rarefied-gas channel $d_p$ controls is a small share of an electrode's through-plane conduction and would dominate only for the gas-filled separator.

\paragraph{Cell-scale check: a bottom-up 18650 stack.}\label{sec:results:stack}
At cell scale we use the 18650 stack as a consistency check. The prediction extrapolates well below the calibration window and the interface resistance is inferred. The quantity cell designers actually need is the through-plane conductivity of the full layer stack. Steinhardt et al.\ \cite{steinhardt2021} published both the complete unit-cell structure of a commercial 18650 cell (per-layer thicknesses and porosities; anode $\varepsilon=0.216$, cathode $\varepsilon=0.171$, separator $\varepsilon=0.45$) and its measured jelly-roll conductivity, $k_\perp = 1.122\WmK \pm 6.2\%$, which permits a bottom-up check. We predict each electrolyte-filled coating with the zero-fit closure at \emph{their} porosities, assemble the repeating unit in series with separator and collector foils, and compare. With perfect interfaces the prediction is $1.29\WmK$, $15\%$ above the measurement, as expected when the layer-only model neglects interface resistance. Matching the measurement exactly requires only $\approx6\,\mu\mathrm{K\,m^2\,W^{-1}}$ of contact resistance per coating/separator interface, two orders of magnitude below the $420\,\mu\mathrm{K\,m^2\,W^{-1}}$ measured on an unbonded, torn-down stack \cite{vishwakarma2015}. A jelly roll wound under tension presses its interfaces together, so the implied value is physically coherent. This inferred resistance is itself loosely determined. It is a small difference of two larger conductivities, so propagating the $\pm6.2\%$ measurement uncertainty together with a $\sim$$12\%$ closure-prediction uncertainty (from the per-layer porosity and composition inputs) gives $\approx6\pm6\,\mu\mathrm{K\,m^2\,W^{-1}}$, i.e.\ anywhere from near-perfect contact to $\sim$$12$, but firmly two orders of magnitude below the torn-down $420$. The qualitative reading (tight jelly-roll interfaces) is robust, but the precise magnitude is not pinned, which is why we use the 18650 as a scale check rather than a calibration. Sweeping the per-interface resistance from torn-down to perfect contact spans predictions from $0.11$ to $1.29\WmK$, close to the $0.15$--$1.4\WmK$ spread of the published cell-level literature \cite{steinhardt2022}. Steinhardt et al.\ \cite{steinhardt2020} also measured a single prismatic cell under controlled external compression and found its through-plane conductivity rises by $11.9\%$ for a $37.1\,$kPa pressure increase, an effect they attribute to inter-layer thermal contact resistance. A pressure law $R_{\mathrm{int}}(P)\propto P^{-n}$ calibrated to that sensitivity gives $k_\perp$ from $\sim$$0.2$ at light contact to the $1.29\WmK$ layer-only limit. Propagated through a 1D-radial cell thermal model this changes the predicted core temperature rise by about a factor of five (e.g.\ $3\to16\,$K at $3\,$W generation as contact loosens). As an independent cross-model anchor, microstructure-resolved simulations \cite{oehlerdiss} place electrolyte-filled anode coatings at $\approx1.9$--$2.3$ and cathode coatings at $\approx0.95$--$1.1\WmK$ at comparable compositions, and the closure's values sit at the same level, two unrelated modeling routes agreeing within $10$--$25\%$.

\section{The data-driven methods as the enabling engine}
\label{app:engine}
The capabilities that separate this work from a forward closure are each supplied by one data-driven method. \emph{Automatic differentiation} (the closure is written in JAX) provides the exact gradients behind both the Newton inverse of Sec.~\ref{sec:results:qc} and the Hamiltonian-Monte-Carlo posterior sampling of Section~\ref{app:bayes}. Without it the inverse would rely on finite differences and the Bayesian step would be impractical. \emph{Gaussian-process regression with split-conformal prediction} supplies the uncertainty layer, the $2\%$ model-form floor entering Eq.~\eqref{eq:sigma} and the distribution-free intervals whose leave-one-family-out coverage flags an out-of-distribution recipe. \emph{Symbolic regression}, run freely over $\{+,-,\times\}$ on the closure-inverted contact fractions, returns a low-order polynomial in $\Pi$ with a constant term matching the VDI sphere value, an unsupervised confirmation that the $\varphi(\Pi)$ form was recovered from the data. This is a \emph{hybrid neuro-symbolic} step. A search over symbolic expressions extracts a physical governing form directly from empirical thermal data and feeds it back into the analytic closure, the same data-driven equation-discovery approach used in neuro-symbolic scientific machine learning, here applied to electrode manufacturing. Finally, a \emph{physics-informed neural network} extends the scalar inverse to a \emph{field}. From synthetic flash-thermography frames it recovers a spatially varying interface-conductance map, locating a delamination defect to within one grid spacing. Each method is demonstrated in the accompanying notebooks, and together they support inverse use, uncertainty quantification, and process-resolved prediction.

\paragraph{PINN field-inverse implementation and training.} The field demonstrator (notebook \texttt{Electrode\_AI.ipynb}) recovers a one-dimensional interface-conductance field $H(x)$ from synthetic flash-thermography frames. The forward physics is the nondimensional transient flash response $\partial_t\theta=\alpha\,\partial_{xx}\theta-H(x)\,\theta$ (with $\alpha=0.05$, a uniform initial flash $\theta(x,0)=1$, and insulated edges $\partial_x\theta=0$ at $x=0,1$). A delamination is a localized dip in $H(x)$ that lets heat linger. Two $\tanh$ multilayer perceptrons (each three hidden layers of $32$ units, double precision) represent $\theta(x,t)$ and $H(x)=\mathrm{softplus}(\cdot)$ (positivity-enforced). The composite loss $L_{\mathrm{PDE}}+10\,L_{\mathrm{IC}}+L_{\mathrm{BC}}+50\,L_{\mathrm{data}}$ combines the PDE residual on $1500$ random $(x,t)$ collocation points (gradients by automatic differentiation), the initial condition on $300$ points, the insulated-edge Neumann penalty, and a data term matching four $2\%$-noise frames at $t=\{0.1,0.3,0.6,1.0\}$ (weight $50$). Training is Adam (learning rate $2\times10^{-3}$, $6000$ iterations, seed $0$) on CPU in $\sim$$200$\,s, with no restarts needed. The differentiable thermal closure supplies the \emph{constitutive} interpretation of the recovered field (mapping the local interface/bridge state to an effective conductivity), whereas using the closure itself as the PDE residual on real coatings is the prospective soft-sensor extension below, not what this synthetic demonstrator does. On this problem the PINN reaches a field relative $L_2$ error of $0.106$ and locates the defect centre to within one grid spacing. \emph{Limitation:} the result is sensitive to loss weighting and to frame sparsity, and the recovered dip \emph{depth} is underestimated (minimum $H\approx1.9$ vs.\ true $0.8$) because the network regularizes toward smooth fields. The demonstrator therefore answers \emph{where}/\emph{whether} a delamination is present reliably, but not \emph{how severe} without finer frames or a sharper prior.

\paragraph{Data splits for the data-driven layer.} To avoid leakage, the fitting, calibration, and evaluation sets are kept disjoint. The symbolic search is fit to the closure-inverted $\varphi(\Pi)$ points as a \emph{form-discovery} step (it predicts no held-out quantity). The Gaussian-process residual model and its split-conformal wrapper use disjoint fit and calibration sets, and the headline coverage test is \emph{leave-one-family-out} (the conformal quantile is calibrated on three families and evaluated on the unseen fourth, giving $87\%$ mean coverage at a $90\%$ target). The residual-corrected hybrid closure is scored by leave-one-state-out, so the discrepancy model is never evaluated on points used to fit it.

\paragraph{From offline inverse to an online soft sensor (prospective).} Because the closure is written in JAX, the derivatives it contributes to a learning objective are exact and analytic, so when the closure is the physics residual (the governing-equation loss) of a physics-informed neural network, training never relies on step-size-sensitive finite-difference approximations of the physics. This is what would let the thermal closure act as a real-time \emph{soft sensor} rather than a post-hoc analysis. From flash-thermography frames the PINN recovers a spatially varying interface-conductance map and localises delamination or localized coating defects to within a grid spacing, fast enough to run \emph{during} calendering and so to monitor interface quality inline (the field-level counterpart of the scalar monitor of Sec.~\ref{sec:results:qc}). This matters most for semi-dry, granulate-based processing, the GranuGoIn target. Semi-dry routes (such as PTFE-fibrillated or low-solvent granule feeds) omit the long wet-mixing and drying steps that homogenise a conventional slurry, so the granules entering the calender carry significant batch-to-batch and along-web variation in porosity and structural density. To hold a strict porosity and interface-quality target the roll gap cannot run open-loop but must be trimmed dynamically. The differentiable soft sensor closes exactly this loop. Using the exact analytical derivatives of the thermal closure as the physical loss of a PINN, it supports an active-learning loop that maps upstream extruder settings (e.g.\ solid content, screw speed/rpm) directly to downstream electrode interface quality, automatically choosing the next set-point to sample without grid-search experiments. The construction is not specific to the thermal field or the calender, so a differentiable physical closure supplying the loss of a physics-informed model is a reusable template for the AI-driven optimisation of semi-dry continuous electrode manufacturing. This is a forward-looking application enabled by the closure, not a validated result of the present paper.

\section{Generalization to fuel-cell gas-diffusion layers}
\label{app:gdl}
Proton-exchange-membrane gas-diffusion layers (GDLs) are a fibrous carbon medium compressed in the cell, with a $\sim100$\,nm-pore microporous layer (MPL). Using published through-plane conductivities of dry GDLs versus compaction load \cite{burheim2010}, all three diagnostics transfer. The point-contact closure underpredicts by $6$--$8\times$ (the electrode/separator signature). A sub-percent contact fraction reproduces the measured conductivities and, for every GDL, \emph{rises with compaction load} (Toray TGP-H-060: $\varphi=0.006\to0.010$ over $4.6$--$13.9$\,bar), so this is the $\varphi(\Pi)$ mechanism with compaction in the role of calendering, and the GDL literature independently identifies inter-fibre contact as controlling \cite{burheim2010}. Finally Knudsen splits the component exactly as before: negligible for the micrometre fibre paper ($\mathrm{Kn}=0.003$, $99\%$ retained) but removing $\sim70\%$ of pore-gas conduction in the $100$\,nm MPL ($\mathrm{Kn}=0.67$).

\section{Proposed experimental campaign (full protocol)}
\label{app:campaign}
The campaign summarised in Sec.~\ref{sec:experiments}, in full.
\begin{enumerate}
\item \textbf{Sample matrix.} Six calendering states from $\Pi=0$ to the production maximum (single batch, single-side coatings); duplicate as-coated sheets for the $\varphi_0$ re-anchor; each state measured dry and electrolyte-soaked.
\item \textbf{Porosity ground truth.} Areal mass; flat-anvil micrometer thickness at $\le0.3\,\mathrm{N\,mm^{-2}}$ (not a ball-tip dial gauge, which compresses the coating \cite{gandert2023}); helium-pycnometric solid density; GUM uncertainty budget.
\item \textbf{Conductivity.} Through-plane laser flash with penetration-model evaluation \cite{mcmasters}, recording the purge gas (it is the pore gas), or a guarded hot plate; $20\,^\circ$C; $n=4$ repeats per sheet, bringing $3\%$ single-shot noise to $1.5\%$.
\item \textbf{Two-pressure Knudsen discrimination} (most informative for separators; recommended for the most porous electrode state): repeat the dry measurement at about $1000$ and at or below $100$\,mbar chamber pressure. By Eq.~\eqref{eq:knudsen}, pore-gas conduction depends on pressure through $\Kn\propto1/p$, while skeleton and bridge conduction do not. The pressure split therefore isolates the gas channel, pins $\beta$, and deliberately breaks the error cancellation identified for continuum sphere-pack separator models in Sec.~\ref{sec:results:baselines}. A differentiable identifiability analysis supports the ``most informative for separators'' point (module \texttt{design\_identifiability.py}). Using the exact $\partial\ln\leff/\partial\beta$ from automatic differentiation, the Fisher information about $\beta$ rises monotonically with $\Kn$ and saturates at high $\Kn$, so the fine-pored separator in helium ($\Kn\!\gg\!1$ even near ambient) constrains $\beta$ several-fold better than a coarse electrode ($\Kn\!\ll\!1$) or any measurement in air; a two-pressure separator design shrinks the $\beta$ posterior multiple-fold versus its prior, whereas an ambient-air electrode point barely moves it. Because the relative sensitivity saturates past $\Kn\!\approx\!1$, moderate reduced pressure already captures most of the information, so deep vacuum is unnecessary. This parameter-identifiability design complements the model-discrimination (texture) design above: one sizes which \emph{parameter} a measurement pins, the other which \emph{mechanism} it selects. Operationally it recommends a dedicated separator coupon measured along a helium pressure ladder ($\sim$$1000/500/250/100$\,mbar) rather than a single two-point split, with the most porous electrode state added to extend the $\Kn$ lever; the ladder is then fit to a single $\beta$ through Eq.~\eqref{eq:knudsen}, and the saturation means the lowest rungs can be coarse.
\item \textbf{Adhesion correlation.} Pull-off adhesion strength at every calendering state. Equation~\eqref{eq:phi} claims the conductivity-relevant contact fraction tracks interface damage and interlocking; then $\varphi(\Pi)$ and adhesion must share sign structure and dip location.
\item \textbf{Cross-section image quantification.} Ion-milled SEM cross-sections at each calendering state, with the contact and bridge areas quantified by image segmentation. This yields a direct, microscopy-based estimate of $\varphi(\Pi)$ that is independent of any thermal measurement, and therefore the sharpest available test of the contact interpretation: the image-derived and the thermally fitted contact fractions must agree in magnitude and in the location of the dip.
\item \textbf{Flake-orientation texture (the reorientation test).} For graphite anodes, X-ray diffraction texture analysis (the graphite $(002)$ pole figure) at each calendering state gives the Hermans orientation factor $S(\Pi)$ of Eq.~\eqref{eq:reorient} directly; the same ion-milled SEM cross-sections give an independent flake-angle distribution. The reorientation mechanism makes two sharp, thermal-measurement-independent predictions: $S$ rises and saturates near the conductivity minimum (for the calibration data, $S\!\to\!0.94$ at $\Pi=0.20$, the measured anode minimum), and the quasi-isotropic NMC cathode shows \emph{no} conduction-relevant texture change. This test separates reorientation from contact damage, which the conductivity data alone cannot.
\item \textbf{Joint multi-property calibration (thermal, ionic, electronic).} On the \emph{same} calendered sheets, add a through-plane ionic measurement (symmetric-cell electrochemical impedance or limiting-current polarization, giving the through-plane tortuosity $\tau_\perp(\Pi)$) and a through-plane electronic measurement (four-probe or micro-ohmmeter across the stack, giving $\sigma_{\mathrm{el}}(\Pi)$). The joint closure of Sec.~\ref{sec:results:coupling} predicts the \emph{shape} of both from latents fixed on thermal data alone, $S(\Pi)$ for the ionic tortuosity and $\varphi(\Pi)$ for the electronic resistivity; these two measurements promote it from a trade-off-geometry check to a calibrated multi-property closure by pinning the absolute magnitudes it does not yet fix, the orientation-dependent ionic Bruggeman exponent $\alpha(S)$ and the carbon-network electronic conductance. They also test its sharpest structural prediction directly: the \emph{same} $S(\Pi)$ that drives the thermal downswing must drive the ionic-tortuosity rise, while the \emph{same} contact form $\varphi(\Pi)$ that makes the thermal u-shape must make the non-monotonic electronic resistivity. This is the measurement set that would convert the shared-latent coupling from inference to calibration.

\item \textbf{Multi-recipe extension and informative design.} Once the first recipe is complete, the same matrix should be repeated across recipes that vary one factor at a time (binder system, conductive-additive type and content, particle size) and across processing route (wet-slurry versus dry-processed), the latter testing whether the same closure describes both routes with the contact parameters re-anchored, and whether a dry graphite anode exhibits the predicted flake-reorientation signature (Sec.~\ref{sec:results:dryproc}). This builds the data basis for predicting contact parameters from recipe descriptors rather than calibrating each recipe separately. Because the closure is differentiable, the calendering states and chamber pressures of later campaigns need not be evenly spaced: they can be chosen to maximize the expected information gain. A D-optimal analysis targets the reorientation-versus-contact question directly. The optimal conductivity states cluster near the conductivity minimum ($\Pi\approx0.2$), the one place the two mechanisms' predictions diverge enough ($\sim$$15\%$, above the noise) to separate by conductivity alone, and the D-optimal XRD-texture points fall at $\Pi=0$ and at that same minimum ($\Pi^{*}\approx0.22$), where two measurements at texture noise $0.05$ pin the alignment rate to $\mathrm{std}\approx0.33$ (from a prior width of $\sim$8). Conductivity and texture thus point to the same critical compression, so the campaign should sample densely around the minimum rather than uniformly.
\item \textbf{Closed-loop extrusion control (GranuGoIn deliverable).} Implement the differentiable soft sensor on a continuous semi-dry granule-extrusion setup (TU Braunschweig / VARTA) to actively control the calendering line loads based on upstream solid-content variations, validating set-point-resolved inline sensing on a moving web beyond the static calendered sheets studied here.
\item \textbf{Acceptance criteria.} (i)~Calibrated residuals at or below the measurement uncertainty at all $\Pi$; (ii)~the pressure split consistent with Eq.~\eqref{eq:knudsen} within the $\beta$ band; (iii)~the $\varphi(\Pi)$--adhesion correlation as predicted; (iv)~the image-derived contact fraction consistent with the thermally fitted one; (v)~the graphite $S(\Pi)$ from texture rises and saturates near the conductivity minimum while the cathode shows no conduction-relevant texture change; and, where the joint multi-property measurements are taken, (vi)~the joint closure reproduces the measured through-plane tortuosity $\tau_\perp(\Pi)$ and electronic resistivity $\sigma_{\mathrm{el}}(\Pi)$ from the thermally fixed latents within their measurement scatter. Failure of (ii) falsifies the Knudsen term as implemented; failure of (iii) or (iv) falsifies the contact interpretation and demotes $\varphi(\Pi)$ to a curve fit; failure of (v) falsifies the reorientation mechanism and returns the anode minimum to a contact-only account with an unexplained low $\lams$; failure of (vi) refutes the shared-latent coupling and demotes the joint closure to a thermal-only model.
\end{enumerate}

\section{Bridging processing routes: wet-slurry and dry-processed electrodes}
\label{sec:results:dryproc}
A second generalization is across the manufacturing route. Dry-processed electrodes omit the solvent, fibrillating a dry binder into a network and densifying the powder between heated rolls. They build a different microstructure (fine fibrils, no drying-induced binder/carbon-black migration) from wet slurry-cast coatings \cite{gandert2026}. Yet from the closure's standpoint both are the same particle bed, and the route merely selects the contact-parameter regime, where dry directly-calendered NMC622 falls on the \emph{same} $\leff(\varepsilon)$ trend as wet electrodes of identical composition, extending it to lower porosity \cite{gandert2026}. The route-dependent physics localizes in $\varphi_0$ and $\lamb$, and because dry processing removes the migration step it removes a dominant source of the interface heterogeneity the bridge/core inverse (Sec.~\ref{sec:results:qc}) is built to flag. Reorientation transfers unchanged (heated-roll densification aligns flakes), so the $S(\Pi)$ prediction should hold for dry anodes too, a test the present cathode comparison cannot yet make.

\section{Coupling to ionic transport: the thermal inverse as a tortuosity proxy}
\label{sec:results:coupling}
The reorientation mechanism has a consequence beyond heat. The same flat flake alignment $S$ that turns the through-plane thermal path onto the $c$-axis also obstructs the through-plane \emph{ionic} path. Ebner et al.\ \cite{ebner2014} show by X-ray tomography and effective-medium theory that horizontal flake alignment raises through-plane tortuosity, worsening with compaction. As $S$ rises at $\varepsilon=0.35$, the model gives a through-plane thermal conductivity decrease from $3.1$ to $1.3\WmK$ and an ionic tortuosity increase from $1.7$ to $8$, inside the measured flake-graphite range \cite{ebner2014}, and the two normalized through-plane transports are positively correlated ($r=+0.85$). A microstructure that removes heat poorly through-plane is therefore also rate-limited, and the levers align. Spheroidized or vertically aligned graphite \cite{billaud2016} improve both, whereas calendering's flat alignment degrades both. Because the differentiable inverse recovers $\lams^{\perp}\!\to\!S$ and $S$ sets the tortuosity, a fast thermal reading can serve as a proxy for the rate-limiting through-plane tortuosity.

\begin{figure}[tb]\centering
\includegraphics[width=\textwidth]{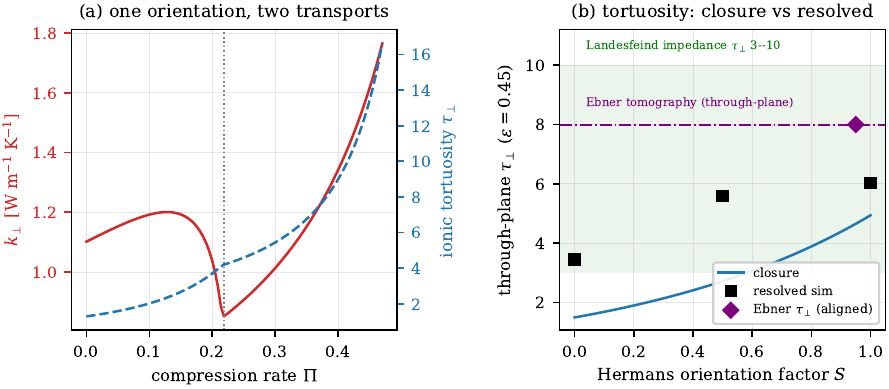}
\caption{Thermal--ionic coupling via flake orientation. (a)~The single orientation evolution $S(\Pi)$ drives two transports at once: the through-plane thermal conductivity $k_\perp$ (the reorientation u-shape, dotted line at its minimum) and the through-plane ionic tortuosity $\tau_\perp$ (rising). (b)~The predicted $\tau_\perp(S)$ (closure) and the resolved pore-scale simulation (squares) against explicit measured graphite-anode values: the Landesfeind impedance range $3$--$10$ (shaded) \cite{landesfeind2016} and the Ebner tomography through-plane $\tau_\perp\approx8$ for aligned graphite (dash-dot, diamond) \cite{ebner2014}. Both model curves fall in the measured band.}
\label{fig:coupling}
\end{figure}

A resolved pore-scale solve on oriented-platelet structures checks the ionic side independently of the closure. The through-plane tortuosity rises from $3.4$ to $6.0$ as the flakes align ($S:0\!\to\!1$). This is an explicit, not merely qualitative, match to the measured graphite-anode values (Fig.~\ref{fig:coupling}b). X-ray tomography gives a through-plane $\tau_\perp$ up to $\approx8$ for aligned graphite near $40\%$ porosity \cite{ebner2014}, and impedance places graphite-anode $\tau_\perp$ in the $3$--$10$ range \cite{landesfeind2016}. Both the closure's $\tau_\perp(S)$ ($1.7\to8$) and the resolved simulation ($3.4\to6.0$) fall inside this band, with the high-alignment end reaching Ebner's $\approx8$. Regressing the resolved data recovers an effective through-plane Bruggeman exponent of $\sim$$2.6$ rising with alignment, reproducing from flake geometry the high exponents measured for graphite anodes \cite{ebner2014,landesfeind2016}. Two independent generative microstructures agree on this exponent ($\sim$$2.6$--$2.9$), so it is a property of the flake morphology. Since the analytic $\tau_\perp$ above uses the lower sphere-baseline exponent $\alpha(S)=1.5+1.5S$, the quoted tortuosities are conservative relative to the published graphite-anode data.

Finally, because the closure and the coupling are differentiable end-to-end, the design trade-off can be optimised directly. A target through-plane tortuosity (a rate requirement) sets the maximum densification, and multi-start gradient ascent through the coupled model returns the calendering setpoint that maximises energy density and heat removal within it. We use exact-gradient ascent here because the closure is cheap and differentiable, but the same coupled model is the natural objective for \emph{multi-objective Bayesian optimisation}. Where the objective is instead a costly experiment (a measured cell, a wet electrochemical test) rather than a microsecond closure call, a Gaussian-process surrogate over the energy/heat/rate trade-off would map the Pareto front in a handful of measurements, connecting this thermal--ionic setpoint problem to the broader data-efficient materials-optimisation literature.

\paragraph{A joint closure for three transport channels, and how many latents it needs.} The natural next step is a single joint closure $F(\varepsilon,\Pi)=(\lambda_{\mathrm{th}},\,D_{\mathrm{ion}},\,\sigma_{\mathrm{el}})$ that drives thermal, ionic, and electronic transport from shared calendering-induced latents (module \texttt{joint\_transport.py}). The instructive finding is that \emph{one} latent is not enough. Flake orientation $S(\Pi)$ governs the anisotropic pair, the through-plane solid thermal conductivity and the ionic tortuosity, and rises monotonically. It therefore cannot produce the \emph{non-monotonic} electronic resistivity that Lain et al.\ \cite{lain2023} measure (a rise under light calendering as carbon pathways shear, then a fall as they re-form shorter). Electronic conduction rides the carbon/contact network, governed by the contact latent $\varphi(\Pi)$, the same dip-then-recover that makes the thermal u-shape. The minimal joint closure thus carries \emph{two} shared latents, both already fixed by the thermal calibration, each governing a pair of channels, $S(\Pi)\!\to\!\{$thermal-solid, ionic$\}$ and $\varphi(\Pi)\!\to\!\{$thermal-contact, electronic$\}$. With the latents fixed on thermal data alone, $F$ reproduces the \emph{sign and shape} of all three channels, the thermal u-shape, the ionic tortuosity rising into the measured $3$--$10$ band, and the non-monotonic electronic resistivity, with no ionic or electronic fitting. This is a trade-off-geometry unification (monotonicity, sign, location of extrema), not a calibrated prediction of absolute ionic and electronic magnitudes. The carbon network's electronic recovery scale in particular is representative rather than thermally pinned, and direct ionic and electronic measurements on these sheets (Sec.~\ref{sec:experiments}) would promote it to a calibrated multi-property closure. The two-latent structure is the two-network decomposition of Sec.~\ref{sec:results:mechanisms} expressed as one forward map across three transport properties.

The thermal--ionic link is part of a larger, two-network picture, and the conductive additive is what makes it two. Through-plane transport in a composite electrode runs on two interpenetrating networks. The first is the oriented active skeleton, governed by $S$. The same graphite basal/$c$-axis anisotropy the reorientation mechanism invokes for heat is the origin of graphite's electronic anisotropy (both ride the $\mathrm{sp}^2$ basal planes), so as the flakes align the skeleton's through-plane thermal, ionic, and \emph{intrinsic} electronic paths all degrade together (pairwise $r=0.85$--$0.97$). The second is the carbon-binder domain (CBD), governed by the contact fraction $\varphi$ and physically identical to the closure's contact term. Conductive carbon black is the percolating \emph{electronic} highway, while the binder is an electronic insulator that supplies adhesion and the thermal contact bridges (the shear-damaged $\varphi_b$ term of the contact-mechanics account, Section~\ref{app:design}). The CBD dominates electron transport in the cathode, whose active material is a poor electronic conductor and whose calendering signature in our decomposition is contact evolution alone. The two networks respond \emph{oppositely} to calendering for electrons. Alignment would turn the skeleton's electronic path onto the $c$-axis, but compaction tightens the carbon network ($\varphi$ rising), and the CBD wins, so electrode electronic conductivity \emph{rises} with calendering even as the skeleton's heat path and the ionic path degrade. The thermal u-shape combines the skeleton-orientation downswing and the CBD-contact upturn, matching the reorientation/contact decomposition of Sec.~\ref{sec:results:mechanisms}.

\begin{figure}[tb]\centering
\includegraphics[width=0.46\textwidth]{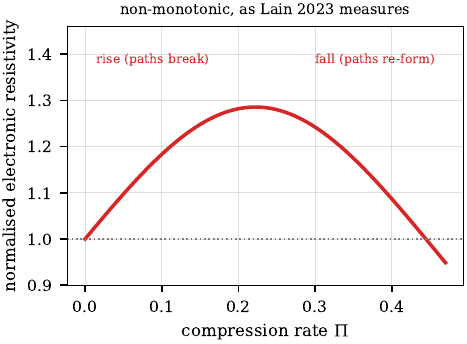}
\caption{Electronic-resistivity trade-off geometry. Treating the through-plane electron path as carried by the carbon/contact network, $\rho_{\mathrm{elec}}\sim1/(\sigma_0+\varphi(\Pi))$ with the contact term's damage-then-recovery shape gives a non-monotonic resistivity (rise then fall), the sign and shape Lain et al.\ \cite{lain2023} measure for graphite-anode volumetric resistivity. One contact mechanism predicts both the thermal u-shape and the electronic non-monotonicity; absolute scale is not calibrated.}
\label{fig:resistivity}
\end{figure}

Calendering does not move all transports together. Instead, it trades heat removal and rate in the oriented active skeleton against electronic percolation in the carbon-binder domain. The differentiable thermal inverse, by recovering $S$ with porosity and the contact state $\varphi$, reads out the geometric state of both networks from one measurement. The rate-limiting ionic tortuosity follows from $S$, while electronic conduction follows mainly from the carbon-binder domain. A literature-anchored additive model quantifies this trade-off. Electronic conduction switches on at a carbon percolation threshold ($\sim$$2$\,vol\%) and is lifted further by calendering as the contact network tightens, so the optimal additive loading is the least carbon that clears the threshold with margin, and extra carbon costs energy density and increases the tortuosity penalty. These network claims regarding electronic transport remain explicitly qualitative. Absolute electronic conductivity and the scaling of the CBD contribution still need to be verified against direct measurements or overlayed against resistivity data. Lain et al.\ \cite{lain2023} provide the relevant comparison data. Cathode electronic conduction is dominated by conductive carbon, the carbon network percolates only above a threshold loading and is anisotropic, graphite single-crystal electronic anisotropy reaches $\sim$$10^{4}$, and calendering moves anode volumetric resistivity non-monotonically, rising under light calendering and falling again at heavier densification as pathways break and re-form. That non-monotonicity is the \emph{same} contact-network damage-then-recovery our thermal contact term $\varphi(\Pi)$ encodes. Treating the through-plane electronic path as carried by the carbon/contact network ($\rho_{\mathrm{elec}}\sim1/(\sigma_0+\varphi(\Pi))$) reproduces the measured rise-then-fall (Fig.~\ref{fig:resistivity}; module \texttt{resistivity\_overlay.py}), so one mechanism predicts both the thermal u-shape and the electronic-resistivity non-monotonicity. This is a trade-off-geometry match in sign and shape, not a calibration of absolute electronic conductivity, which this thermal dataset cannot provide.

\section{Extended limitations}
\label{sec:limitations}
The five scope-defining limitations are stated in the main text where they constrain interpretation (Sec.~\ref{sec:discussion}, ``Scope and limitations''); the complete inventory follows.

\noindent(1)~All conclusions about the \emph{process dependence} rest on one public dataset: four families, three recipes, one laboratory, one metrology chain. We are not aware of a second through-plane thermal series that resolves the calendering trajectory as finely (27 states), so the campaign of Sec.~\ref{sec:experiments} remains necessary before the contact-evolution claims can be generalized. The absolute scale is independently corroborated. Our calibrated coating conductivities sit within the $\sim$$0.1$--$3.5\WmK$ range reported for porous electrodes across independent laboratories and methods \cite{maleki1999,richter2017,werner2017,loges2016,liu2024}, and the electrolyte conductivity we adopt ($0.18\WmK$) brackets independently measured values \cite{werner2017,loges2016}. The anode/cathode ordering varies between studies and metrologies, so we make no universal claim about it. What no published dataset yet supplies, and the campaign does, is a second calendering-resolved thermal series with paired texture and contact-area metrology.

\noindent(2)~With four parameters on six to eight points per family, $\lams$ and $\varphi_0$ are partially confounded at $\Pi=0$ (the lumped bridge scalar of assumption~A4, fitted within the family-fixed set of assumption~A5), so we report the profile-likelihood valley and grouped quantities rather than over-interpreting point estimates. (3)~The quadratic $\varphi(\Pi)$ is phenomenology anchored to adhesion observations, not derived contact mechanics, and $b$ is unconstrained outside each family's tested compression range. The reorientation account of the anode (Sec.~\ref{sec:results:mechanisms}) is mechanistic and bounded, but on conductivity data alone it is degenerate with a contact-only fit of equal accuracy. It is adopted on parsimony, the $c$-axis anchor for the otherwise-unexplained low $\lams$, and the anode/cathode contrast, and it remains a hypothesis until the texture measurement of Sec.~\ref{sec:experiments} confirms it.

\noindent(4)~The in-plane flake conductivity $\lambda_a$ in Eq.~\eqref{eq:reorient} is a literature value. The $S(\Pi)$ trend is robust to it over $\lambda_a\in[100,600]\WmK$ (anode MAPE $1.9$--$3.0\%$) and to the $c$-axis value over its single-digit range, but the absolute as-coated $S_0$ is not. Starting ideal affine compression from a random as-coated state rotates the flakes only weakly ($S\approx0.1$ at the conductivity minimum), below the pure-reorientation fit's $S\approx0.94$. Graphite coatings, however, are textured before calendering: XRD studies report substantial as-coated basal texture that rises with calendering \cite{malifarge2017,baade2017}. Baade et al.\ measure a degree of preferred $(002)$ orientation of $76.5\%$ already in the as-coated state \cite{baade2017}, which anchors $S_0$ in the $0.5$--$0.7$ band quantitatively rather than by assertion. Starting the same affine compression from $S_0=0.5$--$0.7$ reaches $S\approx0.6$--$0.75$ at the minimum, so the robust conclusions need only partial alignment. The residual gap to the pure fit's $0.94$ is the share of the downswing assigned to contact evolution, and the texture measurement remains the way to apportion the two.

\noindent(5)~Until that measurement is available, structural uncertainty is propagated by Bayesian model averaging over three closures: contact, reorientation, and their mixture. It is negligible on the calibration data, where the variants agree to a few percent, but the between-mechanism spread grows to $\sim$$30\%$ in extrapolation (a $90\%$ band of $2.8$--$4.6\WmK$ at $\Pi=0.55$). We weight the three members equally because, on six to eight states, WAIC weights are themselves unstable and over-concentrate on negligible in-sample differences.

\noindent(6)~Stack-derived coating values lump the coating/collector contact resistance; free-standing-film metrology would separate it. (7)~Electrode porosities extrapolate the porosity range of the base closure's packed-bed validation, so the extrapolation is empirical. (8)~The sphere-pack geometry is wrong for stretched-film separators; we claim bounds there, not point predictions. (9)~Anisotropy enters, per assumption~A2, through the reorientation of Sec.~\ref{sec:model:reorient}, and trace conservation lets the closure predict the in-plane direction and the anisotropy ratio as well, but the ZBS unit cell under-counts lateral percolation. A fully anisotropic, spheroidal-particle unit cell that captures in-plane percolation exactly remains future work, and $S(\Pi)$ is inferred from conductivity rather than measured here. (10)~Wet-state predictions assume fully saturated pores; partial wetting during electrolyte filling is untreated.

\noindent(11)~\emph{Measurement-method dependence.} The calibration data were acquired by laser-flash analysis (LFA); the methodological study of Gandert \cite{gandert2026} reports that on these electrodes LFA and the guarded-hot-plate (GHP) method disagree by a factor of $2.3$ to $4.0$, with LFA reading higher because it excludes inter-sample contact resistance. The difference is not a pure scale factor. The u-shape is most pronounced in the LFA data, while the contact- and pressure-sensitive GHP shows a weaker or flatter trend. Which method best reflects an intrinsic bulk coating value is still under investigation. Our calibration and the reorientation interpretation therefore pertain to the LFA-derived apparent through-plane coating conductivity under source subtraction, and the fitted $\lams$ and $\varphi_0$ are method-scoped effective quantities. We turn this caveat into a measured result by calibrating the closure directly on Gandert's GHP data for the same electrodes (in air, at $2$ and $6$\,bar; module \texttt{ghp\_calibration.py}). Three findings follow. The closure fits the GHP states to $\sim$$6.5\%$, comparable to its LFA fit, so its form is metrology-robust. The absolute scale is $1.8\times$ (anode) to $2.4\times$ (cathode) lower than LFA, at or modestly below the lower end of the directly reported $2.3$--$4.0\times$ method gap (the recalibrated whole-closure shift need not equal the raw point-by-point method ratio), and localises in the contact term for the cathode and in the solid conductivity for the anode. Tellingly, the GHP places the graphite through-plane solid conductivity at the $c$-axis floor ($\lams\approx5\WmK$, bound-active), the same anomalously low value the reorientation mechanism derives from the LFA data, now corroborated by an independent method. What remains LFA-led is only the sharpness of the u-shape minimum. The texture measurement of Sec.~\ref{sec:experiments} is independent of any thermal method. The interfacial resistance that GHP includes and LFA omits is the contact physics our 18650 stack check treats explicitly, so measuring one recipe by both methods, as the campaign specifies, pins the scale factor directly. A third, contact-free metrology brackets the question from the other side: photothermal deflection spectroscopy (PDS) measures coating diffusivity with no inter-layer contact resistance \cite{loges2016}. Independent cell-level work confirms the stakes of this contact term. Inter-layer contact pressure controls the stack thermal conductivity, and realistic pack pretension cuts the aging-induced conductivity loss from $\sim$$23\%$ to a few percent \cite{kovachev2021}. Gandert also finds that Oehler's closure with isotropic graphite ($139\WmK$) overpredicts the anode by up to $30\%$ \cite{gandert2026}, consistent with the lower, $c$-axis-weighted through-plane conductivity our reorientation mechanism implies.

\noindent(12)~All predictions are at room temperature with constituent properties fixed at $\sim$$300\,$K; assumption~A1 (radiation and pore convection negligible) is an ambient-temperature scoping, since radiative transfer grows as $T^{3}$ and should be re-checked for elevated-temperature operation. Cell-level measurements show a mild temperature dependence (about $-1\%$ per K \cite{steinhardt2020}), so absolute predictions should be re-evaluated for elevated-temperature operation, although the pore-gas and Knudsen terms already carry their own explicit temperature dependence through Eq.~\eqref{eq:knudsen}.

\section{Model-variant extensions tested}
\label{app:variants}
Several extensions of the calibrated closure were proposed and tested (module \texttt{revision\_experiments.py}); we report which are identifiable and beneficial on the present dataset and which require measurements it does not contain. The recurring theme is the identifiability ceiling of Section~\ref{sec:results:errors}.

\paragraph{Where the calendering dependence lives: $\varphi(\Pi)$ versus $\lambda_b(\Pi)$.} Because only the as-coated product $G_0=\varphi_0\lamb$ is identified at $\Pi=0$, one might carry the process dependence in the bridge conductivity $\lambda_b(\Pi)$ at constant $\varphi_0$ rather than in the area fraction $\varphi(\Pi)$. Refitting with a quadratic $\lambda_b(\Pi)$ and fixed $\varphi_0$ reproduces the u-shape \emph{worse} (in-sample MAPE rising from $1.8/5.4\%$ to $5.3/10.4\%$ for the thin/thick anode, and comparably for the cathodes), which shows the damage-then-recovery signature is carried more naturally by the contact \emph{area} than by the bridge conductivity. The $G_0$ confound is thus an as-coated, single-state degeneracy and does not extend to the calendering trajectory, which selects $\varphi(\Pi)$.

\paragraph{Pore-size distribution in the Knudsen term.} Replacing the single mean-field pore size with a volume-weighted lognormal distribution (geometric standard deviation $1.6$) shifts the rarefied-gas channel by $+5$ to $+9\%$ for the micrometre-pored electrodes but by $\sim$$+66\%$ for the $\sim$$100$\,nm separator. The sign and magnitude depend on the assumed pore-network weighting (parallel/volume weighting raises the gas conductivity by emphasising the coarse tail; a series/bottleneck weighting would lower it), so the correction is model-dependent. Because the gas channel is a minor share of an electrode's through-plane conduction, the effect on electrode $\leff$ is small; for the gas-dominated separator it is large. The two-pressure campaign (Sec.~\ref{sec:experiments}) is the clean experimental discriminator, as already argued.

\paragraph{Orientation-coupled contact $\varphi\,(1-gS)$.} Letting contact effectiveness fall as flakes align improves the in-sample graphite fit (MAPE $1.9\to1.2\%$ thin, $5.2\to2.1\%$ thick, fitted $g\approx0.3$--$0.7$), reaching the contact-quadratic level. But it adds a parameter to an anode fit that is already degenerate (Sec.~\ref{sec:results:mechanisms}): it is a third comparably-fitting member of the contact-versus-reorientation competition, not a resolution of it, and the same-sheet XRD/SEM measurements remain the discriminator.

\paragraph{Variants requiring measurements the dataset lacks.} Six further ideas are well posed but not identifiable from single-pressure, monotonic-compression, conductivity-only data, and are specified as campaign targets (Sec.~\ref{sec:experiments}, Section~\ref{app:campaign}) rather than fitted here. (i)~Splitting the lumped $\varphi$ into intra-coating bridge and coating--collector interface conduction (assumption~A4) needs the paired guarded-hot-plate-plus-adhesion/SEM metrology. (ii)~A coupled (rather than competing) reorientation-plus-contact graphite model is already carried as the three-member Bayesian average (Sec.~\ref{sec:results:bayes}), whose mechanism weight stays under-determined until same-sheet texture is added. (iii)~Path dependence or hysteresis in $\varphi$ cannot be tested without load-cycling (unload) states. (iv)~A wetted-bridge occupancy term for partial saturation needs filling measurements; the existing solid-bridge term already damps the wet-to-dry sensitivity that Sec.~\ref{sec:results:knudsen} flags. (v)~A state-dependent gas--solid accommodation coefficient $\beta$ is degenerate with the other parameters at the single ambient pressure of the dataset and needs the two-pressure ladder. (vi)~A PINN built on an explicit interface-physics formulation, rather than the generic field inverse of Section~\ref{app:engine}, is gated on real flash-thermography frames.


\begin{thebibliography}{99}\small

\bibitem{bandhauer2011}
T.~M. Bandhauer, S.~Garimella, T.~F. Fuller,
A critical review of thermal issues in lithium-ion batteries,
\emph{J. Electrochem. Soc.} \textbf{158} (2011) R1--R25.

\bibitem{richter2017}
F.~Richter, S.~Kjelstrup, P.~J.~S. Vie, O.~S. Burheim,
Thermal conductivity and internal temperature profiles of Li-ion secondary batteries,
\emph{J. Power Sources} \textbf{359} (2017) 592--600.

\bibitem{steinhardt2021}
M.~Steinhardt et al.,
Low-effort determination of heat capacity and thermal conductivity for cylindrical 18650 and 21700 lithium-ion cells,
\emph{J. Energy Storage} \textbf{42} (2021) 103065.

\bibitem{oehlerdiss}
D.~Oehler,
\emph{Bestimmung der thermischen Transporteigenschaften por\"oser Elektroden von Lithium-Ionen Batterien},
Dissertation, KIT Scientific Publishing, Karlsruhe, 2021. DOI 10.5445/KSP/1000136047.

\bibitem{steinhardt2022}
M.~Steinhardt, J.~V. Barreras, H.~Ruan, B.~Wu, G.~J. Offer, A.~Jossen,
Meta-analysis of experimental results for heat capacity and thermal conductivity in lithium-ion batteries: A critical review,
\emph{J. Power Sources} \textbf{522} (2022) 230829.

\bibitem{steinhardt2020}
M.~Steinhardt, E.~I. Gillich, M.~Stiegler, A.~Jossen,
Thermal conductivity inside prismatic lithium-ion cells with dependencies on temperature and external compression pressure,
\emph{J. Energy Storage} \textbf{32} (2020) 101680.

\bibitem{vadakkepatt2016}
A.~Vadakkepatt, B.~Trembacki, S.~R. Mathur, J.~Y. Murthy,
Bruggeman's exponents for effective thermal conductivity of lithium-ion battery electrodes,
\emph{J. Electrochem. Soc.} \textbf{163} (2016) A119--A130.

\bibitem{malifarge2017}
S.~Malifarge, B.~Delobel, C.~Delacourt,
Quantification of preferred orientation in graphite electrodes for Li-ion batteries with a novel X-ray-diffraction-based method,
\emph{J. Power Sources} \textbf{343} (2017) 338--344.

\bibitem{baade2017}
P.~Baade, M.~Ebner, V.~Wood,
Rapid, non-invasive method for quantifying particle orientation distributions in graphite anodes,
\emph{J. Electrochem. Soc.} \textbf{164} (2017) E348--E351.

\bibitem{ebner2014}
M.~Ebner, D.-W. Chung, R.~E. Garc\'ia, V.~Wood,
Tortuosity anisotropy in lithium-ion battery electrodes,
\emph{Adv. Energy Mater.} \textbf{4} (2014) 1301278.

\bibitem{billaud2016}
J.~Billaud, F.~Bouville, T.~Magrini, C.~Villevieille, A.~R. Studart,
Magnetically aligned graphite electrodes for high-rate performance Li-ion batteries,
\emph{Nat. Energy} \textbf{1} (2016) 16097.

\bibitem{landesfeind2016}
J.~Landesfeind, J.~Hattendorff, A.~Ehrl, W.~A. Wall, H.~A. Gasteiger,
Tortuosity determination of battery electrodes and separators by impedance spectroscopy,
\emph{J. Electrochem. Soc.} \textbf{163} (2016) A1373--A1387.

\bibitem{gandert2023}
J.~C. Gandert, M.~M\"uller, S.~Paarmann, O.~Queisser, T.~Wetzel,
Effective thermal conductivity of lithium-ion battery electrodes in dependence on the degree of calendering,
\emph{Energy Technol.} \textbf{11} (2023) 2300259.

\bibitem{burheim2010}
O.~S. Burheim, P.~J.~S. Vie, J.~G. Pharoah, S.~Kjelstrup,
Ex situ measurements of through-plane thermal conductivities in a polymer electrolyte fuel cell,
\emph{J. Power Sources} \textbf{195} (2010) 249--256. (GDL through-plane $k$ vs.\ compaction; compiled with other GDL data by E.~Pfrang et al., IntechOpen, 2011.)

\bibitem{guo2010}
G.~Guo, B.~Long, B.~Cheng, S.~Zhou, P.~Xu, B.~Cao,
Three-dimensional thermal finite element modeling of lithium-ion battery in thermal abuse application,
\emph{J. Power Sources} \textbf{195} (2010) 2393--2398.

\bibitem{jeon2011}
D.~H. Jeon, S.~M. Baek,
Thermal modeling of cylindrical lithium ion battery during discharge cycle,
\emph{Energy Convers. Manage.} \textbf{52} (2011) 2973--2981.

\bibitem{yue2017}
F.~Yue, G.~Zhang, J.~Zhang, J.~Lin, K.~Jiao,
Numerical simulation of transport characteristics of Li-ion battery in different discharging modes,
\emph{Appl. Therm. Eng.} \textbf{126} (2017) 70--80.

\bibitem{chen2017}
C.-F. Chen, A.~Verma, P.~P. Mukherjee,
Probing the role of electrode microstructure in the lithium-ion battery thermal behavior,
\emph{J. Electrochem. Soc.} \textbf{164} (2017) E3146--E3158.

\bibitem{gandert2026}
J.~C. Gandert,
\emph{Production-related Characterization of the Thermal Transport Properties of Battery Electrodes via Laser Flash Analysis and Guarded Hot Plate Method},
Dissertation, Karlsruher Institut f\"ur Technologie (KIT), 2026 (m\"undliche Pr\"ufung 15 April 2026).

\bibitem{guk2026}
E.~Guk, M.~Faraji-Niri, M.~Farhadi~Tolie, J.~Marco,
Dataset of ultrasonic frequency-domain signals from lithium-ion battery electrodes before and after calendering,
\emph{Data in Brief} \textbf{64} (2026) 112433.

\bibitem{hidalgo2023}
M.~Faraji-Niri, M.~F.~V. Hidalgo, et al.,
A dataset of calendered {NMC622} electrodes and cells: roll temperature, porosity and loading with adhesion, density and microscopy,
\emph{Data in Brief} \textbf{52} (2023) 109798.

\bibitem{oehler2021}
D.~Oehler, P.~Seegert, T.~Wetzel,
Modeling the thermal conductivity of porous electrodes of Li-ion batteries as a function of microstructure parameters,
\emph{Energy Technol.} \textbf{9} (2021) 2000574.

\bibitem{sangros2017}
C.~Sangr\'os Gim\'enez, B.~Finke, C.~Schilde, L.~Frob\"ose, A.~Kwade,
Numerical simulation of the behavior of lithium-ion battery electrodes during the calendaring process via the discrete element method,
\emph{Powder Technol.} \textbf{349} (2019) 1--11.

\bibitem{vishwakarma2015}
V.~Vishwakarma, C.~Waghela, Z.~Wei, R.~Prasher, S.~C. Nagpure, J.~Li, F.~Liu, C.~Daniel, A.~Jain,
Heat transfer enhancement in a lithium-ion cell through improved material-level thermal transport,
\emph{J. Power Sources} \textbf{300} (2015) 123--131.

\bibitem{zehner1970}
P.~Zehner, E.~U. Schl\"under,
W\"armeleitf\"ahigkeit von Sch\"uttungen bei m\"a\ss{}igen Temperaturen,
\emph{Chem. Ing. Tech.} \textbf{42} (1970) 933--941.

\bibitem{bauer1978}
R.~Bauer, E.~U. Schl\"under,
Effective radial thermal conductivity of packings in gas flow. Part II,
\emph{Int. Chem. Eng.} \textbf{18} (1978) 189--204.

\bibitem{vdi}
E.~Tsotsas,
Thermal conductivity of packed beds,
in: \emph{VDI Heat Atlas}, 2nd ed., Ch.~D6.3, Springer, Berlin, 2010.

\bibitem{wiener1912}
O.~Wiener,
Die Theorie des Mischk\"orpers f\"ur das Feld der station\"aren Str\"omung,
\emph{Abh. Math.-Phys. Kl. K\"onigl. S\"achs. Ges. Wiss.} \textbf{32} (1912) 509--604.

\bibitem{hashin1962}
Z.~Hashin, S.~Shtrikman,
A variational approach to the theory of the effective magnetic permeability of multiphase materials,
\emph{J. Appl. Phys.} \textbf{33} (1962) 3125--3131.

\bibitem{maxwell1891}
J.~C. Maxwell,
\emph{A Treatise on Electricity and Magnetism}, 3rd ed., Vol.~1,
Clarendon Press, Oxford, 1891.

\bibitem{eucken1932}
A.~Eucken,
Die W\"armeleitf\"ahigkeit keramischer feuerfester Stoffe: ihre Berechnung aus der W\"armeleitf\"ahigkeit der Bestandteile,
\emph{VDI-Forschungsheft} \textbf{353}, VDI-Verlag, Berlin, 1932.

\bibitem{bruggeman1935}
D.~A.~G. Bruggeman,
Berechnung verschiedener physikalischer Konstanten von heterogenen Substanzen. I,
\emph{Ann. Phys.} \textbf{416} (1935) 636--664.

\bibitem{kennard1938}
E.~H. Kennard,
\emph{Kinetic Theory of Gases},
McGraw-Hill, New York, 1938.

\bibitem{kaganer1969}
M.~G. Kaganer,
\emph{Thermal Insulation in Cryogenic Engineering},
Israel Program for Scientific Translations, Jerusalem, 1969.

\bibitem{burheim2013}
O.~S. Burheim, M.~A. Onsrud, J.~G. Pharoah, F.~Vullum-Bruer, P.~J.~S. Vie,
Thermal conductivity, heat sources and temperature profiles of Li-ion secondary batteries,
\emph{ECS Trans.} \textbf{58} (2014) 145--171 (224th ECS Meeting, Abs.~1190).

\bibitem{marconnet2018}
Y.~Sun, R.~Kantharaj, A.~Marconnet,
Characterization of thermal conductivity and thermal transport in lithium-ion batteries,
Thermal \& Fluids Analysis Workshop (TFAWS), NASA JSC, Houston, TX, 2018.

\bibitem{maleki1999}
H.~Maleki, S.~Al~Hallaj, J.~R. Selman, R.~B. Dinwiddie, H.~Wang,
Thermal properties of lithium-ion battery and components,
\emph{J. Electrochem. Soc.} \textbf{146} (1999) 947--954.

\bibitem{werner2017}
D.~Werner, A.~Loges, D.~J. Becker, T.~Wetzel,
Thermal conductivity of {Li}-ion batteries and their electrode configurations -- a novel combination of modelling and experimental approach,
\emph{J. Power Sources} \textbf{364} (2017) 72--83.

\bibitem{marconnet2024}
A.~Marconnet, S.~Herberger, S.~Paarmann, P.~Seegert, T.~Wetzel,
Impact of aging on the thermophysical properties of lithium-ion battery electrodes,
\emph{J. Power Sources} \textbf{603} (2024) 234367.

\bibitem{liu2024}
J.~Liu, L.~Wang, G.-B. Liu, L.-W. Fan,
Determination of the inherent thermal conductivity and thermal contact resistance of individual thin-layer materials in {Li}-ion batteries,
\emph{Int. J. Heat Mass Transfer} \textbf{230} (2024) 125741.

\bibitem{lain2023}
M.~J. Lain, G.~Apachitei, D.-E. Dogaru, W.~D. Widanage, J.~Marco, M.~Copley,
Measurement of anisotropic volumetric resistivity in lithium ion electrodes,
\emph{RSC Adv.} \textbf{13} (2023) 33437--33445.

\bibitem{loges2016}
A.~Loges, S.~Herberger, D.~Werner, T.~Wetzel,
Thermal characterization of {Li}-ion cell electrodes by photothermal deflection spectroscopy,
\emph{J. Power Sources} \textbf{325} (2016) 104--115.

\bibitem{kovachev2021}
G.~Kovachev, A.~Astner, G.~Gstrein, L.~Aiello, J.~Hemmer, W.~Sinz, C.~Ellersdorfer,
Thermal conductivity in aged {Li}-ion cells under various compression conditions and state-of-charge,
\emph{Batteries} \textbf{7} (2021) 42.

\bibitem{li2026}
S.~Li, C.~Lin, Y.~Tian, Z.~Tao, P.~Xie,
Layered electro-thermal modeling and self-heating optimization for large-capacity {Li}-ion batteries,
\emph{eTransportation} \textbf{28} (2026) 100544.

\bibitem{choi2026}
C.~Choi, S.-Y. Choe,
{AI}-driven optimization of {SOC}-dependent parameters for reduced order electrochemical thermal model of lithium-ion batteries,
\emph{J. Power Sources} \textbf{663} (2026) 238849.

\bibitem{mcmasters}
R.~L. McMasters, J.~V. Beck, R.~B. Dinwiddie, H.~Wang,
Accounting for penetration of laser heating in flash thermal diffusivity experiments,
\emph{J. Heat Transfer} \textbf{121} (1999) 15--21.

\end{thebibliography}
\end{document}